%% file: main.tex
\documentclass[conference]{IEEEtran}
\pdfoutput=1 % to submit to arXiv
\usepackage[hidelinks]{hyperref}

\usepackage{cite}
\usepackage{amsmath,amssymb,amsfonts}
\usepackage{algorithmic}
\usepackage{textcomp}
\usepackage{xcolor}
\usepackage{multicol}

\usepackage{enumerate}
\usepackage[linesnumbered, ruled, vlined,resetcount]{algorithm2e} % For algorithms
\usepackage{graphics}
\usepackage{balance}
\usepackage{graphicx}
\usepackage{subfig}
\usepackage{amsthm, amssymb,amsthm, amsfonts,color}
\usepackage{bm}
\usepackage{listings}
\usepackage{enumitem}
\usepackage{verbatim} % for comment -by yusong
\usepackage{booktabs}
\usepackage{color}
\usepackage{bbding} % 邮箱符号\Envelope
\input{commands}
\clearpage
\newcommand{\red}[1]{{#1}}

\newcommand{\green}[1]{\textcolor[rgb]{0.2,0.8,0.2}{#1}}
\newcommand{\gray}[1]{\textcolor{gray}{#1}} % will be deleted
\newcommand{\lframe}{\kw{Layph}}
\newcommand{\oursys}{\kw{Layph}}

\newenvironment{proofS}{
         \vspace{1ex}
        {\noindent\bf Proof sketch:\ }}{\eop\vspace{1ex}}

\newcommand{\shortcut}{shortcut\xspace}

\newcommand{\TLA}{TLA\xspace}

\def\BibTeX{{\rm B\kern-.05em{\sc i\kern-.025em b}\kern-.08em
    T\kern-.1667em\lower.7ex\hbox{E}\kern-.125emX}}

\IEEEoverridecommandlockouts

\begin{document}

%%%%%%%%%%%%%%%%%%%%%%%%
% \input{responce}
%%%%%%%%%%%%%%%%%%%%%%%%

\title{Layph: Making Change Propagation Constraint in \eat{Asynchronous} Incremental Graph Processing %through Graph Sketching
by Layering Graph}

\author{
Song Yu$^{\dagger}$, Shufeng Gong$^{\dagger \vdash}$, Yanfeng Zhang\textsuperscript{\Envelope}$^{\dagger}$\thanks{\textsuperscript{\Envelope}Yanfeng Zhang is the corresponding author.}, Wenyuan Yu$^{\S}$,  Qiang Yin$^{\ddagger}$, Chao Tian$^{\mathparagraph}$, Qian Tao$^{\S}$, \\
Yongze Yan$^{\dagger}$, Ge Yu$^{\dagger}$, Jingren Zhou$^{\S}$
% }
% \affiliation{%
%   \institution{
  \\
  \textit{
  ${\dagger}$ Northeastern University
    % \hspace{3ex}
    ${\S}$ Alibaba Group %\hspace{3ex}
    ${\ddagger}$ Shanghai Jiao Tong University %\hspace{3ex}
    $\mathparagraph$ Chinese Academy of Sciences} \\ %\hspace{3ex}
    \textit{$\vdash$ 
    % Key Laboratory of Medical Image Computing (Northeastern University), Ministry of Education
    Key Laboratory of Intelligent Computing in Medical Image of Ministry of Education, Northeastern University
    }
%  }
% }
% \affiliation{
% \institution{
    \\ 
    $\{$yusong, yanyongz$\}$@stumail.neu.edu.cn,
    $\{$zhangyf, gongsf, yuge$\}$@mail.neu.edu.cn,
    $\{$wenyuan.ywy, qian.tao, \\ jingren.zhou$\}$@alibaba-inc.com, 
    $\{$q.yin$\}$@sjtu.edu.cn,
    $\{$tianchao$\}$@iscas.ac.cn
    % }
% }
}

\eat{
Author1, Author2\footnotemark[1], %对应同一个脚注符号
Author3\footnotemark[2], %\Envelope
Author4, 
Author5\footnotemark[2],%\Envelope
}

\maketitle

%\newcommand{\thefootnote}{\fnsymbol{footnote}} %将脚注符号设置为fnsymbol类型，即特殊符号表示
%\footnotetext{\textsuperscript{\Envelope}Yanfeng Zhang is the corresponding author.} %对应脚注[1]
% \footnotetext[2]{Corresponding authors.} %对应脚注[2]

\eat{
\IEEEauthorblockN{Qiange Wang$^*$,  Xin Ai$^*$, Yanfeng Zhang\textsuperscript{\Envelope} \thanks{ Qiange and Xin contributed equally. Yanfeng is the corresponding author.}, Jing Chen, Ge Yu}

\IEEEauthorblockA{\textit{School of Computer Science and Engineering}\\\textit{Northeastern University,}
Shenyang, China \\
\{wangqiange,aixin0,chenjing\}@stumail.neu.edu.cn;
\{zhangyf,yuge\}@mail.neu.edu.cn}
}

%\maketitle

%% article.
\begin{abstract}
Real-world graphs are constantly evolving, which demands
% demanding 
updates of the previous analysis results to accommodate graph changes. By using the memoized previous computation state, incremental graph computation can reduce unnecessary recomputation. %Existing approaches capture dependencies between (intermediate or final) vertex states. 
However, a small change may propagate over the whole graph and lead to large-scale iterative computations. 
% To address this problem, we propose a layered graph framework, \lframe.  
% It divides the graph into two layers. 
To address this problem, we propose Layph, a two-layered
graph framework. 
The upper layer is a skeleton of the graph which is much smaller than the original graph, and the lower layer has some disjoint subgraphs. 
% In \lframe, the iterative computations are only performed on the upper layer skeleton and a few the subgraphs in the lower layer. 
% \lframe通过将代价高昂的迭代计算限制在少数被更新影响的子图和小skeleton上进行，有效限制了更新消息在全图范围内迭代传播，进而有效减少了增量图处理过程中的活跃边的数量，提升了增量图处理的性能。
Layph limits %expensive 
costly global iterative computations on the original graph to the small graph skeleton and a few subgraphs updated with the input graph changes. %It effectively limits the iterative propagation of update messages in the whole graph,
In this way, many vertices and edges are not involved in iterative computations, which significantly reduces the computation overhead and improves the performance of incremental graph processing.
% The vertices in subgraphs that do not change can be updated with no iterative computation. %It relies on a few independent
%small-scale local computations and a global computation on a graph skeleton so that a small number of local updates only affect a few local computations and may (or may not) trigger incremental computation on the skeleton. To support generalized graph sketching, we build a framework \oursys on top of Ingress, which is a message-driven incremental graph processing system with different memoization policies. Furthermore, \oursys is equipped with several necessary optimizations for graph sketching. 
Our experimental results %in the real-world 
show that Layph outperforms current state-of-the-art incremental graph systems by $9.08\times$ on average (up to $36.66\times$) in response time. % 这是所有系统所有算法的
\end{abstract}


\input{1introduction}

\input{2preliminary}
%\input{3auto}
\input{3overview}
% \input{3sketch_based_framework}
\input{4skeleton}
\input{4sketch_based_framework}

\input{6expr}
\input{7related}

\bibliographystyle{IEEEtran}  %声明选择的格式: IEEEtran, acm, unsrt
\bibliography{sample} %bib文件名，需要放在同一个文件夹下，否则要在filename前说明路径

\end{document}

%% file: commands.tex
\usepackage{newfloat}
\usepackage{multirow}
\DeclareFloatingEnvironment[fileext=frm,placement={!ht},name=Frame]{myfloat}
\makeatletter
\newcommand{\removelatexerror}{\let\@latex@error\@gobble}
\makeatother
\newcommand{\kw}[1]{{\ensuremath{\mathsf{#1}}}\xspace}
\newcommand{\eat}[1]{}

\newcommand{\stitle}[1]{\vspace{1.6ex}\noindent{\bf #1}}
\newcommand{\etitle}[1]{\vspace{1ex}\noindent{\underline{\em #1}}}

\newcommand{\beqn}{\begin{eqnarray*}}
\newcommand{\eeqn}{\end{eqnarray*}}

\newcommand{\ie}{\emph{i.e.,}\xspace}
\newcommand{\eg}{\emph{e.g.,}\xspace}

\newcommand{\aka}{\emph{a.k.a.}\xspace}

\newcommand{\note}[1]{\textcolor{blue}{#1}}

\newcommand{\changys}[1]{\textcolor{red}{#1}}

\newcommand{\eop}{\hspace*{\fill}\mbox{$\Box$}}     % End of proof
\newcounter{example}%[section]
\renewcommand{\theexample}{\arabic{example}}
\newenvironment{example}{
        \vspace{1ex}
        \refstepcounter{example}
        {\noindent\bf Example \theexample:}}{
        \eop\vspace{1ex}}

\newcounter{theorem}
\renewcommand{\thetheorem}{\arabic{theorem}}
\newenvironment{theorem}{\begin{em}
        \refstepcounter{theorem}
        {\vspace{1.5ex} \noindent\bf  Theorem  \thetheorem:}}{
        \end{em}\eop\vspace{1.5ex}} %\hspace*{\fill}\vspace*{1ex}}

\newtheorem{definition}{Definition}

\newcommand{\bi}{\begin{itemize}}
\newcommand{\ei}{\end{itemize}}
        {\end{itemize}} %\vspace{-0.5ex}}

\newcommand{\PageRank}{\kw{PageRank}}
\newcommand{\SSSP}{\kw{SSSP}}

\newcommand{\PHP}{\kw{PHP}}

\newcommand{\PR}{\kw{PageRank}}
\newcommand{\BFS}{\kw{BFS}}
\newcommand{\G}{G}
\renewcommand{\Pr}{\kw{PR}}

\newcommand{\summ}{\kw{sum}}

\newcommand{\ra}{\rightarrow}

\newcommand{\A}{\mathcal{A}}
\newcommand{\AS}{\vec{\mathcal{A}}}
\newcommand{\PAR}{\mathcal{P}}

\newcommand{\SK}{\mathcal{S}}
\newcommand{\I}{\mathcal{I}_{\mathcal{A}}}

\newcommand{\GE}{\mathcal{F}} %%propogation function
\newcommand{\AGG}{\mathcal{G}} %aggregation function

\newcommand{\Paragraph} [1] {\smallskip\noindent{\bf #1. }}

\newcounter{lemma}%[section]
\renewcommand{\thelemma}{\arabic{theorem}}
\newcommand{\sr}[1]{\blue{#1}}

%% file: 1introduction.tex
\vspace{-0.16in}
\section{Introduction}
\label{sec-intro}
\vspace{-0.4ex}

Iterative graph algorithms, \eg single source shortest path (\SSSP) and \PageRank, have been widely applied in many fields \cite{tang2009efficient, page1999pagerank,cho2005page, ahn2013heterogeneous, berger2008graph}. %, such as social network analysis \cite{tang2009efficient}, web ranking \cite{page1999pagerank,cho2005page}, commodity recommendation \cite{ahn2013heterogeneous}, and biology \cite{berger2008graph}. %In response to the increasing need for processing large-scale graphs, great efforts have been put into developing highly-optimized big graph processing systems, such as Gemini~\cite{zhu2016gemini}, Grape~\cite{fan2017grape}, Galois \cite{}, and GraphScope \cite{}. 
Real-world graphs are continuously evolving with structure changes, 
where vertices and edges are inserted or deleted arbitrarily. These changes are 
usually small, %and relative calm \cite{gabert2021elga}
\eg there were 6.4 million articles on English Wikipedia in $2021$~\cite{wiki}, but the average number of new articles per day was only $580$. 
Traditional classical graph processing systems \cite{malewicz2010pregel,  gonzalez2012powergraph, wang2020powerlog, zhang2013maiter, zhang2011priter, zhu2016gemini, fan2017grape} have to recompute the updated graph from scratch. 
However, there are considerable overlaps between computations before and after the graph updates.
It is desirable to adopt \textit{incremental graph computation} to cope with these small changes efficiently. 
That is, a batched iterative algorithm is applied to compute the result over the original graph $\G$ till convergence, 
and then an incremental algorithm is used to adjust the result in response to the input changes $\Delta \G$ to $\G$. 
\eat{
By making use of the memoized previous results, incremental computation can reduce unnecessary recomputation and is often more efficient than restarting computation on the updated graph from scratch. The benefits of incremental computation have led to the development of many incremental graph processing systems, such as Tornado~\cite{shi2016tornado}, KickStarter~\cite{vora2017kickstarter}, GraphBolt~\cite{mariappan2019graphbolt}, Ingress \cite{gong2021ingress}, DZiG \cite{DZiG}, and RisGraph~\cite{feng2021risgraph}.}

\begin{figure}
\vspace{-0.10in}
		\centering
		\subfloat[\SSSP]{\label{fig:update-prop_sssp}
		\includegraphics[width=0.48\linewidth]{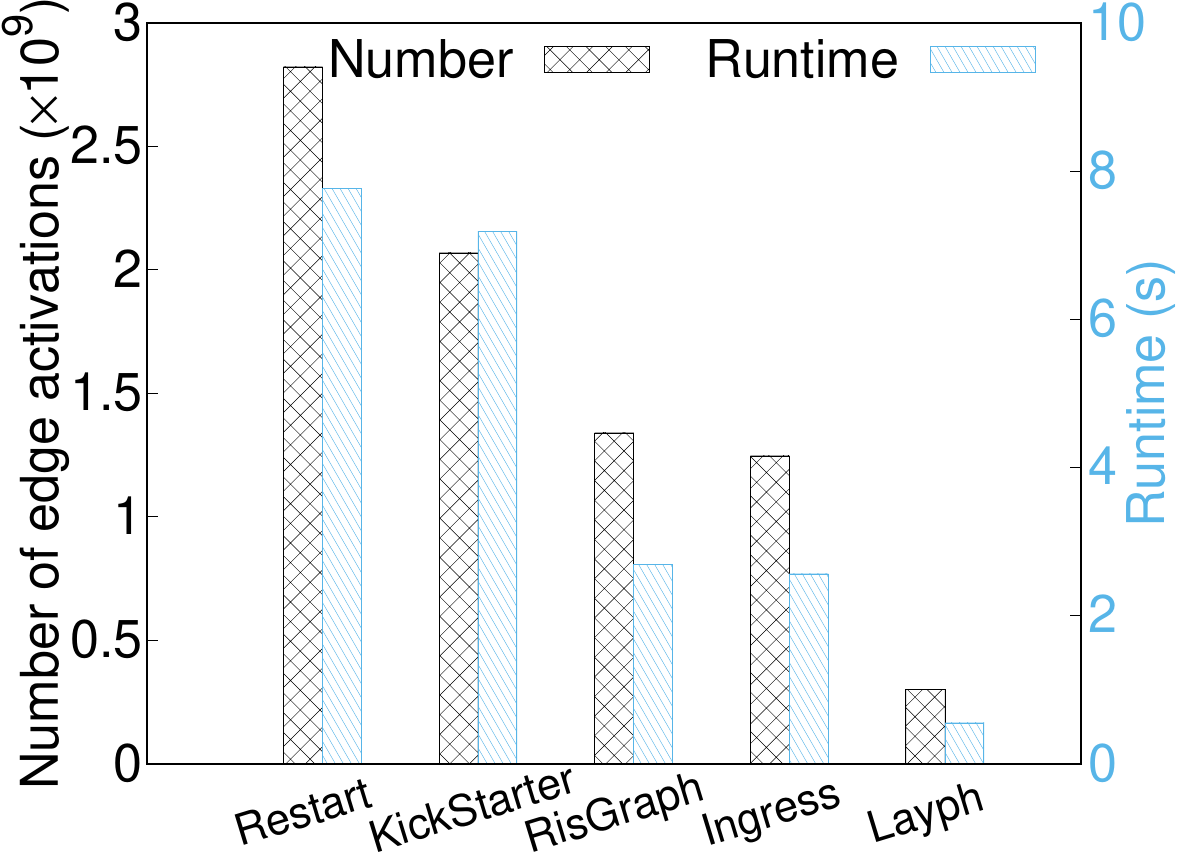}}
		\subfloat[\PageRank]{\label{fig:update-prop_pr}
		\includegraphics[width=0.48\linewidth]{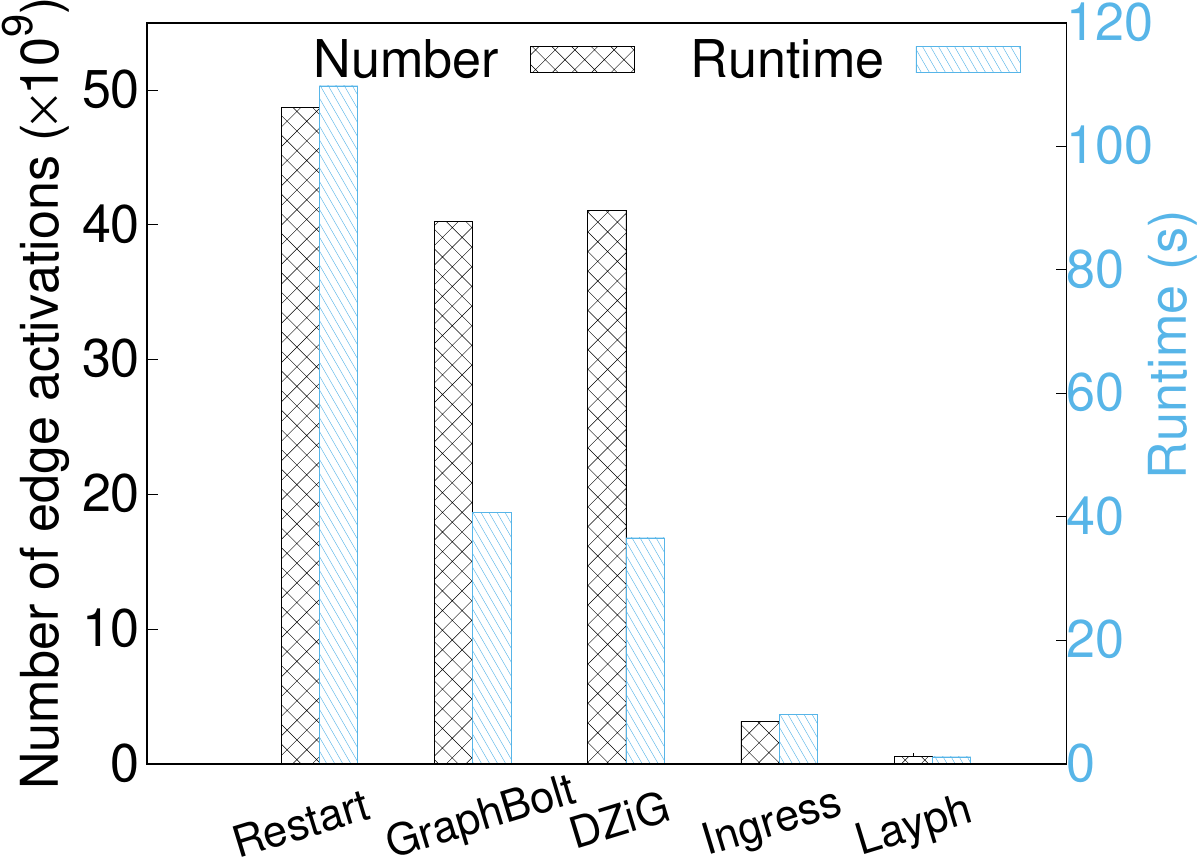}}
		\vspace{-0.05in}
		\caption{Number of edge activations and runtime of different incremental graph processing systems for SSSP and PageRank.}
		\label{fig:update-prop}
		\vspace{-0.17in}
\end{figure}

% By making use of the memoized previous iterative computation state, \eg intermediate vertex states or messages, the incremental computation can reduce unnecessary recomputation. 
The incremental graph computation can reduce unnecessary recomputation by using the memoized iterative computation state, \eg intermediate vertex states or messages. 
{The benefits of incremental graph computation have led to the development of many incremental graph processing systems, such as KickStarter \cite{vora2017kickstarter}, GraphBolt \cite{mariappan2019graphbolt}, Ingress \cite{gong2021ingress}, DZiG \cite{DZiG}, and RisGraph \cite{feng2021risgraph}. They memoize (intermediate or final) vertex states and organize them in a data structure that captures result dependencies, such as a tree (for critical path) \cite{vora2017kickstarter}, \cite{feng2021risgraph} or a multilayer network (for per-iteration dependencies) \cite{mariappan2019graphbolt}, \cite{DZiG}. With such a structure, the update of a vertex/edge will be propagated for updating the memoized intermediate/final states of vertices iteratively.
However, an upstream vertex/edge update may incur a large number of updates to the downstream vertex/edge states in existing incremental
graph processing systems}. That is, a small change may propagate over the entire graph and lead to large-scale iterative computations. %an upstream vertex/edge update can be propagated to the vertices/edges that are affected by the update. %may still incur a large number of updates of the downstream vertex/edge states.

With 5000 random edge updates on the UK graph (see Table \ref{tab:data} for details), we run \SSSP and \PR on five state-of-the-art incremental graph processing systems (KickStarter \cite{vora2017kickstarter}, GraphBolt \cite{mariappan2019graphbolt}, DZiG \cite{DZiG}, RisGraph \cite{feng2021risgraph}, and Ingress \cite{gong2021ingress}) and a Restart system that starts computations on the updated graph from scratch. The number of edge activations and runtime of these systems are reported in Figure \ref{fig:update-prop}. Even though the amount of updates is small ($|\Delta G|/|G|=5000/(9.4\times 10^8) 
 \textless 0.001\%$), 
these updates propagate widely and iteratively on the graph, resulting in a large number of edge activations in some systems, which is almost approaching the number in restarting iterative computations.

\begin{figure*}[tbp]
%\hspace{-0.3in}
\vspace{-0.2in}
    \centering
    \includegraphics[width=6.5in]{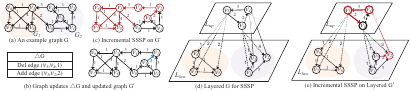}
    \vspace{-0.05in}
    \caption{An illustrative example of a layered graph for incremental \SSSP, where $v_0$ is source vertex, and $G_1$ and $G_2$ are two dense subgraphs. 
    %In (a), an original graph and the updates that take place on the graph.
    The dashed lines are the shortcuts between two vertices, through which the shortest distance from a vertex to another one can be directly obtained.
    The number labeled on each link represents the weight of the edge or \shortcut. 
    In (c) and (e), the red links or circles represent the activated edges/shortcuts or vertices involved in iterative computations. 
    %\TLA constructed from the original graph, where $H_0$ consists of two sugraphs, distinguished by different color regions. And $H_1$ consists of the skeleton of the original graph. 
    %In (b) and (e), the red circles and links define the scope of iterative \SSSP computation.
    }
    \label{fig:contraction_SSSP_inc}
    \vspace{-0.2in}
\end{figure*}

We empirically illustrate this observation with an example in Figure \ref{fig:contraction_SSSP_inc}. Figure \ref{fig:contraction_SSSP_inc}b shows an updated graph based on graph $G$, 
where the edge ($v_3$, $v_4$) is deleted and a new edge ($v_3$, $v_2$) is added.
%by deleting edge $(v_3, v_4)$ and adding edge $(v_3, v_2)$.
As shown in Figure \ref{fig:contraction_SSSP_inc}c, 
when running \SSSP, 
% these systems \note{which systems?} 
existing incremental graph processing systems \cite{vora2017kickstarter, gong2021ingress,feng2021risgraph}
activate most of the vertices and edges. % as shown in Figure \ref{fig:contraction_SSSP_inc}(b). 
%And because 
%And the activated vertices may be updated several times with iterations, 
As the iteration proceeds, the activated vertices may be updated several times, \eg $v_4$ and its downstream vertices are updated twice due to the update messages from $v_2$ at different iterations. 
% 我想表达的是不同时刻到达v3，故没有聚合消息，以及后面的G0都被激活两次，所以至少得体现不同时刻没有聚合吧
%reaches $v_3$ through $v_1 \ra v_3$ and through $v_1 \ra v_2 \ra v_3$ update time is different, resulting in vertices $\{v_3, v_4, v_5, v_6, v_7\}$ are updated twice to reach the final convergence. 
\eat{When running \PageRank, %they
these systems eventually activate all vertices and edges on the graph, and iterate many times to reach a convergence state, 
\eg Ingress \cite{gong2021ingress} requires 6 iterations over the entire graph to converge.
}
% 最后三句，逻辑好像有gap

%上一章举例说明了增量图计算在执行增量计算时可能会引起大面积的点和边参与迭代计算

% new version
\stitle{Challenge}. 
Based on the above observations and illustration, %discussions, 
we can see that very small graph changes can also lead to a large number of iterative computations, even on the basis of previous memoized vertex/edge states provided by incremental processing systems. The main reason is that, in real-world graphs, vertices are either directly or indirectly connected in several hops, which makes it hard to constrain the affected area. The native properties of real graphs fundamentally limit the effectiveness of incremental graph computation. Is it possible to reconstruct the graph structure to boost the performance of incremental graph computation?

%eat

\stitle{Intuition}.
% Before introducing our solution, 
% We first provide the intuition behind it. 
In incremental graph computation, the messages initiated by graph updates are propagated iteratively to update the states of vertices.
When an update message enters into a \textit{dense subgraph} %(\aka \textit{densely connected subgraph}) 
from \emph{entry vertices}, 
%where the entry (resp. exit) vertex of the subgraph has an incoming edge (resp. an outgoing edge) outside the subgraph, 
a large number of internal vertices and edges within the subgraph will be activated and involved in the iterative computation.  
The incoming messages probably require many iterations to get out of this dense subgraph from \emph{exit vertices}. 
A natural idea is to extract the \emph{entry} and \emph{exit} vertices of the dense subgraph, and construct shortcuts between them to propagate messages directly through the dense subgraph, which can
%Therefore, using shortcuts to send the external messages received by entry vertices directly to exit vertices can 
avoid the activations of a large number of internal vertices and edges. 
As shown in Figure 2d, we extract the entry vertex $v_0$ and exit vertex $v_4$ of $G_2$ and construct a shortcut between them. Then the messages can be propagated directly through $G_2$ via the shortcut. %from $v_0$ to $v_4$.  
Furthermore, %in asynchronous iterative computation mode, 
we construct a shortcut between the entry vertices and the internal vertices in each subgraph. %we can delay the propagation of messages until enough messages are received. Thus 
The entry vertices can accumulate the incoming messages % that should be sent to internal vertices, 
and eventually assign them to the internal vertices at a time via the shortcuts. As shown in Figure 2d, after $v_5$ accumulates all incoming messages, $v_5$ will send the update messages to $v_6$-$v_8$ at a time. %These are the intuitions behind our layered graph structure, where the dense subgraphs locate at the lower layer, the entry/exit vertices constructing a graph skeleton locate at the upper layer, and the shortcuts connect the entry vertices (at the upper layer) to the internal vertices (at the lower layer) or exit vertices.
% 这也就是说，每个子图内仅仅需要入口和出口顶点参与全局迭代计算，而其它的内部顶点仅仅的收敛状态仅仅通过入口顶点及其shortcut来直接获得。
In this way, only the entry and exit vertices of subgraphs and outliers participate in the global iterative computations. %The other internal vertices in each subgraph can be directly calculated from the entry vertices via shortcuts.

% 很自然我们可以原图上的顶点按照是否参与全局迭代计算划分为两层。
\stitle{Our Solution}.
%To answer this question, %solve this problem, 
{Based on the above intuition, we propose an incremental graph processing framework by \underline{lay}ering the  gra\underline{ph}, \lframe.}
As shown in Figure 2d, %\lframe divides the graph into two layers, the upper layer ($L_{up}$) and the lower layer ($L_{low}$). % is constructed by some key vertices, and the other vertices are \green{in the lower layer.} 
%The upper layer 
\lframe divides the graph into two layers, the upper layer ($L_{up}$) %that participates in the global iterative computation 
and the lower layer ($L_{low}$). $L_{up}$ is a skeleton of the original graph $G$ composed of the boundary vertices of subgraphs and outliers, 
the size of which is much smaller than that of $G$. $L_{low}$ is composed of some disjoint subgraphs. 
% The vertices between $L_{up}$ and $L_{low}$ are connected by shortcuts (dashed lines) or edges. 
Vertices on $L_{up}$ and vertices on $L_{low}$ are connected by shortcuts (dashed lines) or edges. 
After $G$ is updated by $\Delta G$,
we first update the layered graph accordingly.
The revision messages are generated and propagated only within the subgraphs on $L_{low}$ that are updated by $\Delta G$. 
As shown in Figure 2e, revision messages are generated from $v_3$ and are propagated within $G_2$. %needs to iteratively propagate the update message internally, for example $G_2$, the internal vertex state of the subgraph will be updated.
Then the messages are uploaded to $L_{up}$, \eg the messages are propagated from $v_2$ to $v_4$ in Figure 2e. The global iterative computations are performed on $L_{up}$. % and the vertices on $L_{up}$ will be updated. 
Compared with 10 edges participating in the iterative computation in Figure 2c, only 2 edges/shortcuts are involved on $L_{up}$ in Figure 2e. Therefore, the global iterative computations on $L_{up}$ are much faster than that on graph $G'$. 
Finally, the updates are assigned to the other subgraphs on $L_{low}$, \eg $G_2$. Vertex states are updated directly through shortcuts without iterative computations.
%Therefore, \lframe mainly aims to replace the expensive large-scale iterative computations on the original graph with the small-scale iterative computations on the skeleton. 
We can see that \lframe performs iterative computations only on the upper layer small skeleton 
and a few subgraphs (on $L_{low}$) that are %affected by the updates defined in 
updated by $\Delta G$. Most vertices and edges on $L_{low}$ are not involved in iterative computations. 
Thus \lframe is able to accelerate the incremental graph computation efficiently.

To sum up, we make %have made 
the following contributions.% in this paper.
%分层增量计算框架，提出了一个分层框架来减少参与迭代计算的边和点的数量，从消息传播的角度建模了增量计算和分层计算，以及计算的高效性。。
%设计了一种图数据分层方法和不同层之间shortcut自动构建和更新策略。分层要求第一层中的数据尽量少，才能减少开销，然而第一层的构建并不是trival的，shortcut的构建则直接影响着结果的正确性，且不同算法shortcut构建不同。
%高效的执行引擎的实现。

\begin{itemize}[leftmargin=*]
    \item \textbf{Layered Incremental Graph Processing Framework}. 
    % old 版本
    % It constraints the incremental iterative computation to a small area, \ie the changed subgraphs and the  skeleton. Therefore, incremental graph processing can be accelerated by replacing the large-scale iterative computation on original graph with small-scale local computation on changed subgraphs and iterative computation on skeleton.
    It constraints the incremental iterative computation to a small area, \ie a few subgraphs affected by the graph update and a small skeleton, 
    %It effectively limits the iterative propagation of update messages in the whole graph, 
    thus greatly reducing the number of edge activations in the iterative process. %, and improving the performance of incremental graph processing.
    (Section \ref{sec:overview} \& \ref{sec:inc_framework})
    \item \textbf{Effective Skeleton Extraction and Automated Shortcut Deduction}. 
    We design an effective skeleton extraction method that reduces the size of the skeleton by replicating vertices. Based on the input vertex-centric program, our proposed framework can 
    deduce the weight of shortcuts automatically. % and propose a construction and update scheme that can adapt to different graph analysis algorithms. 
    (Section \ref{sec:sketch})
    
    \item \textbf{High-Performance Runtime Engine}. %Based on the incremental graph processing framework of the two-layer architecture, 
    We implement our runtime engine {\lframe} based on Ingress \cite{gong2021ingress} and Alibaba's libgrape-lite \cite{libgrape}. Comparing with current state-of-the-art incremental graph processing systems, {\lframe} can achieve 3.13-15.82$\times$ speedup over Kickstarter \cite{vora2017kickstarter}, 2.54-8.49$\times$ speedup over RisGraph \cite{feng2021risgraph}, 2.99-36.66$\times$ speedup over GraphBolt \cite{mariappan2019graphbolt},  2.92-32.93$\times$ speedup over DZiG \cite{DZiG}, and 1.06-7.22$\times$ speedup over Ingress \cite{gong2021ingress}. (Section %\ref{sec:system} \&
    \ref{sec:expr})
\end{itemize}

%% file: 2preliminary.tex
\section{Preliminaries}
\label{sec:prelim}
%想办法把accumulation这个事给说出来。先说GAS，然后说相等的时候可以转换成GF的形式，对于哪些可以转换powerlog已经做了一些研究，本文不对此深入探索。当可以转换成accumulation的形式时，则每个顶点的状态由所有收到的消息组成。且可以异步执行。
This section provides the necessary preliminaries for iterative graph computation and incremental graph computation. % and sketch. %sketch是不是应该放在后边说？这里最好还是说一些增量的东西。

\subsection{Iterative Graph Computation}
\label{sec:prelim:async-algo}

%eat

Given an input graph $\G = (V, E)$, where $V$ is a finite set of vertices and $E \subseteq V \times V$ is a set of edges. %, whose weights are $w_{u,v}$ for each edge $(u,v)\in E$ in a weighted graph or consistent value 1 in an unweighted graph. 
The weight of each edge $(u,v)\in E$ is $w_{u,v}$ in a weighted graph or a consistent value 1 in an unweighted graph. 
In general, an iterative graph algorithm $\A$ that executes in an accumulative model, includes two types of operations, \ie \textit{message generation} $\GE$ and \textit{message aggregation} $\AGG$ \cite{wang2020powerlog, zhang2013maiter, rahman2020graphpulse}. %and \textit{vertex update} $\F$, can be abstracted in a \textit{vertex-centric} format as follows \cite{malewicz2010pregel, gonzalez2012powergraph}.
\begin{equation}\label{eq:iterv}
\begin{aligned}
  &m^i_{u,v} = \GE\big(m_u^{i-1}, w_{u,v}\big),\\
  %m^i_v&=\AGG\big(\{m^i_{u,v}|(u,v)\in E\}\big),\\
  &x^i_v = \AGG\big(x^{i-1}_v, \{m^i_{*,v}|(*,v)\in E\}\big).
  %x^i_v &= \F\big(x^{i-1}_v, m^i_v\big).
\end{aligned}
\end{equation}
where $m^{i-1}_u=\AGG(\{m^{i-1}_{*,u}|(*,u)\in E\})$.

The message generate operation $\GE$ applied on each vertex $u\in V$ prepares the message $m_{u,v}$ for each outgoing edge $(u,v)$ based on %an initial \textit{root message} value 
the aggregation of received message $m_u$ and the edge weight $w_{u,v}$. The aggregation operation $\AGG$ is applied on each destination vertex $v$. It first aggregates the messages that terminate at $v$ to obtain a new message $m_v$, then % Furthermore, $\AGG$ is used to 
aggregates the old vertex state $x_v$ and the aggregated message 
$m_v$ to update the vertex state 
$x_v$. The two-step process is applied iteratively till convergence (when vertex states become stable). To sum up, an iterative graph computation can be expressed as $\A$$=$$(\GE, \AGG, X^0, M^0)$ where $\GE$ and $\AGG$ are the operations that specify the algorithm logic, and $X^0$$=$$\{x_v^0$$\mid$$v\in V\}$ and $M^0$$=$$\{m_v^0$$\mid$$v\in V\}$ are the initial values of vertex states and root messages respectively. A graph computation on the input graph $G$ can be denoted as $\A(G)$.

%Note that not all graph algorithm can be executed asynchronouslly and expressed as euqation \ref{eq:iterv}. 
Suppose $\A$ can be executed asynchronously, then it can be expressed as Equation (\ref{eq:iterv}) naturally, such as \SSSP. %Single Source Shortest Path (\SSSP). 
Otherwise, %we should rewrite the synchronous algorithms as an accumulative mode and execute them asynchronously,
the synchronous algorithms should be rewritten in accumulative mode and executed asynchronously,
such as PageRank. There are some efforts \cite{zhang2013maiter, wang2020powerlog} that rewrite a synchronous algorithm in asynchronous accumulative mode.

\begin{example}
\label{exa-program}
We show two example algorithms.

\etitle{(a) \SSSP}.  \SSSP computes the shortest distance from a given source $s$ to all vertices in a directed and weighted graph $\G$. $\A$ is 
represented as follows
\begin{itemize}
    \item $\GE(m_u, w_{u,v}) = m_u + w_{u,v}$; \ \ \ \
    % \item 
    $\AGG = \kw{min}$;
    \item $x_v^0 = 0$ if $v=s$, otherwise $x_v^0 = +\infty$;
    \item $m_v^0 = 0$ if $v=s$, otherwise $m_v^0 = +\infty$.
\end{itemize}
Here the state $x_v$ of $v$ indicates the shortest distance from source $s$ to $v$
and $w_{u,v}$ represents the length of the edge $(u,v)$. Initially, we have $x_v^0=m_v^0=0$ for $v=s$, and $x_v^0=m_v^0=+\infty$ for all $v \neq s$. 
\eat{Based on the above tuples, the \SSSP can be represented as follows.
\begin{equation}
\begin{aligned}
    m^i_{u,v} &= m^{i-1}_u + w_{u,v},\\
    m^i_v&=\kw{min}\big(\{m^i_{u,v}|(u,v)\in E\}\big),\\
    x^i_v &= \kw{min}(x^{i-1}_v, m^i_v),
\end{aligned}
\end{equation}
}%eat
Each vertex $u$ generates and sends a message $m_{u,v}$ to each neighbor $v$, which represents the current shortest distance from the source. Each destination vertex $v$ aggregates the messages from its incoming neighbors %$u$ 
and updates its state $x_v$ by $\kw{min}$. The algorithm terminates when the shortest distance values of all vertices are not changed.

\etitle{(b) \PageRank}.
\PageRank computes the set of ranking scores $\{\Pr_v = d \times \summ_{(u, v) \in E} \Pr_u / N_u + (1 - d) \mid v\in V\}$. Here $d$ is a constant damping factor and $N_u$ denotes the number of outgoing neighbors of $u$. %As opposed to 
Different from the %standard 
original \PageRank algorithm that exploits the power method, an asynchronous \PageRank algorithm~\cite{zhang2013maiter} that has been proved to be equivalent to the original \PageRank can be represented as follows
%The four tuples of asynchronous \PageRank are shown as follows.
\begin{itemize}
    \item $\GE(m_u, w_{u,v}) = m_u\times d/N_u$; \ \ \ \
    % \item 
    $\AGG = \summ;$ % ()$; %\sum
    % \item $x_v^0 = 0$, $\forall v\in V$;
    % \item $m_v^0 = 1-d$, $\forall v\in V$.
    \item $x_v^0 = 0$, $\forall v\in V$; \ \ \ \  $m_v^0 = 1-d$, $\forall v\in V$.
\end{itemize}
Intuitively, each vertex $v$ uses its state $x_v$ to store its \PageRank score. Initially, we have $x_v^0 = 0$ and $m_v^0 = 1-d$ for all $v \in V$. 
\eat{Based on the above tuples, the \PageRank can be represented as follows.
\begin{equation}
\begin{aligned}
    m^i_{u,v} &= m^{i-1}_u \times d/N_u,\\
    m^i_v&=\sum_{(u,v)\in E}m^i_{u,v},\\
    x^i_v &= x^{i-1}_v + m^i_v.
\end{aligned}
\end{equation}
}%eat
%Each time a vertex $u$ converts the received message $m_u$ to $m_u \times d / N_u$ and propagates it to all its neighbors $v$. 
Every time when a vertex $u$ receives a message $m_u$, it will send $m_u \times d / N_u$ to each neighbor $v$.  
Each neighbor $v$ aggregates the messages from its incoming neighbors by $\summ$ and updates its state by accumulating the aggregated messages. The algorithm terminates when all vertex states are stable. %over two consecutive iterations. 
%until $x_v$ is stable.
\end{example}

Equation (\ref{eq:iterv}) defines the vertex-centric format of asynchronous iterative computation. 
%from which 
On this basis, we can define a set-based iterative computation as follows

\begin{equation}\label{eq:iter}
\begin{aligned}
\vspace{-0.08in}
%\overline{M}^i &= \GE(M^{i-1});\\
M^i &= \GE(M^{i-1});\\
%M^i &= \AGG(\overline{M}^i); \\
X^i &= \AGG(X^{i-1}\cup M^i).
\end{aligned}
\vspace{-0.05in}
\end{equation}
$X=\{x_v\mid v\in V\}$ is the set of vertex states. $M^0=\{m^0_v\mid v\in V\}$ is the set of root messages of each vertex and $M^{k\neq 0}=\{m^k_v\}$ %is the set of received messages. %$\overline{M}=\{m_{u,v}\mid (u,v)\in E\}$ 
is the set of generated messages on all edges. %A slight meaning change of $\GE$ and $\AGG$ in set-based format should be noticed. 
It should be noticed that these are slight meaning changes of $\GE$ and $\AGG$ in set-based format. 
$\GE$ is the message generate operation with edge information embedded so it only needs a single parameter $M$. $\AGG$ is the group-by aggregator (group by vertex id). Based on this set-based computation, the vertex states set $X$ after $n$ iterations is
\begin{equation}
\vspace{-0.05in}
\begin{aligned}
\label{eq:iterresult}
X^n=&\AGG\Big(X^0\cup (\AGG\circ \GE)(M^0)\cup\ldots\cup (\AGG\circ \GE)^n(M^0)\Big)\\
=&\AGG\Big(X^0\cup \bigcup_{k=1}^{n}(\AGG\circ \GE)^k(M^0)\Big),
\end{aligned}
\vspace{-0.05in}
\end{equation}
where $\AGG\circ \GE(\cdot)=\AGG(\GE(\cdot))$ and $(\AGG\circ \GE)^k$ denotes $k$ applications of $(\AGG\circ \GE)$.

\eat{
\stitle{The property of asynchronous iterative computation}
%G只是一个聚集操作，因此最后聚集也行，所以
\begin{equation}\label{eq:gfg}
    GFGF=GFF
\end{equation}
%异步计算中，可以优先计算一部分，然后进行全局同步得到的相同的结果。
\begin{equation}\label{eq:ff}
    G_V F_{E_1} F_{E_2}
\end{equation}
}

\Paragraph{Message Passing's Perspective} 
From message propagation's perspective, the final state $x_v$ 
of each vertex $v$ 
is obtained by accumulating the messages $M^0$ initiated from all vertices transferred along different paths. In each iteration, \ie one time application of $\GE$ and $\AGG$, a message is processed and split into several messages from a vertex to its direct neighbors (under the effect of $\GE$). The messages received from different incoming neighbors are aggregated into one message (under the effect of $\AGG$), which will be propagated again in the next iteration. At the same time, the aggregated message is applied to the vertex state (under the effect of $\AGG$). This is exactly the process described in Equation (\ref{eq:iterv}). %This message propagation defined by iterative graph computation motivates our %graph sketching approach. 
%layered graph framework.

%Equation (\ref{eq:iterresult}) shows the result of iterative computation. The function $\GE$ generates messages and propagates them along edges. The function $\AGG$ aggregates messages for next round propagation and accumulates them to vertex states. From the final destination's perspective, the final state $x_v$ of a vertex $v$ depends on the messages $M^0$ initiated from all vertices transferred along different paths. These messages are transformed by $\AGG$ and $\GE$ during propagation, but nonetheless they are arrived at the final destination vertex. This message propagation defined by iterative graph computation motivates our graph sketching approach. 

\vspace{-0.05in}
\subsection{Incremental Graph Computation} 

Given an iterative graph computation $\A$ and its incremental counterpart $\I$,
the problem of incremental computation arises when the input graph $\G$ is updated with $\Delta \G$. Let $\A(\G)$ denote the output of %a batch graph algorithm $\A$ applied on the old graph $\G$. 
an old graph $G$ with the effect of batch graph algorithm $\A$. 
The inputs of incremental computation include %$\G$, 
$\A(\G)$ and graph updates $\Delta \G$. % to $\G$. 
Then we have 
\begin{equation}
    \I(\A(\G), \Delta G)=\A(\G \oplus \Delta \G)
\end{equation}
It means that the incremental computation $\I(\A(\G), \Delta G)$ that is performed based on the old result $\A(\G)$ and the graph updates $\Delta \G$ is expected to output $\A(\G \oplus \Delta \G)$, where $\G \oplus \Delta \G$ denotes applying the updates $\Delta \G$ to $\G$. It is noticeable that the incrementalization scheme $\I$ is algorithm-specific and is deduced from its original algorithm $\A$.

The input batch update $\Delta \G$ consists of a set of {\em unit updates}.  To simplify our discussion, we consider the insertion or deletion of a single edge as a unit update in a sequence, which can simulate certain modifications. For instance, each change to an edge weight can be considered as deleting the edge and followed by adding another edge with the new property. 
% Vertex updates are dual edge updates and should be handled. 
The incremental computation $\I$ will identify the changes to the old output $\A(\G)$ and make corrections of the previous computation in response to $\Delta \G$.

\Paragraph{Message Passing's Perspective} From Equation (\ref{eq:iterresult}) we know that the input changes will affect the message propagation since both $\GE$ and $\AGG$ are correlated with the graph structure, and as a result, will change the final vertex states. 
% 边的添加、删除、更新将会导致原结果不正确。由于边的更新可以理解为删除一条旧边和添加一条新边，故下面我们仅仅分析边的删除和添加两种情况。当从原图$G$中删除一条边时，那么原来计算中沿着该边发出的消息对于新图$G+\Delta G$而言就是非法的。增量计算时通过生成一条回收消息来回撤它的影响。当在原图$G$中中添加一条边时，在新图$G+\Delta G$上应该沿着该边需要发送的消息在原计算中没有发送，因此该消息是缺失的。所以增量计算时需要通过生成一条补偿消息来replay它的影响。
\eat{So the addition and deletion of edges will affect the result on $G$. %the original graph $G$. 
% Since the update of an edge can be understood as deleting an old edge and adding a new edge, we only analyze the deletion and addition of an edge below. 
When an edge is deleted from $G$ %the original graph $G$
, then the messages sent along the edge in the original computation are invalid for the new graph. %$G \oplus \Delta G$. 
When an edge is added to $G$%the original graph $G$
, the messages that need be sent along the edge on the new graph %$G \oplus \Delta G$ 
are not sent in the original computation, so the messages are missing. %so it is necessary to replay its influence by generating a compensation message during incremental calculation. 
}
Due to the insertion, update, or deletion of an edge, a set of messages might become \textit{invalid}, and another set of messages might be \textit{missing}. %\sr{
An old message transmitted during the run over the original graph $G$ is called {\em invalid} if the path for passing the message disappears due to input updates $\Delta \G$. A new message transferred in the run over the $\G \oplus \Delta \G$ is called {\em missing} if it did not appear in the run over $G$. %} 
In incremental computation, we should first discover all the invalid and missing messages and then perform the corrections on the affected areas of $G \oplus \Delta G$ by generating \emph{cancellation messages} (resp. \emph{compensation messages}) to retract %undo 
(resp. replay) effects of the invalid messages (resp. missing messages) %their effects 
\cite{vora2017kickstarter, gong2021ingress, mariappan2019graphbolt, DZiG}. In this paper, the cancellation and compensation messages are collectively called as \emph{revision messages}. 

% atc_stream_graph是ATC上改graphbolt那个，里面提到了回撤(retract old values)
% DZiG提到了加边补发和删边回收

%%% Local Variables:
%%% mode: latex
%%% TeX-master: "paper"
%%% End:

%% file: 3overview.tex
\section{Framework Overview}\label{sec:overview}

\begin{figure}[tbp]
%\vspace{-0.1in}
    \centering
    \includegraphics[width=3.3in]{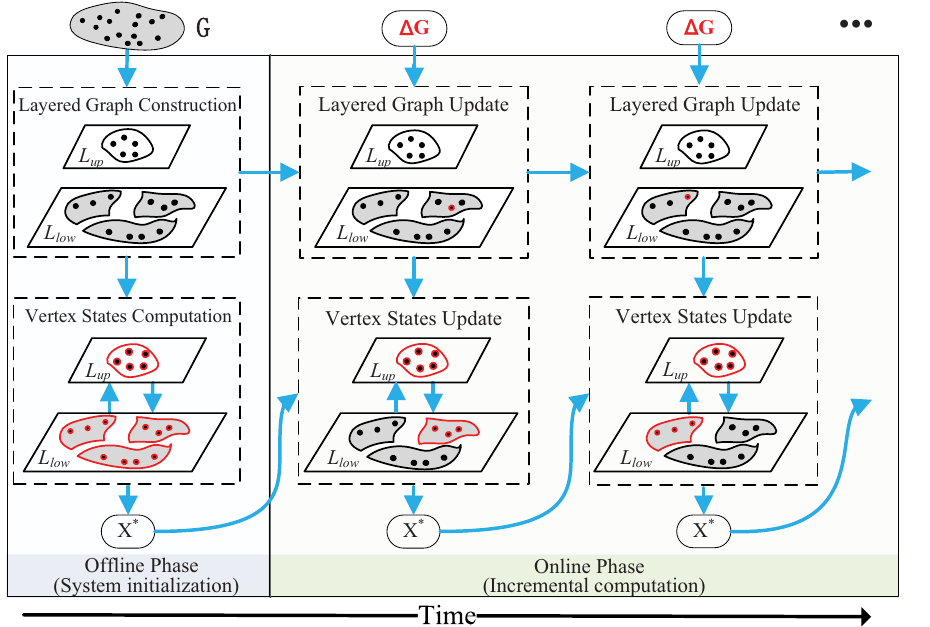}
    \vspace{-0.05in}
    \caption{Workflow of \oursys.}
    \label{fig:overview}
    \vspace{-0.25in}
\end{figure}

In this section, we first present the workflow of the layered graph framework and then analyze the benefits of %overview the workflow of incremental graph processing with 
\lframe.

%eat
\eat{The incremental global iterative computations on original updated graph can be replaced by the iterative computation on upper layer. 
Then the iterative computations are constrained to the upper layer and the subgraphs that are updated by $\Delta G$ in lower layer (used to update shortcuts and upload revision messages). %We constrain the iterative computation in the upper layer. The entry vertices of each dense subgraph cache and accumulate all the received messages, and finally assign them to  the internal vertices of the dense subgraph in the lower layer.
In this way, %the iterative propagation of revision messages on large graph is constrained within the upper layer skeleton, 
incremental graph processing is more efficient due to the fewer vertices and edges involved in iterative computations. %which is more efficient to revise the vertices state.
}
%在增量计算中，顶点的状态修正是通过修正消息【graphbolt，ingress，dzig】来修复。当修正消息进入到一个密集区时，触发大量的顶点和边参与迭代计算，从而造成大量开销。事实上，密集区域中的点收到的所有的修正消息都要经过boundary/entry点，一个很自然的想法是我先不激活内部点，而是通过shortcut直接将修正消息送到出口从而加快修正消息在图中的传播。基于此，我们将这些boundary/entry/exit点从原图中提取出来，构成一个skeleton。这样整个图就分成了两层，upper-skeleton层和lower层。当图发生改变时，首先推导出修正消息，然后将修正消息传播到上层skeleton中。并在skeleton中传播，最后，skeleton中的点收敛后，将其累积的消息发送给内部顶点，为了能够尽快收到，在skeleton中的点与下层中的点，也建立了shortcut，从而快速修正所有受影响的点的状态。

\stitle{Workflow of Layph}.
The overall workflow is illustrated in Figure \ref{fig:overview}. At the beginning of incremental graph processing, given a graph $G$, we first divide the graph into two layers. The upper layer ($L_{up}$) is the skeleton of the graph. $L_{up}$ consists of the entry/exit vertices of all dense subgraphs, vertices that are not in any dense subgraph, and the shortcuts or edges between them. % regions, the dependencies between them, and 
%the edges between different regions, which is called skeleton. 
The lower layer ($L_{low}$) is composed of all disjointed dense subgraphs. The entry vertices (on $L_{up}$) and the internal vertices (on $L_{low}$) of each dense subgraph are connected with shortcuts between $L_{up}$ and $L_{low}$. Please refer to Section \ref{sec:sketch} for the details of constructing the layered graph. %\red{Then we incrementally process the dynamic graph with a layered graph.} %感觉不对，改成下面的:
Then we perform incremental graph computations on the layered graph, which includes two steps, 
%There are two steps in \red{processing the dynamic graph $=>$ handling graph updates}, 
\romannumeral1) the layered graph update (Section \ref{sec:sketch}) and \romannumeral2) the vertex states update (Section \ref{sec:inc_framework}). 

\etitle{Layered Graph Update}.
Given a layered graph $\overline{\G}$ of an original graph $\G$, $\overline{G}$ should be updated, when $\G$ is updated by $\Delta G$. This is because the shortcuts, including the shortcuts on $L_{up}$ and the shortcuts between $L_{up}$ and $L_{low}$, 
may be changed as the graph changes. %the incremental graph processing includes two steps, the update of layered graph and the incremental update of vertices states.
The shortcut update requires iterative computations and is only performed on the subgraphs updated by $\Delta G$. Meanwhile, the shortcut update can be parallelized well as the subgraphs %localized computations 
are independent of each other.
\eat{
\begin{itemize}[leftmargin=*]
    \item \textbf{Update of the Shortcuts on $L_{up}$}. The shortcuts on $L_{up}$ connect the entry and exit vertices of each dense subgraph. The addition or deletion edge within dense subgraph may result in the change of weight of shortcuts. This update can be performed in parallel since the subgraphs are disjoint. \red{Clik Here:} %边的增删等会导致从入口到出口的shortcut的weight改变，入口出口点的消息也可能会导致在新增和删除shortcut 这里强调的并行和不相交和下面的重复，是否可以提到后面注意里面，一起说，节省篇幅,没说为啥更新，
    \item \textbf{Update of the Shortcuts between $L_{up}$ and $L_{low}$}. The shortcuts between $L_{up}$ and $L_{low}$ connect the entry vertices (on $L_{up}$) and the internal vertices (on $L_{low}$) of each dense subgraph. This update can also be performed in parallel as the subgraphs are disjoint.
\end{itemize}
}

%\etitle{两层架构更新}给定一个分层的图结构（构建分层图结构参考sec？），当图发生局部变化后，需要相应更新分层图，因为由于图结构的变化，可能会导致一些shortcut的改变，包括l1中点之间的shortcut和l1l2之间的shortcut。
%\item{shortcut-in-l1}在l1中的shortcut是l2中各个子图的入口和出口之间的shortcut，由于各个子图之间没有交集，因此shortcut的更新可以并行执行
%\item{shortcut between l1 and l2}l1和l2之间的shortcut是l1中入口（在l2中）到l1中其他顶点之间的shortcut，用来最后分配消息，或者无迭代更新l1中受影响的点，同样，由于子图之间相互隔离，因此该操作也可以并行执行

\etitle{Vertex States Update}.
When the graph changes, we first deduce the revision messages based on the memoized information \cite{gong2021ingress, vora2017kickstarter, mariappan2019graphbolt}, then propagate the revision messages on the layered graph to revise the vertex states. The incremental computation on \lframe is performed as follows. %There are three steps in the propagation of revision messages on the layered graph: uploading the messages from $L_{low}$ to $L_{up}$, iterative computation on $L_{up}$, and assigning accumulated messages to $L_{low}$.
%为了使上层图收到修正消息并且更新顶点状态，我们应该首先将下层中的消息上传到上层。
\begin{itemize}[leftmargin=*]
    \item \textbf{Revision messages upload}. %Since revision messages deduced by vertices in $L_{low}$ and the subgraphs are disjoint in $L_{low}$, the revision messages are not able to  propagate between subgraphs. 
    In order to apply the revision messages deduced by vertices on $L_{low}$ to vertices on $L_{up}$,  %propagate the revision messages to the upper layer and revise the sother subgraphs, 
    the revision messages on $L_{low}$ should be uploaded %the revision messages 
    to $L_{up}$. Similar to shortcut updates, the messages upload  can also be performed in parallel and only performed on subgraphs affected by $\Delta G$. %\red{can be performed in parallel}
    \item \textbf{Iterative computation on $L_{up}$}. After receiving the revision messages from $L_{low}$, iterative computations are performed to propagate the revision messages and revise the states of the vertices on $L_{up}$. \eat{Meanwhile, the enter vertices (on $L_{up}$) of subgraphs %in $L_{low}$ 
    % cache and accumulate all the received messages, as these messages need to be distributed to internal vertices (in $L_{low}$) later. % 误解需要缓存所有消息
    cache aggregated messages of all received messages, since these messages need to be assigned to internal vertices later.}
    \item \textbf{Revision messages assignment}. After the iterative computations on $L_{up}$, the entry vertices (on $L_{up}$) of each subgraph accumulate all the revision messages. The accumulated revision messages are assigned from entry vertices to internal vertices (on $L_{low}$) through shortcuts to revise the states of vertices on $L_{low}$.
\end{itemize}

%从上述flow可以看出，在laygraph中，迭代计算只发生在被更新的子图中和上部的skeleton。对于未受影响的子图内的点和边并不参与迭代计算。layph将迭代计算限制在上层$L_{up}$和部分被更新的子图上，有效限制了被激活边的数量，从而提升系统的相应性能。另一方面，在多次增量计算中，未更新的子图上的shortcut可以复用，节省了大量shortcut的计算成本。子图上shortcut的更新和消息上传在子图内部，可以良好的并行，避免了全局迭代计算中的aotmicconflict。

\stitle{Analysis.} From the above workflow of \lframe, we can see that the iterative computations only perform on $L_{up}$ and a few subgraphs on $L_{low}$. The vertices and edges within subgraphs that are not updated by $\Delta G$ are not involved in iterative computations, which saves significant computation overhead. 

%eat

%% file: 4skeleton.tex
\section{Layered Graph Construction And Update}
\label{sec:sketch}
\eat{
lframe的执行基于分层图，本章将介绍如何构建一个分层图根据原始图，以及如何更新分层图，当原始图数据发生变化后。
4.1 Layered Graph Construction

4.1.1 Key Vertices Extraction

4.1.2 Shortcuts Calculation

4.2 Layered Graph Update

}

% This section presents how to extract the graph skeleton and how to incrementally update it. The basic idea of graph sketching is to make global computations localized by establishing shortcuts in each dense area. The dense areas (\aka densely connected subgraphs) can be detected by existing community detection algorithms, which will be elaborated in Section \ref{sec:system:cluster}. We first provide the formal definition of \textit{entry/exit vertices} of these dense areas.
\lframe is performed on a layered graph. This section presents how to construct a layered graph and update it incrementally.

% \vspace{-0.1in}
\subsection{Layered Graph Construction}
\label{sec:layer:build}

In this section, we first introduce how to extract vertices on the upper layer. Then we provide an automated shortcut calculation method. %that can be used to calculate the shortcuts in $L_{up}$ and the shortcuts between the $L_{up}$ and $L_{low}$. 
% The basic idea is that the special areas (\aka densely connected subgraphs) can be detected by existing community detection algorithms, and then we further design a vertex replication method to obtain the specific subgraph we need, so as to extract key vertices. 
%Then, we provide an automated shortcut deduced method between vertices in perform local computations for each obtained subgraph to obtain the shortcuts in the upper layer and the shortcuts between the lower layer and the upper layer.

% 在layer graph中，关键顶点作用是为了替换原图上的顶点参与迭代计算。正如前面所介绍的关键顶点是一些区域或者子图的边界顶点，因此我们的目标就是希望通过子图上少量的边界顶点替换调子图参与计算。子图的边界顶点有入口顶点和出口顶点组成，我们首先提供形式化的定义。
%正如sec1和sec3中所述，上层是由原图中的关键点组成，那么应该提取哪些关键点才能使上层中的图数据较小，此外那么关键点应该具备可以构建shortcut后避免大量通信的目的，因此应该选择一些密集子图的边缘点。为了发现密集子图，我们采用了***方法。

%\vspace{1em}
\subsubsection{Upper Layer Vertices Extraction}
\label{sec:layer:build:key-vertex}

% It is worth noting that the construction of the lower graph $H_2$ does not depend on the specific graph analysis algorithm, in other words, $H_2$ only needs to be constructed once for different algorithms.
%To build the layered graph based on the given graph, we should first extract the key vertices to construct the upper layer. 
As we have presented the intuition behind \lframe in Section \ref{sec-intro} and the workflow in Section \ref{sec:overview}, we should extract the entry and exit vertices of the dense subgraphs and the vertices that are not in any subgraphs into the upper layer to construct the skeleton of the graph. %With the shortcut, the messages can be propagated directly from the entry vertex to the exit vertex without activating a larger number of vertices and edges inside the subgraph. 
This requires us to discover all the dense subgraphs from the original graph. %Based on the above discussion, key vertices extraction becomes \red{dense subgraph discovery $=>$ a dense subgraph discovery problem}. 
Before introducing the method of dense subgraph discovery, we first provide the formal definition of \textit{entry/exit/internal vertices} and \textit{dense subgraph}.
%In the layered graph, the key vertices are used to replace the vertices on the original graph to participate in the iterative computation. As mentioned above, the key vertices are the boundary vertices of some subgraphs, so our goal is to replace the subgraph to participate in the iterative computation through the boundary vertices of the subgraph. The boundary vertices of a subgraph are composed of entry vertices and exit vertices. 
\begin{definition}[Entry/Exit/Internal Vertices]
\label{def:entryexit}
Given a subgraph $G_i(V_i, E_i)$ of the graph $G(V, E)$, 
where $V_i\subseteq V$ and $E_i\subseteq E$.
The entry vertices of $G_i$ are defined as $V_i^{I}=\{v\mid (u,v)\in E, u\in V\setminus V_i, v\in V_i\}$, the exit vertices of $G_i$ are defined as $V_i^{O}=\{v\mid (v,w)\in E, v\in V_i, w\in V\setminus V_i\}$, and the internal vertices of $G_i$ are defined as $\hat{V}_i=V_i-V_i^I-V_i^O$.
\end{definition}
\begin{definition}[Dense Subgraph]
\label{def:dsubg}
Given an input graph $G(V, E)$, the subgraph $G_i(V_i, E_i)$ of $G$ is a dense subgraph such that the product of the number of entry vertices and that of exit vertices is smaller than the number of edges in $G_i$, \ie $|V_i^I| \times |V_i^O|< |E_i|$.
\eat{\begin{itemize}
    \item the product of the number of entry vertices and that of exit vertices is smaller than the number of edges in $G_i$, \ie $|V_i^I| \times |V_i^O|< |E_i|$, %入口点的数量*出口点的数量小于内部边的数量
    %\item the number of vertices is smaller than $K$, \ie $|V_i| < K$.
\end{itemize}}
\end{definition}

%解释为啥densesubgraph那么定义。我们在建立从入口到出口的消息时，针对每个入口，我们都需要建立对应出口应该发送的消息，因此需要建立的shortcut数量为入口数量乘以出口数量。如果图内部的边本身就非常少，即shortcut的数量大于子图的内部边的数量，那么通过shortcut传播消息的速度反而不如内部边快，因此，密集图需要满足性质（1）。此外，在极端情况下，当子图中点越多，密集子图内部边越多，但是子图太大容易导致更新和计算shortcut的代价太大。比如在计算sssp时，如果将整个图看为子图，而source点看为外部点，因此子图内的规模不宜过大。
Our definition of the dense subgraph is based on the following observation. For each entry vertex $v \in V_i^I$ of subgraph $G_i$, it is required to connect $v$ with all exit vertices using shortcuts. Thus, the number of the shortcuts in $G_i$ is the product of the number of entry and exit vertices, \ie $|V_i^I| \times |V_i^O|$. If there are only a few edges in $G_i$, \eg $|V_i^I| \times |V_i^O|> |E_i|$, then propagating messages from entry to exit vertices through the shortcuts is slower than that through the edges in $G_i$, because more shortcuts result in more message generation operations and aggregation operations. %2) If there are too many vertices in the subgraph, it will take too much time to update the shortcut from the entry vertex to the inner vertex. \red{ 2)need an experiment to verify}

\eat{后边要引出找密集子图的本质是在划分/社区发现等行为，即找与外界联系比较少的团体。为简单期间，我们并未引入新的密集子图方法，鉴于密集子图的寻找与图划分或社区发现的目的类似，而已经有非常多关于图划分和社区发现的工作，比如？？？，我们采用其中的图划分或社区发现方法来寻找关键点。但是由于密集子图的目的和图划分或社区发现等方法仍然存在区别，因此我们在采用图划分或社区发现方法时，添加了限制，1）入点和出点的数量相乘小于内部边，内部点数量小于k。

ys:
 由定义\ref{}和上述分析知道，密集子图需要具备入口顶点和出口顶点尽量少和内部边多的特性。这让我们很自然联想到社区发现的目标，即在一个图中找出外部连接少且内部连接密集的子图。显然，我们的目标与社区相似，只是我们的要求比其更严格，这意味着我们的密集子图将是社区发现获得社区的子集。故，我们修改现有的社区发现算法Louvain, 在前进行社区发现的过程中加入大小限制，并将获得的社区集合作为密集子图的候选集，最后利用密集子图的定义要求从候选集中我们需要的密集子图。
 %故，我们可以先采用社区发现的方法获得社区集合，即密集子图的候选集，然后利用密集子图的定义从候选集中找出我们需要密集子图。

（引出新的问题，解决新的问题）虽然上述的方法已经能够发现非常好的效果，但是我们发现在所找到的密集子图中存在如图\ref{}(a)所示的情况（多个入点或出点可以合并），因此我们基于寻找到的密集子图进行了二次处理，replication}

% ys
From Definition \ref{def:dsubg}, a dense subgraph requires %needs to have the characteristics of 
as many internal edges as possible and as few boundary (entry/exit) vertices as possible. %few entry vertices and exit vertices as possible 
We found that the requirements of a dense subgraph are similar to that of the community. % if the boundary vertices as few edges. $=>$ We find that the requirements for dense subgraphs are similar to those for communities if the number of edges at the boundary vertices is small.} 
The community requires as many internal edges as possible and as few external edges as possible. This inspired us to adopt a community discovery algorithm to discover dense subgraphs. %This leads us naturally to the goal of community discovery, which is to find communities with few external connections and dense internal connections in a graph. 
Therefore, in this paper, we %do not provide a novel dense subgraph discovery method, but 
use the community discovery algorithm to find dense subgraphs, such as Louvain \cite{blondel2008louvain}.
However, the community discovery algorithms may find extremely large subgraphs, which decreases the performance of our system since extremely large graphs may result in an imbalance workload. %We found that the extremely large graph may decrease the performance of \lframe. This is because updating shortcuts built on large graphs requires more time and large graphs result in serious workload imbalance. 
Therefore, we add a threshold $K$ to limit the size of each subgraph when discovering the subgraphs, \ie the number of vertices in each subgraph is smaller than $K$. %However, a value of $K$ that is too small or too large will decrease the performance of \lframe, as shown in Table \ref{tab:vary_K}. This is because too large $K$ may result in extremely large subgraphs. When $K$ is too small, some subgraphs are forcibly split. This leads to a large number of small subgraphs that have some entry and exit vertices, which increases the number of shortcuts and the size of the skeleton on $L_{up}$. The performance of \lframe will be decreased. 
As a rule of thumb, $K$ is set around 0.002-0.2\% of the total number of vertices. We also employ the work stealing technique to handle the imbalance workload, in which an idle processing thread will actively search out work for it to complete. %然而k的值太大或太小都会影响layph性能， 当k太大表明对最大subgraph没限制，从而导致系统性能下降，而当K太小时，会强制将一些subgraph划分开，从而导致大量较小子图，且每个子图中的入口和出口点相对较多，因此导致在Lup出现大量的边，从而增加增量迭代计算的时间开销。
%Furthermore, due to the difference between dense subgraph and community discovery, we add the size limitation when discovering community to get . After finding all the communities, we use Definition \ref{def:dsubg} to filter out the invalid communities. Finally, the remaining communities are the dense subgraphs we want. % 修改前
%Furthermore, 
A community may not be a dense subgraph. % according to Definition \ref{def:dsubg}. %due to the difference between dense subgraphs and communities, we first obtain a set of communities as the dense subgraphs candidate set by the community discovery algorithm. 因此我们在
%Thus, 
We select the dense subgraphs according to Definition \ref{def:dsubg}, \ie $|V_i^I| \times |V_i^O|< |E_i|$, from the dense subgraphs candidate set discovered by the community discovery algorithm. %according to Definition \ref{def:dsubg}, \red{\ie $|V_i^O| \times |V_i^I|< |E_i|$}, from the candidate set. % 这是修改后的

After discovering the dense subgraphs, the internal vertices and edges within them are put into the lower layer, the other vertices and edges \ie entry/exit vertices of subgraphs and the vertices that are not in any dense subgraphs and their edges are extracted into the upper layer. %获得密集子图后，将密集子图中的点放置在第二层而密集子图的入口点和出口点和其他链接密集子图的边和点作为重要顶点放在第一层中。
\begin{figure*}[tbp]
% \hspace{-0.3in}
\vspace{-0.3in}
    \centering
    \includegraphics[width=6in]{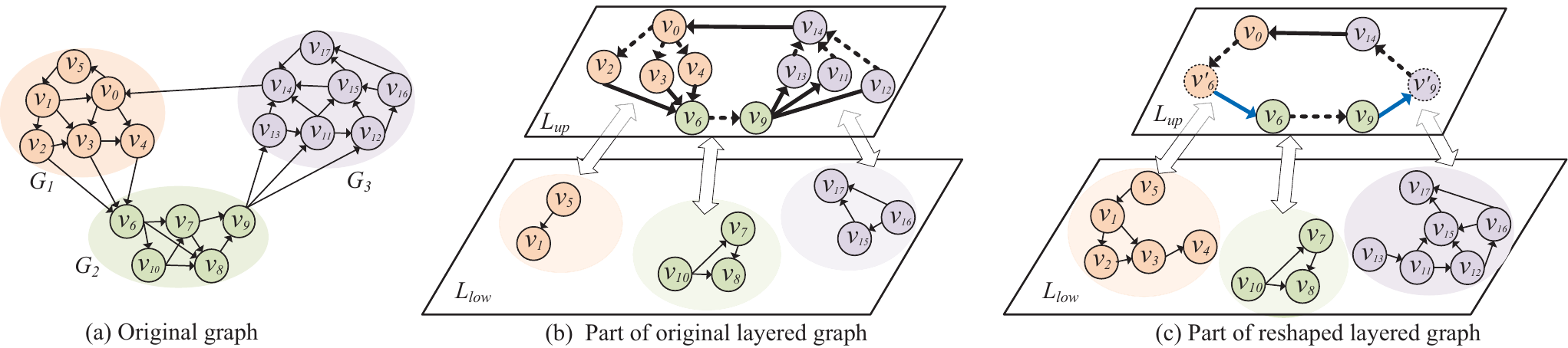}
    \vspace{-0.05in}
    \caption{An illustrative example of the upper layer reshaping. A dotted circle is a proxy vertex. A bold black link is a weighted/unweighted edge on original graph. A dotted link is a shortcut from an entry vertex to an exit vertex in a subgraph. A blue link is a connection between a vertex and its replicated proxy vertex. In (b) and (c), for simplicity, we use two-way hollow arrows to represent the set of shortcuts and edges between $L_{up}$ and $L_{low}$.}
    \label{fig:mirror}
    \vspace{-0.18in}
\end{figure*}

%In most cases, good results are obtained following the above method, but in some cases, the subgraphs are not very profitable. We therefore conducted a detailed study of the under-performing cases, as described below.

\Paragraph{Problem Study} 
Although we can discover dense subgraphs by using the above method, 
it suffers from a key limitation: the shortcuts that need to be established are still numerous due to the massive number of entry/exit vertices. As shown in Figure \ref{fig:mirror}, we find that most boundary vertices (entry/exit vertices) have high degrees and are likely to have many connections to/from other subgraphs, leading to many entry/exit vertices in the target/source subgraphs. For example, vertex $v_9$ is with high out-degree and has 3 out-edges connected to subgraph $G_3$, leading to 3 entry vertices in subgraph $G_3$, and vertex $v_6$ with a high in-degree and has 3 in-edges originating from subgraph $G_1$, leading to 3 exit vertices in subgraph $G_1$. 
%We randomly pick a high in-degree vertex in the XXX graph (see Table \ref{} for dataset statistics) and show the top 20 highest number of exit vertices in a subgraph connected to this vertex in Figure \ref{}. The number of exit vertices connected to this vertex ranges from XX to YY. We have similar findings for high out-degree vertices. 
\eat{We analyzed the problem in several real graphs. 
%We measure the number of entry/exit vertices in a subgraph that share the same source/target vertex. 
Table \ref{tab:degree} shows the average of the top 1\%, 5\%, and 10\% largest numbers of entry/exit vertices that share the same source/target vertex. We can see that the problem is serious and common in these real graphs. %most graphs except for road networks (\eg Euro-Road). 
}
A large number of entry/exit vertices incurs a large skeleton of the upper layer 
as shown in Figure \ref{fig:mirror}b, which will hurt the performance of iterative computation and increase the computation cost for shortcut calculations/updates.

\Paragraph{Solution: Vertex Replication}
Figure \ref{fig:mirror} demonstrates that there exist some entry/exit vertices in a subgraph that share the same source/target vertex. This inspires us to propose a vertex replication approach for reducing the number of entry/exit vertices and shortcuts. The idea is illustrated in Figure \ref{fig:mirror}c. After dense subgraph discovery, if the number of entry/exit vertices in a subgraph $G_i$ that share the same source/target vertex $v$ is larger than a threshold, the source/target vertex $v$ (host vertex) will be replicated in subgraph $G_i$ as a \textit{proxy vertex} $v'$. A high-degree vertex could have many proxy vertices in multiple different dense subgraphs. Both entry and exit vertices can have proxy vertices in other dense subgraphs. For example, in Figure \ref{fig:mirror}c, entry vertex $v_6$ has a proxy vertex $v'_6$ acting as a new exit vertex in subgraph $G_1$. Originally, there are 3 exit vertices in subgraph $G_1$ linking to the entry vertex $v_6$, while now there is only one exit vertex $v'_6$. Exit vertex $v_9$ has a proxy vertex $v'_9$ in $G_3$ as a new entry vertex. There are supposed to be 3 entry vertices in $G_3$ all originating from vertex $v_9$, but now there is only one entry vertex $v'_9$. 

By replicating exit or entry vertices between subgraphs, some boundary vertices of dense subgraphs become internal vertices and move from $L_{up}$ to $L_{low}$. The size of the graph skeleton on $L_{up}$ is greatly reduced.% as the number of vertices and edges in $L_{up}$ are both greatly reduced. %entry/exit vertices %is greatly reduced, such that the number of key vertices 
%and the number of internal shortcuts are both greatly reduced.

\eat{\Paragraph{Computation on Reshaped Skeleton}
By vertex replication, a new connection type is introduced, i.e., the connection between the original vertex and the proxy vertex, which is referred to as \textit{replication link}. There are three types of connections, \ie the original edge (with or without weight), the weighted shortcut, and the replication link that is used to synchronize messages between host vertex and proxy vertex. The local shortcuts from the proxy vertices that act as entry vertices are established during local iteration phase. During global computation on skeleton, the messages collected from multiple proxy vertices (acting as exit vertices in different subgraphs) are sent to their corresponding host vertex for aggregation. During incremental computation, the computation logic on skeleton with proxy vertices is similar to that on original skeleton. It is beneficial for saving incremental computation cost since it is likely to invoke a small amount of updates on the reshaped skeleton and less number of shortcut updates due to less number of entry vertices.}

\vspace{1em}
\subsubsection{Shortcuts Calculation}
\label{sec:layer:build:shortcut}
\eat{要说有两种shortcut和为什么要有这两种shortcut，在上层中，没有密集子图的内部点和边，只有入口和出口点，对于每一个密集子图的入口和出口，为了能够从入口快速正确接受到入口发过来的消息，需要建立shortcut，因此来完成上层中的迭代计算。上层中的迭代过程中，密集子图的入口并么有往下传播消息给内部顶点，因此需要缓存发给内部顶点的消息。而这些消息发给内部顶点仍然可能需要多次迭代才能发给内部所有的顶点，为了能够高效将这些消息发送给内部顶点，使内部顶点能够快速更新状态，我们在每个入口与内部顶点也建立的shortcut，由于入口点在}

On the upper layer, there are only entry and exit vertices of each subgraph. It is required to connect them with shortcuts for propagating messages from entry vertices to exit vertices correctly and quickly. During the iterative computations on $L_{up}$, the entry vertices send messages to exit vertices directly through shortcuts and do not propagate the messages down to internal vertices. %until the iterative computations terminate. 
In order to revise the states of vertices on $L_{low}$, the entry vertices cache these messages that should be propagated to internal vertices, then propagate them down to internal vertices after the iterative computations terminate. However, these messages spread to all internal vertices may require iterative computations. \eat{since the internal vertices may not connect with entry vertices directly} In order to propagate the messages from the entry vertices to internal vertices efficiently, we also connect them with shortcuts.

Based on the above discussion, there are two kinds of shortcuts in the layered graph, 1) the shortcuts from entry vertices to exit vertices of the dense subgraph, and 2) the shortcuts from entry vertices to internal vertices of the dense subgraph. Essentially, both of these shortcuts connect the entry vertices and other vertices of the dense subgraph. Therefore, they can be calculated simultaneously with the same method. Before introducing the shortcut calculation method, we first provide the formal definition of the \textit{shortcut}.
%In a subgraph $G_i$, an entry vertex that receives a message outside $G_i$ will propagate it along the internal edges. After many iterations, the effect of this message will finally be applied to all the vertices in $G_i$. Within $G_i$, the effect of an input message on an arbitrary vertex can be directly obtained through a \textit{shortcut}, which is formally defined as follows.

{
\begin{definition}[Shortcut]
\label{def:shortcut}
Given a subgraph $G_i(V_i, E_i)$ and the input messages vector $M=\{m_u\mid u\in V_i^I\}$ arriving at entry vertices $V_i^I$, the shortcuts $S_i$ are the direct connections from entry vertices $V_i^{I}$ to all vertices $V_i$, i.e., $S_i=\{\vec{w}_{u,v}\mid u\in V_i^I, v\in V_i\}$ where $\vec{w}_{u,v}$ is the weight of a shortcut from vertex $u$ to vertex $v$, such that 
\begin{equation}
\label{eq:shortcut}
    \AGG_{V_i}\big(\GE_{S_i}(M)\big)=\AGG_{V_i}\Big(\bigcup_{k=1}^{\infty}(\AGG_{V_i}\circ \GE_{E_i})^k(M)\Big),
\end{equation}
where $\GE_{S_i}$ and $\GE_{E_i}$ indicate the message propagation through the shortcuts $S_i$ and the original edges $E_i$ respectively, and $\AGG_{V_i}$ indicates the message aggregation on vertex set $V_i$.
\end{definition}

%Equation (\ref{eq:shortcut}) indicates that a single invoking of message propagation and aggregation through the shortcuts will make the identical effect of iterative invoking of message propagation and aggregation through the original edges. The expensive local iterative computation inside a dense subgraph is greatly saved by using shortcuts.

%shortcut的weight可以通过如下公式计算。首先设置一个单位消息对于每个入口顶点u，然后执行迭代计算使该消息传播给内部顶点和出口顶点，直到所有的点不再收到消息，或者收到的消息可以忽略不计。最后每个顶点收到的消息值可以作为从入口到改点的shortcut的weight。其中单位消息必须是一个F操作的单位元，如例2所示当F为+时，单位元为0，那么在SSSP中，每个点收到的消息则为入口点到该点的最短距离，入例2所示。
\red{The shortcut weight $\vec{w}_{u,v}$ from entry vertex $u$ to vertex $v$ in $G_i$ can be calculated by the following equation} 
%The shortcut is calculated according to Definition \ref{def:shortcut}. We let a \textit{unit message} enter into an entry vertex of $G_i$. The unit message is then iteratively propagated and aggregated (by iteratively applying $\GE_{E_i}$ and $\AGG_{V_i}$) inside $G_i$ until all vertices inside $G_i$ no longer receive any messages or the received messages can be ignored. The unit message should be the identity element of the $\GE$ operation to make initiation. Finally, the accumulated received messages on each vertex are aggregated by $\AGG_{V_i}$ to obtain the weight of the shortcut from that entry vertex. Specifically, assuming a unit message $m_u$ is initiated from the entry vertex $u$, the shortcut weight $\vec{w}_{u,v}$ from entry vertex $u$ to an arbitrary vertex $v$ in $G_i$ is calculated by
\begin{equation}
\label{eq:shortcut:calculation}
    \vec{w}_{u,v}=\AGG_{v}\Big(\bigcup_{k=1}^{\infty}(\AGG_{V_i}\circ \GE_{E_i})^k(m_u)\Big),
\end{equation}
\red{where $\AGG_{v}$ is the group-by aggregation on vertex $v$, $m_u$ is the unit message. It means that we first initialize a unit message $m_u$ for entry vertex $u$. Then we perform iterative computation on the subgraph $G_i$ to propagate messages from $u$ to $v$ until all the vertices in $G_i$ no longer receive any messages or the received messages can be ignored. Finally, the aggregated value of messages received by $v$ can be treated as the weight of the shortcut from $u$ to $v$, \ie $\vec{w}_{u,v}$. The unit message $m_u$ should be the identity element of the $\GE$ operation to make initiation. As shown in Example 2, the identity element of `+' is 0. Then, in SSSP, the \kw{min} value of the messages received by $v$ originated from $u$ is the shortest path from $u$ to $v$, \ie the weight of the shortcut from $u$ to $v$.}
%The shortcuts from other entry vertices can be established in a similar way. The shortcuts in other subgraphs can be calculated concurrently. The shortcut calculation is algorithm-specific. 
To alleviate the burden of users, \lframe can %be automated 
\textit{automatically} complete the shortcut calculation 
by invoking the user-defined $\GE$ and $\AGG$ functions without the user's intervention (see \ref{sec:prelim:async-algo}).

\begin{example}\label{exa-shortcut}
%考虑例1中的sssp在图2上执行，在a-d所构成的cluster内部建立shortcut时，先在入口点a处设置一个单位消息1，然后将该消息采用G和F函数在a-d组成的子图中传播。当迭代计算终止时，得到b-e收到消息的集合，然后执行消息的聚集。
% Shortcut calculation for two representative algorithms.
% \etitle{(a) \SSSP}.
Consider running \SSSP on the graph as shown in Figure \ref{fig:contraction_SSSP_inc}a. When computing the shortcuts inside subgraph $G_2$, a unit message $m_{v_0} = 0$ (as the identity element of `+' since $\GE=m_u + w_{u, v}$ containing `+') is input into entry vertex $v_0$. We iteratively perform $\GE=m_u + w_{u, v}$ to propagate messages and use $\AGG=\kw{min}$ to aggregate the received messages for each vertex. Finally, as shown in Figure \ref{fig:contraction_SSSP_inc}d, the aggregated values of the received messages on $\{v_1, v_2, v_3, v_4\}$ are $\{1, 4, 1, 2\}$ respectively, \ie the weights of shortcuts are $\vec{w}_{v_0,v_1}=1,\vec{w}_{v_0,v_2}=4,\vec{w}_{v_0,v_3}=1, \vec{w}_{v_0,v_4}=2$. 
% 在上面的例子可以看出，shortcut本质就是从入口到出口顶点的最短距离值，因此明显能够满足Equation (\ref{eq:shortcut})。
\eat{
As can be seen from the above example, the weight of the shortcut is essentially the shortest distance from the entry vertex to other vertices, so it can obviously satisfy Equation (\ref{eq:shortcut}).
}
\end{example}

%After subsection \ref{sec:layer:build:key-vertex} and \ref{sec:layer:build:shortcut} to get key vertices and shortcuts, 
Finally, we give the formal %extraction method of skeleton below.
definition %construction method 
of the layered graph.

% 构建Layered Graph
% 外部点叫啥？ outlier, single
% 经过前面部分的介绍，我们能够通过给定一个输入图G,获得一个L个密集子图的{G1G2}的集合,以及以及每个子图$i$所包含的shortcuts$S_i$。下面基于以上结果给出layered graph的能够被形式化构建。
\stitle{Layered Graph}. %After the introduction in the previous section, 
% \section{
Given an input graph $G(V, E)$, a set of $N$ dense subgraphs $\{G_1(V_1,E_1),$ $\ldots, G_N(V_N,E_N)\}$, 
%the single vertices \red{$V^-=\{V-\cup_{i=1}^NV_i\}$ that are not in any dense subgraphs. }
\eat{and the remaining part $\overline{G}=(\overline{V}, \overline{E})$, where $\overline{V}=V-\cup_{i=1}^NV_i$ represents the set of vertices that are not in any dense subgraph, 
%\red{$\overline{E}=\overline{V}\times \overline{V} \cap E$} represents the set of edges that are in $\overline{V}$
\red{$\overline{E}=E-\cup_{i=1}^NE_i$}
represents the set of edges that are not in any dense subgraph,} 
the layered graph is formed by the upper layer $L_{up}=(L_V^{up}, L_E^{up})$, the lower layer $L_{low}=(L_V^{low}, L_E^{low})$ and the edges between $L_{up}$ and $L_{low}$, where $L_V^{up}$ (resp. $L_V^{low}$) is the vertex set on the upper layer (resp. the lower layer) and $L_E^{up}$ (resp. $L_E^{low}$) is the edge set on the upper layer (resp. the lower layer). %a set of single vertices $V^{-}=V\setminus \{V_1, V_2, \ldots, V_L\}$, and the shortcuts $S_i$ contained in each subgraph $G_i$. The layered graph can be formally constructed based on the above results. }
\begin{itemize}[leftmargin=*]
% Upper layer ($L_{up}$). Upper layer 由一个图骨架构成，表示为$G_S(V_S, E_S)$, 其中$V_S$和$E_S$分别表示图骨架上的顶点集合和边集合。
\item \textit{Upper layer ($L_{up}$)}. %The upper layer consists of %a graph skeleton, represented as $G_S(V_S, E_S)$, where $V_S$ and $E_S$ represent a vertex set and edge set on the graph skeleton, respectively. The details are as follows:
    \begin{itemize}[leftmargin=*]
         \eat{\item Vertex set
        $L_{up}^V=\{V_i^I \cup V_i^O \cup \overline{V}\}$ 
        is composed of the entry and exit vertices of all dense subgraphs $G_i$ and the vertices that are not in any dense subgraphs. }
        \item Vertex set
        $L_{up}^V$$=$$\bigcup_{i=1}^{N}\{V_i^I, V_i^O\}\cup \{V-\cup_{i=1}^NV_i\}$
        % \overline{V}$
        is composed of the entry and exit vertices of all dense subgraphs and the vertices that are not in any dense subgraphs. 
        %, \ie $V_S=\bigcup_{i=1}^{N}\{V_i^I, V_i^O\}\cup V^-$;
        % \item Edge set $L_{up}^E=\{L_{up}^V\times L_{up}^E \cap E\}\bigcup_{i=1}^{N}\{\vec{w}_{u,v}\mid \vec{w}_{u,v}\in S_i, u\in V_i^I, v\in V_i^O\}$ 
        \eat{\item Edge set $L_{up}^E=\{L_{up}^V\times L_{up}^V \cap E\}\bigcup_{i=1}^{N}\{\vec{w}_{u,v}\mid \vec{w}_{u,v}\in S_i, u\in V_i^I, v\in V_i^O\}$ 
        is composed of the edges between vertices in the upper layer and the shortcuts from entry vertices to exit vertices in each dense subgraph.}
        \item Edge set $L_{up}^E$$=$$\bigcup_{i=1}^{N}\{\vec{w}_{u,v}$$\mid$$\vec{w}_{u,v}\in S_i, u\in V_i^I, v\in V_i^O\} \cup \{E - \bigcup_{i=1}^{N}E_i\}$ 
        is composed of the shortcuts from entry vertices to exit vertices in each dense subgraph and the edges that are not in any dense subgraphs.
        
        \eat{follows:%上层点互联的边和出入口之间的shortcut其他点与出入口连接的边
        \begin{itemize}[leftmargin=*]
            \item The edges between single vertices, \ie $\{(u,v)\mid (u,v)\in E, u\in V^-, v\in V^-\}$;
            \item The shortcuts that connect from entry vertices to exit vertices in each dense subgraph, $\bigcup_{i=1}^{N}\{\vec{w}_{u,v}\mid \vec{w}_{u,v}\in S_i, u\in V_i^I, v\in V_i^O\}$;
            \item The edges from single vertices to entry vertices, $\{(u,v)\mid (u,v)\in E, u\in V^-, v\in \bigcup_{i=1}^{N}V_i^I\}$;
            \item The edges from exit vertices to single vertices, $\{(u,v)\mid (u,v)\in E, u\in \bigcup_{i=1}^{N}V_i^O, v\in V^-\}$;
            \item The edges from the exit vertices of $G_i$ to the entry vertices of $G_j$, $\{(u,v)\mid (u,v)\in E, u\in V_i^O, v\in V_j^I, i \neq j\}$.
        \end{itemize}}
    \end{itemize}
% low layer: low layer由L个独立的密集子图构成,其顶点集合表示为$\{v\mid v\in \bigcup_{i=1}^{N}V_i\}$, 其边集表示为$\{(u,v)\mid (u,v)\in \bigcup_{i=1}^{N}E_i\}$。
\item \textit{Low layer ($L_{low}$)}. 
    \begin{itemize}
        % \item Vertex set $L_{low}^V=\bigcup_{i=1}^N \{V_i-(V_i^I\cup V_i^O)\}$ is composed of all the internal vertices of dense subgraphs. 
        \item Vertex set $L_{low}^V=\bigcup_{i=1}^N \{\hat{V}_i\}$ is composed of the internal vertices of all dense subgraphs.
        \item Edge set 
        $L_{low}^E$$=$$\bigcup_{i=1}^{N}\big\{E_i 
        - 
        % \{(u,v) \in E_i | u \in V^I_i \cup V^O_i, v\in\hat{V}_i\}  \cup 
        \{(u,v) \in E_i | u\in \hat{V}_i, v \in V^I_i \cup V^O_i\}\big\}
        $
        is composed of the edges within each subgraph, except the edges %connected with entry or exit vertices.
        from internal vertices to entry/exit vertices.
    \end{itemize}
    %The low layer consists of L separate dense subgraphs, and its vertex set is represented by $\{v\mid v\in \bigcup_{i=1}^{N}V_i\}$, and its edge set is represented by $\{ (u,v)\mid (u,v)\in \bigcup_{i=1}^{N}E_i\}$
% \item \textit{Edges between $L_{up}$ and $L_{low}$}. $L_{\red{up\_low}}^E=\bigcup_{i=1}^{N}\big\{\{ \vec{w}_{u,v}\in S_i \mid u\in V_i^I, v\in \{V_i-(V_i^I \cup V_i^O)\}\}\cup\{(u,v)\in E_i \mid u\in\{V-(V_i^I\cup V_i^O)\}, v\in V_i^I\cup V_i^O\}\cup\{(u,v)\in E_i \mid u\in V_i^O, v\in\{V-(V_i^I\cup V_i^O)\}\}\big\}$ is composed of the shortcuts from entry vertices to internal vertices within each dense subgraph \red{and the edges from internal vertices to entry/exit vertices.}
\item \textit{Edges between $L_{up}$ and $L_{low}$}. $L_{up\_low}^E=\bigcup_{i=1}^{N}\big\{\{
\vec{w}_{u,v}\in S_i \mid u\in V_i^I, v\in \hat{V}_i\} 
\cup\{(u,v)\in E_i \mid u\in\hat{V}_i, v\in V_i^I\cup  V_i^O\} 
% \cup\{(u,v)\in E_i \mid u\in V_i^O, v\in\hat{V}_i\}
\big\}$ 
is composed of the shortcuts from entry vertices to internal vertices and the edges %between internal vertices and entry/exit vertices, 
from internal vertices to entry/exit vertices within each dense subgraph.% and the 
% edges from exit vertices to internal vertices.
\end{itemize}
 
\eat{
\stitle{Layered Graph Construction}. Given an input graph $G(V, E)$, a set of $L$ dense subgraphs $\{G_1(V_1,E_1), G_2(V_2,E_2), \ldots, G_L(V_L,E_L)\}$, and a set of outlier vertices $V^{-}=V\setminus \{V_1, V_2, \ldots, V_L\}$, the graph skeleton $G_S(V_S, E_S)$ of $G(V, E)$ is formed by the node set $V_S$ and edge set $E_S$. 
\begin{itemize}[leftmargin=*]
    \item Node set $V_S$ is composed of the entry and exit vertices of all $G_i$ and the outlier vertices, \ie $V_S=\bigcup_{i=1}^{N}\{V_i^I, V_i^O\}\cup V^-$;
    \item Edge set $E_S$ is composed as follows:
    \begin{itemize}[leftmargin=*]
        \item The edges between outlier vertices, \ie $\{(u,v)\mid (u,v)\in E, u\in V^-, v\in V^-\}$;
        \item The shortcuts that connect from entry vertices to exit vertices in each dense subgraph, $\bigcup_{i=1}^{N}\{\vec{w}_{u,v}\mid \vec{w}_{u,v}\in S_i, u\in V_i^I, v\in V_i^O\}$;
        \item The edges from outlier vertices to entry vertices, $\{(u,v)\mid (u,v)\in E, u\in V^-, v\in \bigcup_{i=1}^{N}V_i^I\}$;
        \item The edges from exit vertices to outlier vertices, $\{(u,v)\mid (u,v)\in E, u\in \bigcup_{i=1}^{N}V_i^O, v\in V^-\}$;
        \item The edges from the exit vertices of $G_i$ to the entry vertices of $G_j$, $\{(u,v)\mid (u,v)\in E, u\in V_i^O, v\in V_j^I, i \neq j\}$.
    \end{itemize}
\end{itemize}
}

The size of the upper layer (with respect to $|L_{up}^V|$ and $|L_{up}^E|$) is expected to be much smaller than that of the original graph (with respect to $|V|$ and $|E|$). For example, in Figure \ref{fig:contraction_SSSP_inc}, the upper layer contains 3 vertices and 3 edges/shortcuts, which is smaller than the original graph with 9 vertices and 14 edges. 
% 我们已经在修订版本中添加空间开销的分析。相比较于原图而言，一个分层的图的空间开销主要来源于两部分，第一部分是存储每个顶点所属的子图以及顶点类型，该部分的开销是O(|V|)，第二部分则是需要额外存储所有子图shortcut，所以其空间开销是O(\sum(|S_i|))，其中$S_i$是每个子图需要建立的shortcut集合，它的上限由定义\ref{3}可以得到为$V^I_i \times V_i$。此外，正如实验中图\ref{fig:space_cost}所展示的，分层图所带来的额外开销将比原图要小。

\stitle{Analysis.} %说一下这一部分是离线操作，layeredgraph的构建包括寻找子图和计算shortcut，这会占用一部分时间开销，但是该离线操作只需要在初始化系统时执行一次。此外由于shortcut的引入，导致整个系统的space开销变大。从构建过程可以看出增加的shortcut的数量为入口到出口，和入口到内部点，因此总体为入口到其他非入口点的数量，既O($\sum^{N}_{i=0}(|V^I_i|\times V_i))$.
%The construction of the layered graph requires two operations finding subgraphs and calculating shortcuts, which results in some time cost. However, these two operations are offline preprocessing operations, which only occur ONE time at system initialization while online incremental computations occur at each graph update $\Delta G$, as shown in Figure 3. Also, 
Due to the introduction of shortcuts, \lframe will require more space. The additional space overhead includes the shortcuts from entry vertices to %exit vertices and the shortcuts from entry vertices to internal vertices, 
all vertices within each subgraph,
\ie $O(\sum^{N}_{i=1}(|V^I_i|\times |V_i|))$, where $|V^I_i|$ is the number of entry vertices of subgraph $G_i$ and $|V_i|$ is the number of all vertices in $G_i$. 
In practice, the additional space overhead is always smaller than that of the original graph, as shown in Figure \ref{fig:space_cost} (in Section \ref{sec:expr:offline_time}).

\eat{Compared with the original graph, the space overhead of a layered graph mainly comes from two parts. The first part is to store the dense subgraph to which each vertex belongs and the vertex type. The cost of this part is O($|V|$). The second part needs to additionally store the shortcuts of all dense subgraphs, so its space overhead is O($\sum^{N}_{i=1}(|V^I_i|\times V_i))$. 
Where $S_i$ is the shortcut set that needs to be created for each subgraph, 
and its upper limit can be obtained by Definition \ref{def:dsubg} as $V^I_i \times V_i$. In addition, as shown in Figure \ref{fig:space_cost} in the experiment, the actual additional overhead brought by the layered graph will be smaller than the original graph size.
}

\subsection{Layered Graph Update}
\label{sec:layer:update}

%\subsubsection{Key vertices update} 这个其实本质上是顶点的更新，也就是说上层顶点的更新，从顶点的更新说，另一个是shortcut的更新。
The layered graph needs to be updated when $G$ is updated with $\Delta G$. The vertices may move between %the upper and lower layer, 
the two layers, due to the generation or disappearance of dense subgraphs, \eg the internal vertices of the newly generated subgraph move from $L_{up}$ to $L_{low}$. In order to avoid the expensive overhead caused by repeated subgraph discovery, we incrementally update the dense subgraphs %vertex set in upper layer 
with incremental community detection methods, such as C-Blondel \cite{zhuang2019dynamo} or DynaMo \cite{seifikar2020c}. %Specifically, we first obtain the subgraphs affected by the edges of $\Delta G$, and finally checked whether these subgraphs satisfy the Definition \ref{def:dsubg}. When the subgraph satisfies the definition, it will continue to be used as a subgraph, otherwise, the subgraph will be dissolved. In addition, over time, in some extreme cases, a large number of original subgraphs may gradually degenerate and be dissolved. Therefore, we detect whether the total benefit $Q = \sum_{i=0}^{L}(|E_i| - |V^{I}_i| \times |V^{O}_i|$ is higher than a specified threshold \changys{$|E|\times \alpha$, where $\alpha$ represents the minimum income ratio set by the user}. If so, \lframe continue to use the architecture, otherwise the clustering process needs to be restarted to obtain high-quality subgraphs. When all subgraphs are updated, we update the key vertices on the upper layer. 
%If the entry vertex or exit vertex of a subgraph is added, it needs to be added to the upper layer. Correspondingly, if the exit vertex or exit vertex of a subgraph becomes an internal vertex, it needs to be deleted from the upper layer.
%In practical，dense grapph的生成和消息并不会常常发生，因为deltaG非常小，而是经过多次deltaG之后才会发生densesubgraph的改变，因此我们会隔一段时间执行增量的算法，而不是一有delta就更新,。因此，但是每次有更新的时候可能会导致shortcut更新。
In practice, the size of $\Delta G$ is very small compared with $G$. A small $\Delta G$ does not have a large effect on existing dense subgraphs. Thus we update the dense subgraphs only when enough $\Delta G$ are accumulated. However, even a very small $\Delta G$ can still change the weight of a number of shortcuts of the layered graph.  %更新密集子图不是一有delta就更新，而是累积多个delta一起更新。但是即使是很小的delta仍然会对现有的分层图产生影响，主要体现在shortcut上，

\stitle{Shortcuts update}. %When the graph $G$ is updated with $\Delta G$, we only need to update the shortcuts of the subgraphs affected by $\Delta G$. Specifically, the shortcut $S_i$ of the subgraph $G_i$ should be updated if $G_i$ is affected by $\Delta G$. 
There are three kinds of shortcut updates.
\romannumeral1) \textit{Deletion}. If all of an entry vertex's in-edges from outside are deleted, \ie the connections from outside are cut off, this entry vertex will become an internal vertex, and the shortcuts originated from it should be removed. 
\romannumeral2) \textit{Addition}. If an in-edge from outside is added to an internal vertex, this internal vertex will become an entry vertex. The shortcuts from it to other vertices in the subgraph should be calculated.
\romannumeral3) \textit{Weight update}. If there are addition or deletion edges within a subgraph, the weight of the shortcuts should be updated. 

%shortcut建立在每个密集子图内部，且从等式(\ref{})可以看出每个子图$G_i$上的shortcut的权重仅仅依赖于子图$G_i$上的点和边。我们只需要更新被$\Delta G$所影响的子图的shortcut即可，且每个子图的shortcut可以相互独立进行更新。
The shortcut is built inside each dense subgraph according to the Definition \ref{def:shortcut}. Moreover, from the Equation (\ref{eq:shortcut:calculation}), we can see that the weight of each shortcut on $G_i$ only depends on the edges and vertices in $G_i$, and the shortcuts on the different subgraphs are independent of each other. %computation of shortcut weights only performs locally on each dense subgraph. %, that is, the shortcuts of different subgraphs are mutually exclusive independent. 
Therefore, we only need to update the shortcuts on the subgraphs affected by $\Delta G$, and the shortcuts for each subgraph can be updated in parallel. For the shortcut deletion or addition, they can be done directly within the subgraph by removing or calculating the shortcut. For the weight update, in order to avoid redundant computation, we use an incremental method to update. 
% 具体而言，i)Deletion，旧的shortcut直接删除即可。ii)Addtion，新入口顶点按照\ref{}节介绍的方法建立shortcut。iiii)weight update，我们需要更新旧的shortcut权重。 
%Specifically, \romannumeral1) \textit{Deletion}. The old shortcut can be deleted directly. \romannumeral2) \textit{Addtion}. The new entry vertex creates a shortcut according to the method introduced in the \ref{sec:layer:build:shortcut} section. \romannumeral3) \textit{Weight update}. We need to update the weight of old shortcut. 
% 从等式(\ref{})可以看出，从$u$到$v$的shortcut权重等于$v$经过所有从$u$到$v$的路径收到的消息的聚集值。密集子图内部的边发生变化后，顶点v收到的消息可能会发生变化。消息的传播通过迭代计算完成。因此shortcut的更新可以采用与迭代图计算相同的增量计算方式。记录一些计算shortcut时的信息，推导出compensation消息和cancellation消息来redo或者undo增加或删除的消息that caused by the addition or deleltion path from u到vdue to the add or deletion edges within subgraph.。

According to Equation (\ref{eq:shortcut:calculation}), the weight of the shortcuts is calculated by iterative computations, and the weight of the shortcut from $u$ to $v$ is equal to the aggregate %value of 
all the messages received by $v$ through all paths from $u$ to $v$.  
After the edge addition or deletion within the dense subgraph, some messages received by $v$ may become invalid or missing.  %compared with previous received messages when run over $G$.
Thus, the update of the shortcut can adopt the existing incremental  computation methods \cite{gong2021ingress, mariappan2019graphbolt,vora2017kickstarter}. %That is, deduce 
The compensation and cancellation messages can be deduced based on the memoized information when calculating the old shortcut. 
%There are some missing or invalid meaasges in the received messages of $v$ due to the edge addition or deletion, so we can use compensation and cancellation messages to redo and undo the effect of missing and invalid messages on vertex $v$.
% Use compensation and cancellation messages to 
These messages will be used to redo and undo the effect of missing and invalid messages on vertex $v$, in which there are some missing and invalid messages in the received messages of $v$ due to the addition and deletion edges within the dense subgraph.

\begin{example}\label{exa-update}
% Shortcut update for two representative algorithms.
% \etitle{(a) \SSSP}.
Consider running \SSSP on the updated graph as shown in Figure \ref{fig:contraction_SSSP_inc}b. Since %the graph updates 
$\Delta G$ only changes $G_2$, the %vertices in $L_{up}$ and 
shortcuts related to $G_1$ do not need to be updated. 
For $G_2$, the vertices on $L_{up}$ do not need to change, since only the inner edges change, and the shortcuts can be updated incrementally. 
Therefore, we can get the weights of the old shortcuts as the initial weights of the new shortcuts,
\ie %$\{\vec{w}_{v_0,v_1}=1,\vec{w}_{v_0,v_2}=4,\vec{w}_{v_0,v_3}=5\}$. 
the initial values of
$\{\vec{w}_{v_0,v_1},\vec{w}_{v_0,v_2},\vec{w}_{v_0,v_3}\}$ are set to $\{1,4,1,2\}$. 
Since the edge $v_3 \ra v_4$ is deleted and the state of $v_4$ depends on $v_3$, it is necessary to generate a cancellation message $m_{v_3,v_4}$$=$$\bot$ ($\bot$ means the vertex needs to be reset to the default state, \ie $\infty$ for \SSSP), and $m_{v_3,v_4}$ sets the state of $v_4$ to $\infty$  \cite{vora2017kickstarter, gong2021ingress, feng2021risgraph}.  
Meantime, $v_4$ will get a message $m_{v_2,v_4}=5$ from its neighbor $v_2$. 
In addition, since the edge $v_3 \ra v_2$ is added, it is necessary to generate a compensation message 
$m_{v_3,v_2}$$=$$3$.
Then all these 
revision messages will be propagated inside $G_2$. %, and  update all vertex states, i.e. $\{x_{v_0}=0,x_{v_1}=1,x_{v_2}=2,x_{v_3}=3\}$.
 Finally, as shown in Figure \ref{fig:contraction_SSSP_inc}e, the aggregated values of the received messages on $\{v_1, v_2, v_3,v_4\}$ are $\{1, 3, 1, 4\}$ respectively, \ie the new weights of the shortcuts are $\vec{w}_{v_0,v_1}$$=$$1,\vec{w}_{v_0,v_2}$$=$$3,\vec{w}_{v_0,v_3}$$=$$1,\vec{w}_{v_0,v_4}$$=$$4$.
\end{example}

%% file: 4sketch_based_framework.tex
\section{Incremental processing with Layered Graph%framework based on TLA
}\label{sec:inc_framework}

This section will introduce how \lframe performs incremental graph processing 
%based 
on the layered graph.% when the graph is updated.

%点出图变化，顶点状态发生变化的本质，图发生变化后，顶点收到的消息可能会发生变化，因为一些消息从原点到目的顶点的路径发生变化，导致一些消息变成无效的或者缺失一些消息。
\stitle{Revision messages deduction}. As shown in Equation (\ref{eq:iterresult}), the final vertex state is determined by the received messages that are from ALL vertices and transferred along different paths. When the graph is updated, the messages received by vertices may change due to the changes in the paths that messages propagate. The incremental graph processing framework can automatically \cite{vora2017kickstarter, gong2021ingress} or manually \cite{mariappan2019graphbolt, DZiG} obtain the revision messages \ie compensation messages and cancellation messages, and propagate them to revise the effect of the missing and invalid messages on vertex states \cite{gong2021ingress, mariappan2019graphbolt}. For the revision messages, we can deduce them by employing the method proposed in our previous work \cite{gong2021ingress}. 

After deducing the revision messages, we propagate them efficiently with the help of \lframe. As we have introduced in Section \ref{sec:overview}, the propagation of revision messages on \lframe is in three steps, 1) messages upload, 2) iterative computation, and 3) messages assignment.

%消息上传

%ingress能正确，我只要达到和ingress一样的效果就可以。

\subsection{Messages Upload}
\label{sec:framework:upload}

\eat{
In this subsection, We first will introduce how to use the incremental strategy to generate revision messages on the lower layer, and then upload all subgraph internal revision messages to the vertices of the upper layer.
}

% 在图发生更新时，为了避免冗余的计算，增量图处理系统往往会利用原图上的收敛结果作为新图上的初始值。同时，为了保证基于该初始状态能够获得正确的计算结果，增量系统采用一些记忆策略生成一些校正消息。\oursys采用Ingress的记忆策略结合图更新来生成修正消息。针对原图中不应该传播的消息，生成一组回收消息$M^{-}$，针对新图中应该发送而原图没有传播的消息，生成一组补偿消息$M^{+}$。
%\stitle{Initializa revision messages}.
\eat{
When the graph is updated, in order to avoid redundant computation, the incremental graph processing system often uses the convergence result on the original graph as the initial state $X^0$ on the new graph. At the same time, in order to ensure that the correct results can be obtained based on the initial state, the incremental systems \cite{mariappan2019graphbolt, gong2021ingress, vora2017kickstarter} adopt some memory strategies to generate some revision messages. \oursys uses the memory strategy of Ingress \cite{gong2021ingress} combined with graph updates $\Delta G$ to generate revision messages. The basic process is to generate a set of \textit{cancellation messages} $M^{-}$ for messages that should not be propagated in the original graph, and generate a set of \textit{compensation messages} $M^{+}$ for messages that should be sent in the new graph but not propagated in the original graph.
}

\eat{Revision messages upload}
%As shown in Equation (\ref{eq:iterresult}), the final vertex state depends on the messages $M^0$ initiated from ALL vertices transferred along different paths.
%没说出有的修正消息在low层，或者从low层中的顶点。
The upper layer $L_{up}$ only contains a subset of vertices, and the internal vertices inside each subgraph on $L_{low}$ do not participate in iterative computation on $L_{up}$. To ensure that all vertices on $L_{up}$ converge with the effects of internal vertices, the iterative computation on 
$L_{up}$ %(composed of a subset of vertices) 
should collect not only the revision 
messages deduced by the vertices on $L_{up}$ but also those by internal vertices. 
Thus, it is required to upload the revision messages deduced by the internal vertices of the dense subgraphs on $L_{low}$ to $L_{up}$. Since the entry/exit vertices of each dense subgraph are on $L_{up}$, messages upload can be done by propagating the revision messages to entry/exit vertices.  %的说是传到入口和出口。

% 因此，我们需要进行一个校正消息上传的操作，即把每个子图内部的校正消息上传给上层的skeleton顶点。由于有的内部点并不是和entryexit点直接相连，因此为了将这些点生成的修正消息上传到上层点，我们针对被图更新$\Delta G$所影响的子图进行进行一个局部迭代计算，将内部的校正消息累积到子图的入口顶点和出口顶点上。这个过程是可以高效并行的，因为每个子图相互独立，不会有任何并发冲突。当每个子图完成上传操作后，
Not all the internal vertices within each dense subgraph have connections with the entry/exit vertices, thus, %in order to propagate all internal vertices' messages to entry/exit vertices, %
%Specifically, we perform a local iterative computation to convergence for the subgraph affected by the graph update $\Delta G$, and accumulate the internal revision messages on the entry and exit vertices of the subgraph. 
we perform a local iterative computation to propagate internal revision messages to the entry/exit vertices of the subgraph. The iterative computation terminates when the messages received by entry/exit vertices can be ignored.  %, and there will be no concurrency conflicts. 消息上传后，入口和出口点得到了所有内部点经过所有路径传过来的消息，那么上层中的初始消息由入口点收到的消息和其他点的消息。表示成如下形式。
After the upload of the messages, %When each subgraph completes the upload operation, 
the accumulated messages on the entry vertices $V_i^I$ and exit vertices $V_i^O$ can be treated as their initial revision messages %of $V_i^I\cup V_i^O$ on $L_{up}$, i.e., 
\ie, $\mathbb{M}^0_{V_i^{I}\cup V_i^{O}}=\AGG_{V_i^{I}\cup V_i^{O}}\big(\bigcup_{k=1}^{\infty}(\AGG_{V_i}\circ \GE_{E_i})^k(\mathbb{M}^0_{V_i})\big)$. %这里有问题，Ei和Vi有问题
\eat{The single vertices $V^-=\{V-\cup_{i=1}^NV_i\}$ that are not in any dense subgraphs. }
Together with the initial messages of vertices that are not in any dense subgraph on $L^V_{up}$, \ie $\mathbb{M}^0_{V-\cup_{i=1}^NV_i}$, % vi包括了入口点 % Hence, when the graph is updated, 
the initial messages of vertices on $L_{up}$ %(including the entry/exit vertices of all subgraphs and the updated single vertices) 
can be expressed as follows

%eat

\vspace{-0.15in}
\begin{equation}
\begin{aligned}\label{eq:inc_init}
    \mathbb{M}_{L_{up}^V}^0 &= \Big(\bigcup_{i=1}^{N}\AGG_{V_i^{I}\cup V_i^{O}}\big(\bigcup_{k=1}^{\infty}(\AGG_{V_i}\circ \GE_{E_i})^k(
    \mathbb{M}^0_{V_i})\big)\Big) \\&
    % \bigcup 
    \cup
    % M^0这里应该是revision message?
    \mathbb{M}^0_{V-\cup_{i=1}^N V_i},
\end{aligned}
\end{equation}
where $\mathbb{M}^0_{V_i}$ represents the initial revision messages. 

\stitle{Note}. It is unnecessary to perform messages upload on all subgraphs on $L_{low}$, because the revision messages are only generated on subgraphs that are affected by $\Delta G$ %according to the revision messages deduction method in 
\cite{vora2017kickstarter, gong2021ingress, mariappan2019graphbolt, DZiG}. In general, since the size of $\Delta G$ is small, the number of affected subgraphs is small, too. % 为什么是ingress和dzig, revision message deduction一段里面是kickstater和ingress,需要对应吗？
For all subgraphs affected by $\Delta G$, messages upload can be efficiently performed in parallel since each subgraph is independent of the other. %Furthermore, during the iterative computation of messages uploading, the messages only are propagated from internal vertices to boundary vertices \ie entry and exit vertices, and no messages enter into the subgraph. The local iterative computation for messages uploading converges very fast. 
%不需要在L2层中所有的子图上执行迭代计算，因为对于没有受DG影响的子图，不会有修正消息产生、cite{ingress，graphbolt，DZig}。此外，由于只往外边传消息，并没有新消息传入内部，因此该过程非常快。有实验吗？
\eat{where, $U_G$ represents the set of subgraph Ids affected by the graph update, $U_V$ represents the set of single vertices affected by the graph update, and there is $U_V \in V^{-}$.
 It can be seen from the Equation (\ref{eq:inc_init}) that the effects of multiple update operations in a subgraph will eventually accumulate at the exit vertex. This will avoid the problem of repeated activation of $G_0$ due to multiple updates of $G_1$ in Figure \ref{fig:contraction_SSSP_inc}(a).
}

\begin{example}\label{exa-local}
% Consider 
Running \SSSP to convergence on the layered graph with $v_0$ as the source vertex in Figure \ref{fig:contraction_SSSP_inc}d. 
When the graph changes as shown in Figure \ref{fig:contraction_SSSP_inc}b, the layered graph is updated as shown in Figure \ref{fig:contraction_SSSP_inc}e. At this time, the convergence states of all vertices on the original graph are taken as the initial states of the vertices on the updated graph, \ie 
% $\{x_{v_0}=0,x_{v_1}=1,x_{v_2}=4,x_{v_3}=5,x_{v_4}=8,x_{v_5}=6,x_{v_6}=7,x_{v_7}=7\}$.
%$\{x_{v_0},x_{v_1},x_{v_2},x_{v_3},x_{v_4},x_{v_5},x_{v_6},x_{v_7}\}$ are $\{0,1,4,5,8,6,7,7\}$.
$\{x_{v_0},...,x_{v_8}\}$ are $\{0,1,4,1,2,5,6,7,7\}$.
Since $G_1$ is not directly affected by $\Delta G$, there is no need to derive 
revision messages on $G_1$.
On $G_2$, %similar to Example \ref{exa-update}, 
%it will generate 
a cancellation message $m_{v_3,v_4}$$=$$\bot$  and two compensation messages $m_{v2,v4}$$=$$5$ and $m_{v_3,v_2}$$=$$3$ will be generated. %according to $\Delta G$. 
% For $G_2$, since the edge $v_1 \ra v_3$ is added, it is necessary to generate a compensation message $m_{v_1,v_3}=x_{v_1}+w_{v_1,v_3}=4$. At the same time, since the weight of edge $v_1 \ra v_2$ is updated from 3 to 1, 
For the cancellation message $m_{v_3,v_4}$$=$$\bot$, it will cause $v_4$ to be reset to the default state (\ie $\infty$), and all the vertices that depend on $v_4$ will be reset to the default state according to the dependency tree \cite{vora2017kickstarter, gong2021ingress, feng2021risgraph}.
Then all the rest of the revision messages will be propagated inside $G_2$, and finally all messages will also be aggregated to the exit vertex $v_4$ on $L_{up}$, \ie $m_{v_4}$$=$$4$. At this time, $L_{up}$ obtains all the revision messages of $L_{low}$, and $v_2$ and $v_4$ of $G_2$ get new states $x_{v_2}$$=$$3$ and $x_{v_4}$$=$$4$.
\end{example}

\vspace{-0.1in}
\subsection{Iterative Computation On The Upper Layer}
\label{sec:framework:compute}

% 解释这个过程的目的：
% 消息上传完成后，底层各个子图推导出的修正消息已经全部传给上层顶点。而只是在入口和出口点缓存了，接下来要将这些消息在上层传播，通过迭代计算。upper顶点的状态可以表示为：一个公式。
%这些被上传的消息缓存在密集子图的出入口内，为了使其他顶点也能接收到下层消息的影响而修正顶点状态，需要执行迭代计算来传播这些消息。迭代计算值局限在上部，入口点和出口点在上层中也会参与迭代计算。密集子图的入口点收到消息后通过shortcut发送给出口点，而不发送给内部顶点。
%缓存这个事是不是在assignment的时候说比较好。由于入口点并没有给内部点发送消息，而是将收到的消息通过shortcut发送给出口点，而内部点并未收到任何消息，而无法正确更新，因此我们将内部收到且应该发送给内部顶点的消息缓存一起来。最后，该迭代可以表示成如下形式。
After the upload of the messages, the revision messages deduced by internal vertices of the subgraphs on $L_{low}$ have been propagated to $L_{up}$. 
\eat{
After the messages upload, the revision messages of $L_{low}$ have been propagated to $L_{up}$. 
}
However, these uploaded messages are only cached in entry and exit vertices of the dense subgraphs according to Equation (\ref{eq:inc_init}). Iterative computations are required to be performed on $L_{up}$ to propagate the revision messages so that the other vertices on $L_{up}$ can receive all the revision messages to revise their states. 

The iterative computations only perform on $L_{up}$, \ie only $L^V_{up}$ and $L^E_{up}$ are involved in iterative computations, and the entry and exit vertices of dense subgraphs will participate in the iterative computations because they are on $L^V_{up}$. When the entry vertices receive messages, they do not send messages to internal vertices, but propagate messages to exit vertices via shortcuts. 
\eat{and cache the aggregated message of all received messages}
After the iterative computations, the states of vertices on $L_{up}$ can be expressed as follows

\vspace{-0.1in}
\begin{equation}
\label{eq:global}
X^*_{L^V_{up}}=\AGG_{L_{up}^V}\Big(X_{L_{up}^V}^0\cup \bigcup_{k=1}^{\infty}(\AGG_{L_{up}^V}\circ \GE_{L_{up}^E})^k(\mathbb{M}_{L_{up}^V}^0)\Big).
\end{equation}

Based on the following Theorem \ref{theorem:global}, We can see that after the iterative computations on $L_{up}$, the vertices converge to the same state as performing the iterative computation on the original graph.

\begin{theorem}\label{theorem:global}
With initial messages $\mathbb{M}_{L_{up}^V}^0$ defined in Equation (\ref{eq:inc_init}) and initial states $X_{L_{up}^V}^0$, the converge states $X_{L_{up}^V}^*$ of the vertices on the upper layer after iterative computation on the upper layer %$G_{L_{up}}(V_{L_{up}},E_{L_{up}})$
$L_{up}(L_{up}^V, L_{up}^E)$ are equal to that after iterative computation on updated graph $G \oplus\Delta G$.
\end{theorem}

\begin{proofS}
\eat{According to Equation (\ref{eq:iterresult}), we have the following converged states after infinite iterations, \ie
\begin{equation}
    X_{L_{up}^V}^*=\AGG_{L_{up}^V}\Big(X_{L_{up}^V}^0\cup \bigcup_{k=1}^{\infty}(\AGG_{L_{up}^V}\circ \GE_{L_{up}^E})^k(M_{L_{up}^V}^0)\Big),
\end{equation}
where $\GE_{L_{up}^E}$ represents message propagation %on the upper layer $L_{up}(L_{up}^V, L_{up}^E)$ 
and $\AGG_{L_{up}^V}$ represents message aggregation %on  vertices 
on the upper layer
$L_{up}(L_{up}^V, L_{up}^E)$. }
By replacing $\mathbb{M}_{L_{up}^V}^0$ with Equation (\ref{eq:inc_init}), the initial messages from each updated subgraph's internal vertices %, 
%i.e., $\bigcup_{i \in U_G} M^{-}_{V_i} \cup M^{+}_{V_i}$,  
are propagated out via boundary vertices. By iteratively applying $\GE_{L_{up}^E}$ and $\AGG_{L_{up}^V}$, these initiated messages no matter from the internal vertices or from the vertices of $L_{up}$ are propagated on $L_{up}$ and will be finally accumulated to vertices on $L_{up}$.
%according to Conditions ({\bf C1}) and ({\bf C2}). 
%More detailed proof will be found in the supplementary file. 
\end{proofS}

\vspace{-0.08in}
\begin{example}\label{exa-global}
Figure \ref{fig:contraction_SSSP_inc}e has introduced the iterative computation on $L_{up}$. Based on Example \ref{exa-local}, we get the states $\{x_{v_0}$$=$$0,x_{v_4}$$=$$4,x_{v_5}$$=$$\infty\}$ and revision message $\{m_{v_4}$$=$$4\}$ of all the vertices on $L_{up}$. Then the iterative computation is performed on $L_{up}$ based on these initial states. First only $v_4$ is the active vertex because it has revision message $\{m_{v_4}$$=$$4\}$. $v_4$ is an exit vertex, and the message $m_{v_4,v_5}$$=$$m_{v_4}$$+$$w_{v_4,v_5}$$=$$7$ is generated through the outgoing edge $(v_4,v_5)$. $v_5$ is an entry vertex, it aggregates the message $m_{v_4,v_5}$ to $m_{v_5}$ to update its own vertex state from $x_{v_5}$$=$$\infty$ to $x_{v_5}$$=$$7$ , and stores $m_{v_5}$ for 
messages assignment (Section \ref{sec:framework:local}). $v_5$ then generates a message $m_{v_5,v_0}$$=$$m_{v_5}$$+$$w_{v_5,v_0}$$=$$9$ and sends it to $v_0$. Then $v_0$ cannot update the message $m_{v_0}$ after receiving $m_{v_5,v_0}$. Therefore, all vertices on $L_{up}$ reach a convergent state, \ie $\{v^*_0$$=$$0, v^*_4$$=$$4, v^*_5$$=$$7\}$.
\end{example}

% example-4(pagerank): 在skeleton上执行全局迭代计算之前,如果子图里面有初始化消息需要先将子图迭代至收敛,将消息全部传播到skeleton上. 例如, 在batch计算时子图内每个点有一个初始消息,在增量计算时,子图内的点可能包含cancellation and compensation messages, 所以都需要在迭代之前,将包含初始消息的子图迭代至收敛,为了将初始消息传播到skeleton上. 

\vspace{-0.1in}
\subsection{Revision Messages Assignment}\label{sec:framework:local}

Since the iterative computation is only performed on $L_{up}$, the revision messages will not touch the internal vertices of each subgraph on $L_{low}$, \ie, the internal vertices will not receive revision messages from outside. It is essential to launch another step to apply outside messages to internal vertices. 

Though the internal vertices do not receive the revision messages from outside, the entry vertices of each dense subgraph have received all the revision messages from vertices in other dense subgraphs and $L_{up}$ according to Theorem \ref{theorem:global}. In order to enable the internal vertices to receive outside revision messages, 
the entry vertices cache the received messages before propagating them to exit vertices via shortcuts during the iterative computations. %the entry vertices cache the aggregated message of all received message during the iterative computations}. 
After many iterations, the entry vertices may cache a large number of messages, and we only store the aggregated messages.
%$=>$ Even after many iterations, the entry vertex does not cache a lot of messages because we only store aggregated messages.}
The cached messages can be expressed as follows
\vspace{-0.05in}
\begin{equation}
    \mathbb{M}_{V^I} = \bigcup^N_{i=1}\AGG_{V_i^I}\big(\bigcup_{k=1}^{\infty}(\AGG_{L_{up}^V}\circ \GE_{L_{up}^E})^k(\mathbb{M}_{L_{up}^V}^0)\big).
\end{equation}

Finally, we send the messages that have been cached in entry vertices to the internal vertices via shortcuts between entry vertices and internal vertices. The states of the vertices on $L_{low}$ can be expressed as follows %As depicted in Theorem \ref{theorem:global}, the final states and accumulated messages are obtained by the vertices on $L_{up}$ (including the entry vertices) after global computation. The messages $M_{V_i^I}$ from outside are accumulated at the entry vertices $V_i^I$ of subgraph $G_i$, \ie $M_{V_i^I} = \AGG_{V_i^I}\big(\bigcup_{k=1}^{\infty}(\AGG_{L_{up}^V}\circ \GE_{L_{up}^E})^k(M_{L_{up}^V}^0)\big)$. %Thus, it can be done directly through the shortcuts from entry vertices. 
% Therefore, the assignment of accumulated messages can be done directly through shortcuts between the two layers. 
\eat{
Therefore, during the local assignment phase, all updated entry vertices on $L_{up}$ directly use their shortcuts to distribute revision messages to vertices on $L_{low}$. Each vertex on $L_{low}$ may be assigned revision messages by multiple entry vertices. It aggregates all received revision messages and finally updates its own vertex state.
This process can be expressed as follows.
}
% After local iterative computation for uploading revision messages to  vertices of $L_{up}$, the lower layer $L_{low}$ vertex states are $X_{L_{low}^V}=\AGG_{L_{low}^V}\big(X_{L_{low}^V}^0\cup \bigcup_{k=1}^{\infty}(\AGG_{V_i}\circ \GE_{E_i})^k(M_{V_i}^0)\big)$.
\vspace{-0.05in}
\begin{equation}
\begin{aligned}
    X_{L_{low}^V}^*=\AGG_{L_{low}^V}\Big(X_{L_{low}^V}\cup \bigcup_{i=1}^{N}(\AGG_{\hat{V}_i}\circ \GE_{\hat{S}_i})(\mathbb{M}_{V_i^I})\Big), 
    % X_{L_{low}^V}^*=\AGG_{L_{low}^V}\Big(X_{L_{low}^V}\cup (\AGG_{L_{low}^V}\circ \GE_{\red{\hat{S}}})(M_{V^I})\Big)
\end{aligned}
\end{equation}
where $\hat{S}_i=\{\vec{w}_{u,v}\in S_i \mid u\in V_i^I, v\in \hat{V}_i\}$ is a set of shortcuts between two layers, $X_{L_{low}^V}$ are vertex states on $L_{low}$ after local iterative computation for uploading revision messages to  vertices of $L_{up}$, \ie
\vspace{-0.05in}
\begin{equation}\label{eq:xlowv}
    X_{L_{low}^V}=\AGG_{L_{low}^V}\big(X_{L_{low}^V}^0\cup \bigcup_{k=1}^{\infty}(\AGG_{V_i}\circ \GE_{E_i})^k(\mathbb{M}_{V_i}^0)\big).
\end{equation} 

% 这里X_{L^V_{low}}没解释，下面证明到是解释了，这里要么解释一下，要没在公式7一起给出来。
\eat{Note that, the revision messages assignment are only performed on the entry vertices that have accumulated revision messages. If all the entry vertices of a subgraph have not received revision messages, the subgraph will not need to be updated. Because the states of vertices are not affected by $\Delta G$}

We have the following theorem to guarantee correctness.

\begin{theorem}\label{theorem:local}
After iterative computation on the upper layer, by assigning the accumulated messages of entry vertices to internal vertices through shortcuts, the resulted internal vertex states are the same as that after iterative computation on updated graph $G \oplus\Delta G$.
\end{theorem}

\begin{proofS}
According to Equation (\ref{eq:xlowv}), after local iterative computation for uploading revision messages to vertices on $L_{up}$, 
%the internal vertex states in each subgraph $G_i(V_i, E_i)$ are $X^V_{L_{low}}$. \eat{, where $\hat{V}_i=V_i-V_i^I-V_i^O$ are the set of internal vertices.} That is, 
the effects from internal vertices have been applied to each other. The accumulated outside messages $\mathbb{M}_{V_i^I}$ include the effects from all other vertices outside the subgraph, which are accumulated at the entry vertices $V_i^I$. By assigning these outside messages to internal vertices, \ie $\AGG_{\hat{V}_i}\big(\GE_{\hat{S}_i}(\mathbb{M}_{V_i^I})\big)$, % S_i是shortcut
the outside effects are applied on internal vertices. The aggregation results of these outside messages and the internal vertex states $X_{L^V_{low}}$ are equal to that obtained by iterative computation on the entire graph. 
%More detailed proof will be found in the supplementary file. 
\end{proofS}

\begin{example}\label{exa-assign}
Following Example \ref{exa-global}%we show how to do the local assignment. For% 
, for the activated entry vertex $v_5$, it assigns revision messages to internal vertices via shortcuts. Specifically, $m_{v_5,v_6}$$=$$m_{v_5}$$+$$\vec{w}_{v_5,v_6}$$=$$8$, $m_{v_5,v_7}$$=$$m_{v_5}$$+$$\vec{w}_{v_5, v_7}$$=$$9$, and $m_{v_5,v_8}$$=$$m_{v_5}$$+$$\vec{w}_{v_5,v_8}$$=$$9$. Finally, $\{v_6$$,$$v_7$$,$$v_8\}$ get the convergence states $\{x^{*}_{v_6}=8,x^{*}_{v_7}=9,x^ {*}_{v_8}$$=$$9\}$ by the message aggregation operation.
\end{example}

%% file: 6expr.tex
\vspace{-0.1in}
\section{Experiments}
\label{sec:expr}

%This section presents the implementation details of \oursys and a thorough evaluation of \oursys.% thoroughly and compare it with existing state-of-the-art incremental graph processing methods.
% and then evaluate it thoroughly.
\eat{
\begin{figure}[t]
\vspace{-0.1in}
    \centering
    \includegraphics[width=2.9in]{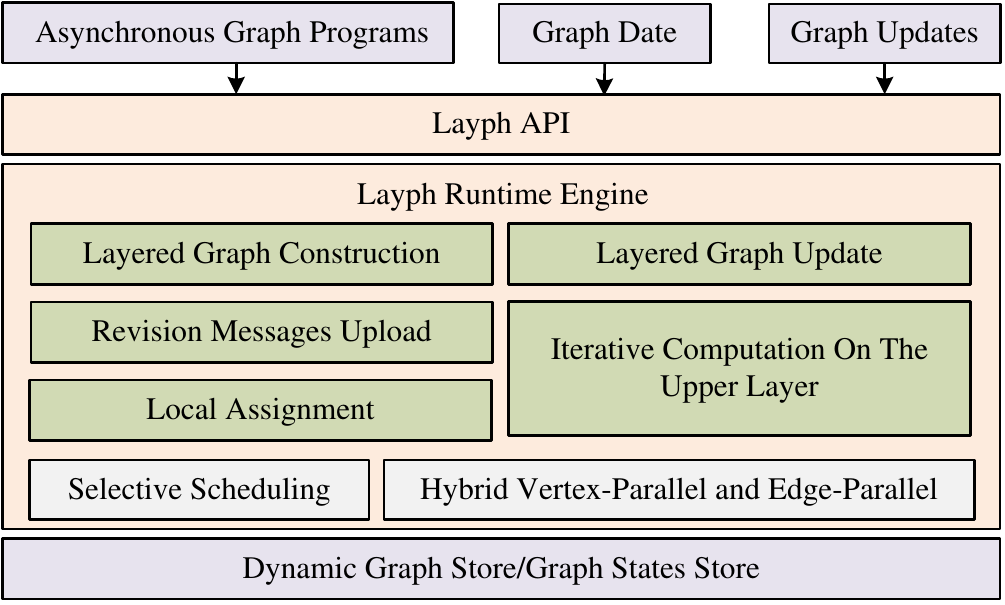}
    \vspace{-0.05in}
    \caption{System overview of \oursys.} % The red circles are the nodes whose states are updated.
    \label{fig:sys_framework}
    \vspace{-0.2in}
\end{figure}
}

%\subsection{Implementation Details}
We implement \oursys on top of %our previous work 
Ingress \cite{gong2021ingress}, an automated incrementalization framework equipped with different memoization policies to support vertex-centric graph computations. In this section, we evaluate \oursys and compare it with existing state-of-the-art incremental graph processing systems. 
\vspace{-0.02in}
\subsection{Experimental Setup}

\begin{table}[!t]
    \caption{Datasets used in the experiments}
    %\vspace{-0.1in}
    \label{tab:data}
    \centering
    \footnotesize
    {\renewcommand{\arraystretch}{1.2}
    \begin{tabular}{l| c |c| c }
        % \toprule
        \hline
        
        \hline
        {\textbf{Graph}} &
        {\textbf{Vertices}} &
        {\textbf{Edges}} &
        {\textbf{Size}} \\
        % \midrule
        % \hline
        %  Euro-Road (ER)~\cite{euroroad} & 50,912,018 & 108,109,320 & 0.94GB \\
         % \hline
         \hline

         \hline
         UK-2005 (UK)~\cite{uk} & 39,459,925 & 936,364,282 & 16GB  \\
         \hline
         IT-2004 (IT)~\cite{it-2004} & 41,291,594 & 1,150,725,436 & 19GB \\
         \hline
         SK-2005 (SK)~\cite{sk-2005} & 50,636,154 & 1,949,412,601 & 33GB \\
         \hline
         Sinaweibo (WB)~\cite{weibo} & 58,655,850 & 261,323,450 & 3.8GB \\
        %  Webbase-2001 (WB)~\cite{webbase} & 118,142,155 & 1,019,903,190 & 18GB \\
        % \bottomrule
        \hline

        \hline
    \end{tabular}
    }
    \vspace{-0.18in}
\end{table}

% This section evaluates \oursys and compares it with existing state-of-the-art incremental graph processing methods. 
We use AliCloud ecs.r6.13xlarge instance (52vCPU, 384GB memory, 64-bit Ubuntu 18.04 with compiler GCC 7.5) for these experiments.

\vspace{-0.02in}
\etitle{Graph Workloads}. We use four typical graph analysis algorithms in our experiments, including  Single Source Shortest Path (\SSSP), Breadth-First Search (\BFS), PageRank (PR), and Penalized Hitting Probability (\PHP) \cite{GuanWZSY11}. \SSSP and \BFS can be written in the form shown in Equation (\ref{eq:iterv}). We also rewrite \PHP and \PR into the form shown in Equation (\ref{eq:iterv}) using the method in \cite{zhang2013maiter, wang2020powerlog}. The former two are %traversal based, and they are 
considered converged when all vertex states no longer change. The latter two are %iterative based, and they are 
considered converged when the difference between the vertex states in two consecutive iterations is less than $1e^{-6}$.

%Among them, for the two algorithms of BFS and SSSP, when all vertex states on the graph no longer change, the convergence state is reached, and for PR and PHP, when the difference between the vertex states in two consecutive iterations is less than $1e^{-6}$, we consider that the algorithm has reached a state of convergence.

% Dataset and updates. 我们在实验中使用了五个真实图(see Table \cite{}), 包括路网图 Euro-Road(ER)\cite{},他是一个无向图, 和四个web graph UK-2005 \cite{}, IT-2005(IT) \cite{}，SK-2005(SK) \cite{}, 和webbase-2001 (WB) \cite{}. 我们通过从原始数据集中选择指定比例或者指定数量的边做作为图的更新$\Delta G$中的插入更新, 剩余的边作为原始图$G$，然后从$G$中选择与插入更新相同的数量的边作为$\Delta G$中的删除更新 。 对于顶点的增删情况，由于对比的部分系统不支持增加顶点和删除顶点，故不考虑增加顶点的情况，但是我们可以通过删除一个顶点的全部边来模拟删除该顶点。

\vspace{-0.02in}
\etitle{Datasets and Updates}. We use four real graphs (see Table \ref{tab:data}) in our experiments, including %a road network Euro-Road (ER) \cite{euroroad}, 
three web graphs UK-2005 (UK) \cite{uk}, IT-2004 (IT) \cite{it-2004}, and SK-2005 (SK) \cite{sk-2005}, 
and a social network Sinaweibo (WB) \cite{weibo}.
We constructed $\Delta G$ by randomly adding new edges to $G$ and removing existing edges from $G$.
The number of added edges and deleted edges are both 5,000 by default unless otherwise specified. 
$\Delta G$ refers to the edge changes by default, besides, we randomly generate a $\Delta G$ with 1,000 changed vertices (including 500 added vertices and 500 deleted vertices) to evaluate the performance of handling vertex updates.

%We randomly select the specified proportion or specified number of edges from the graph dataset as the insert updates forming $\Delta G$, and the remaining edges are used as the original graph $G$. Select the same number of edges from $G$ as delete updates and insert updates in $\Delta G$. For the addition and deletion of vertices, since some of the compared systems do not support adding and deleting vertices, the case of adding vertices is not considered, but we can simulate deleting a vertex by deleting all the edges of the vertex.

% Competitors. 我们比较了{\oursys}和四个先进的以顶点为中心的增量图系统，GraphBolt \cite{}, KickStarter \cite{}, DZIG \cite{}, Ingress \cite{}和RisGraph。其中KickStarter和RisGraph都是基于依赖树实现的单依赖增量图系统，所以它们不能处理PageRank和PHP算法。 In light of this, 我们仅仅测试PageRank和PHP(resp. SSSP和BFS)在GraphBolt和Ingress(resp. KickStarter and RisGraph).

\vspace{-0.02in}
\etitle{Competitors}.
% 这地方先简单介绍下这些系统呗，说核心技术，一句话就行：
% GraphBolt通过记录迭代过程中每轮的所有顶点的状态，在增量阶段通过图的变化来捕捉应该每轮中相关顶点状态的差异，对旧图的收敛结果进行校正，最后得到新图上正确的收敛结果。DZIG采用了一种稀疏感知增量处理技术，在图算法的迭代计算过程中能够自适应的切换增量策略，在存在稀疏计算的情况下保持更好的性能。KickStarter通过在迭代过程中维护一个依赖树，当图变化时利用依赖树来调整旧图上的收敛值，以此来避免结点全部重新初始化。RisGraph也是采用依赖树来支持单调图算的增量系统，其凭借高效的动态图结构、本地化数据访问和并发控制机制实现了高吞吐量和低延迟。Ingress是一个能够根据算法条件进行自动判断并能够采用最佳增量方案的图计算系统，其提出的MF增量策略和GraphBolt相比无须存储任何额外信息就能完成增量计算，具有高效和低存储开销的特性。
\eat{
Since experiments on GraphBolt \cite{mariappan2019graphbolt} and Ingress \cite{gong2021ingress} show that they are already more advanced than Differential Dataflow \cite{mcsherry2013differential}, Tornado \cite{shi2016tornado}, we will mainly focus on comparing
}
We compare {\oursys} with five state-of-the-art incremental graph processing systems, GraphBolt \cite{mariappan2019graphbolt}, KickStarter \cite{vora2017kickstarter}, DZiG \cite{DZiG}, Ingress \cite{gong2021ingress}, and RisGraph \cite{feng2021risgraph}. 
%Compared with GraphBolt, the proposed \textit{memoization-free} strategy in Ingress stores only the converged vertex states rather than the per-iteration vertex states, leading to higher efficiency and lower storage overhead. 
In fact, KickStarter and RisGraph do not support \PR and \PHP due to their single-dependency requirement. GraphBolt and DZiG do not provide the implementations of \SSSP and \BFS. In light of this, we only run \PR and \PHP (resp. \SSSP and \BFS) on GraphBolt and DZiG (resp. KickStarter and RisGraph). 
%\red{Furthermore, among these competitors only Ingress can support vertex updates. }
All of these systems are running with 16 worker threads.

% \begin{table}[t]
% \caption{Real-life graphs}\label{tab:data}
% \vspace{-0.1in}
%   \centering
% %  \begin{scriptsize}
% \begin{small}
% \setlength{\tabcolsep}{3pt}
%   \begin{tabular}{lccc}
%      \toprule
%      % after \\: \hline or \cline{col1-col2} \cline{col3-col4} ...
%      {\bf Graph} & {\bf \#Vertices} & {\bf \#Edges} & {\bf Size} \\
%      \midrule
%      UK-2005 (UK)~\cite{} & 39,459,925 & 936,364,282 & 16GB  \\
%      Euro-Road (UR)~\cite{} & 50,912,018 & 108,109,320 & 906M \\
%      IT-2005 (IT)~\cite{} & 41,291,594 & 1,150,725,436 & 19G \\
%      SK-2005 (SK)~\cite{} & 50,636,154 & 1,949,412,6014 & 33G \\
%      Webbase-2001 (WB)~\cite{} & 118,142,155 & 1,019,903,190 & 18GB \\
%      \bottomrule
%   \end{tabular}
% \end{small}
% \vspace{-0.1in}
% \end{table}

\newcommand{\tw}{0.18\textwidth}
\newcommand{\hsp}{\hspace{0.1in}}
\begin{figure*}[tbp]
\vspace{-0.3in}
\hspace{-0.2in}
  \centering
  %\begin{minipage}{1.50\textwidth}
  \centering
    \subfloat[SSSP]{\includegraphics[width = \tw]{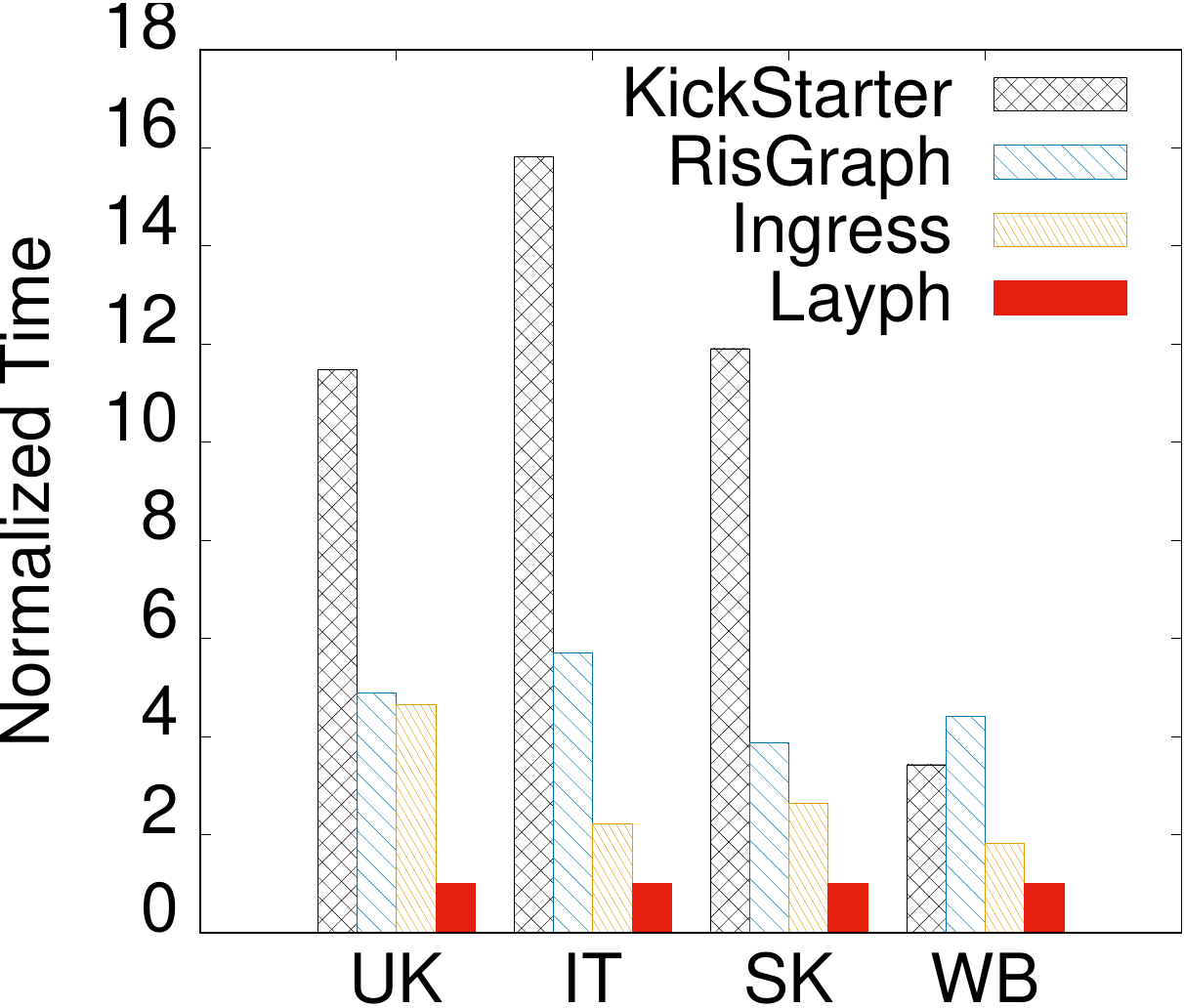}\label{fig:runtime_sssp}}\hsp
  \subfloat[BFS]{\includegraphics[width = \tw]{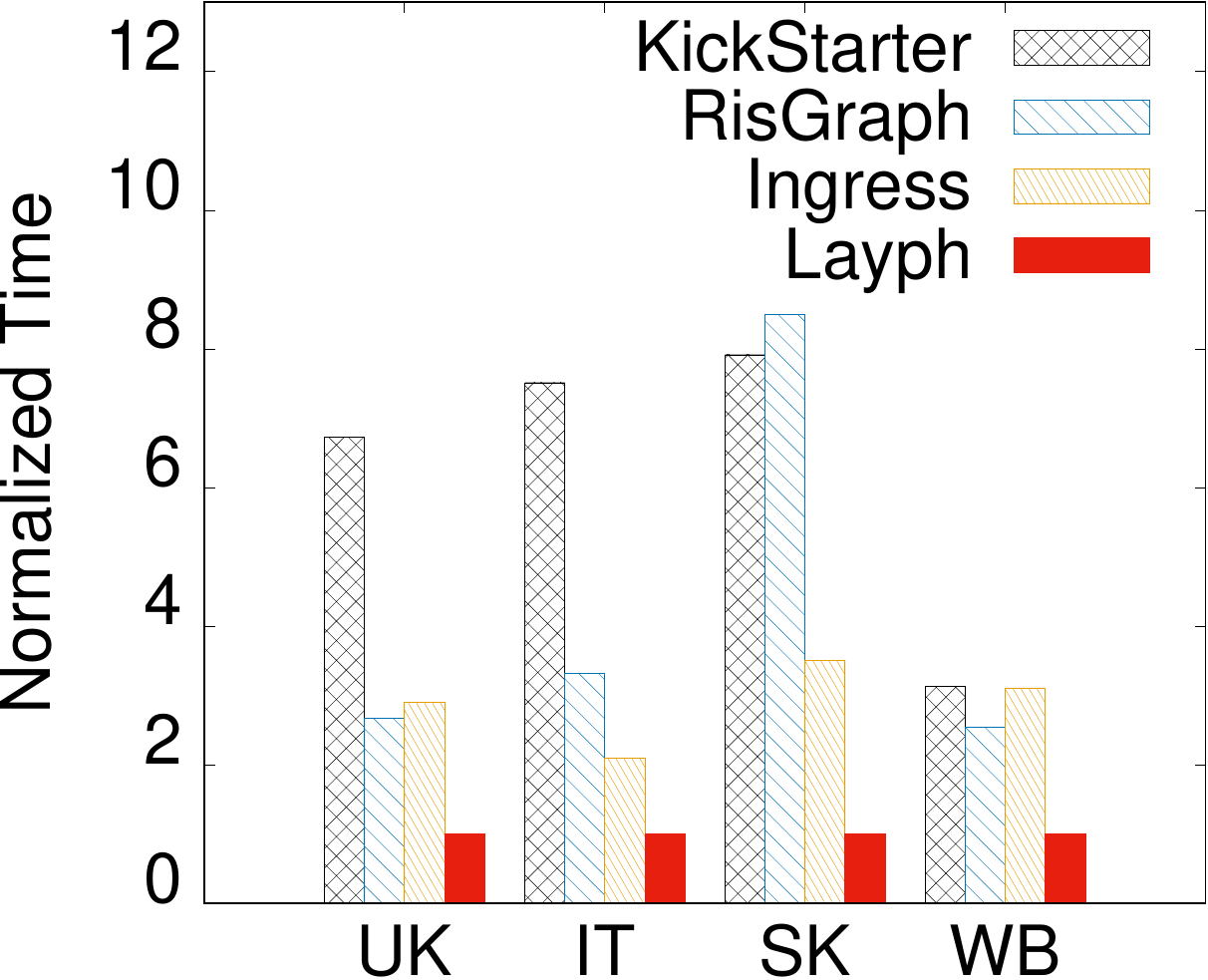}\label{fig:runtime_bfs}}\hsp
  \subfloat[PageRank]{\includegraphics[width = \tw]{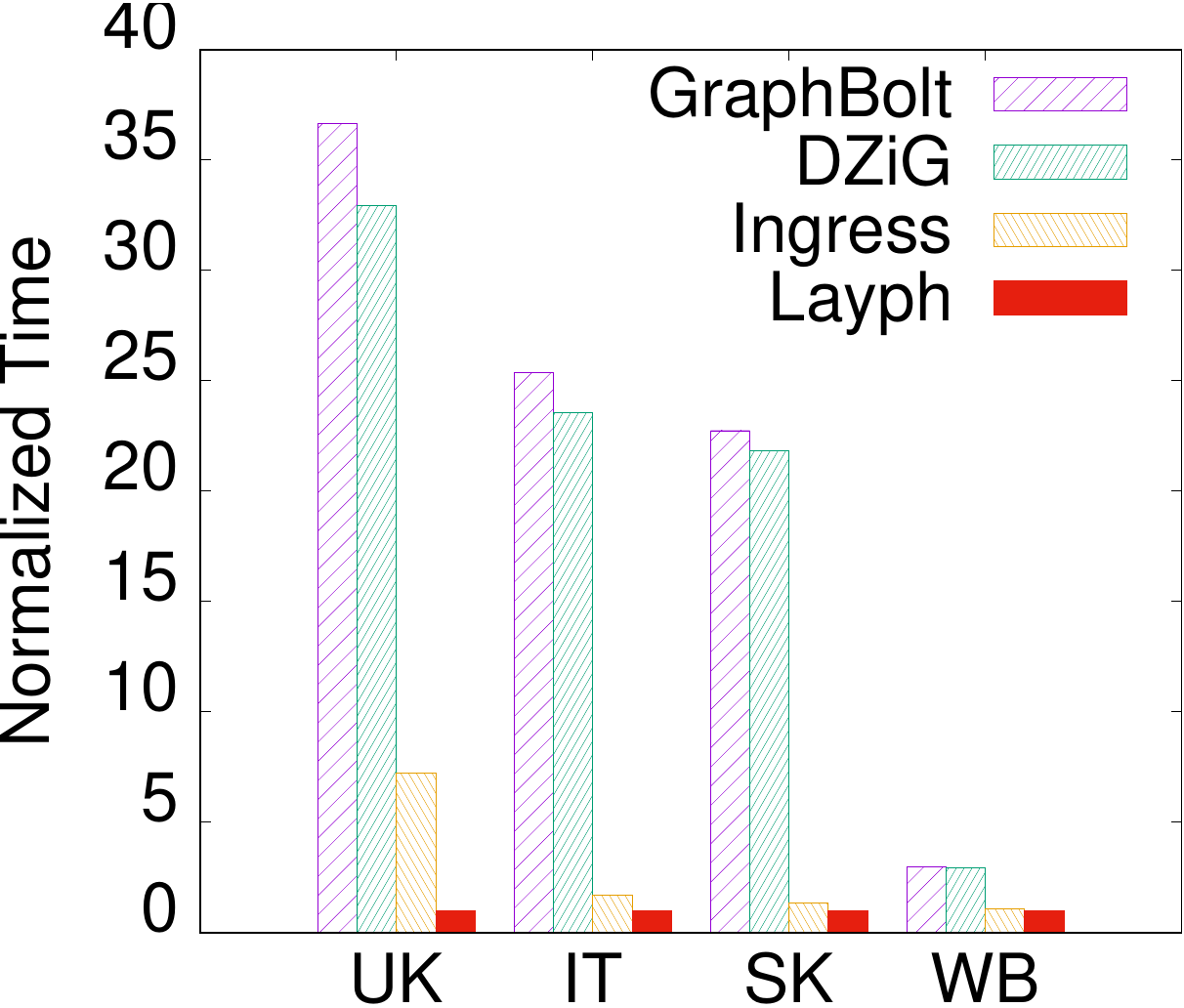}\label{fig:runtime_pagerank}}\hsp
  \subfloat[PHP]{\includegraphics[width = \tw]{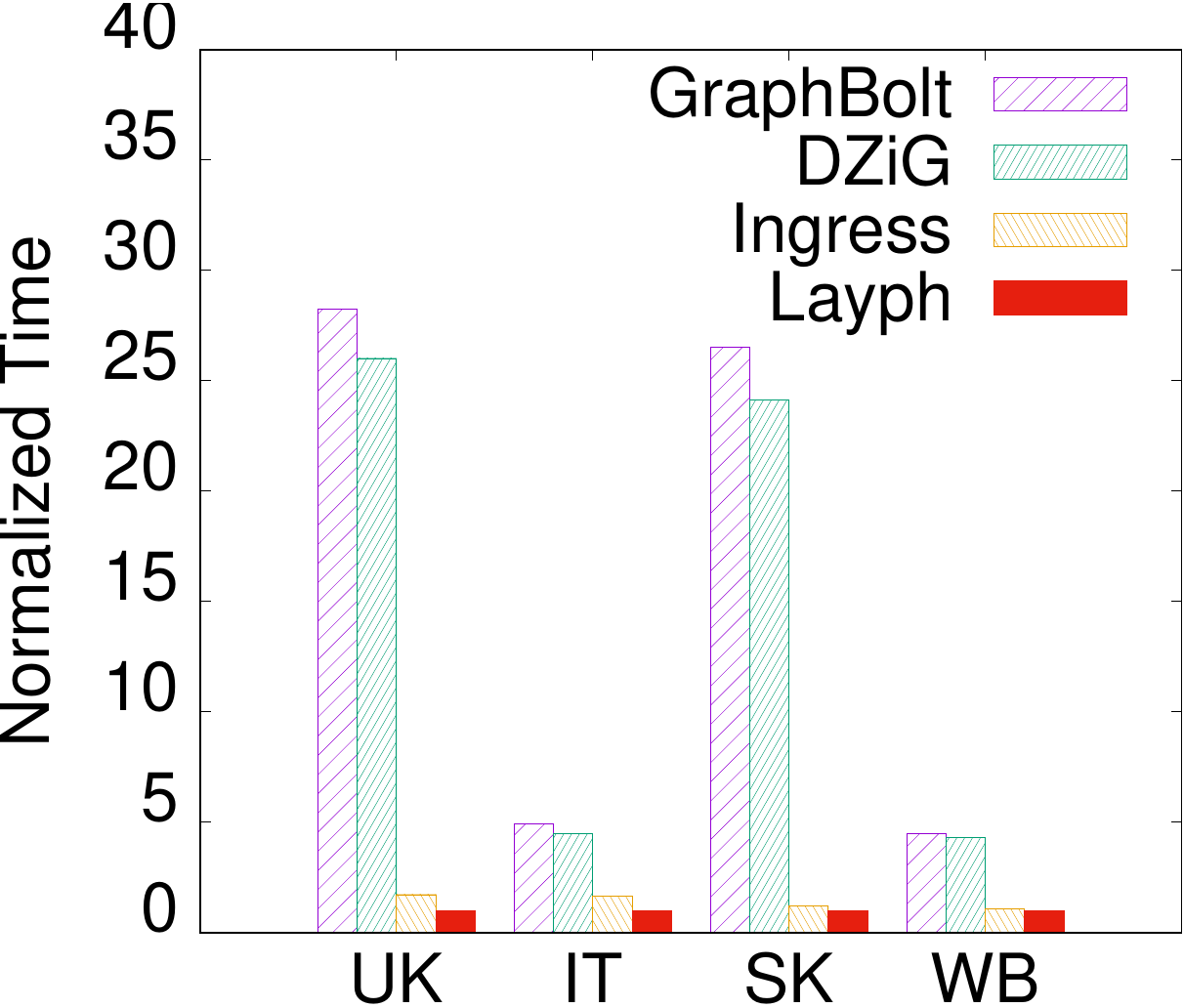}\label{fig:runtime_php}}\hsp
  \subfloat[\red{PageRank(Vertex update)}]{\includegraphics[width = \tw]{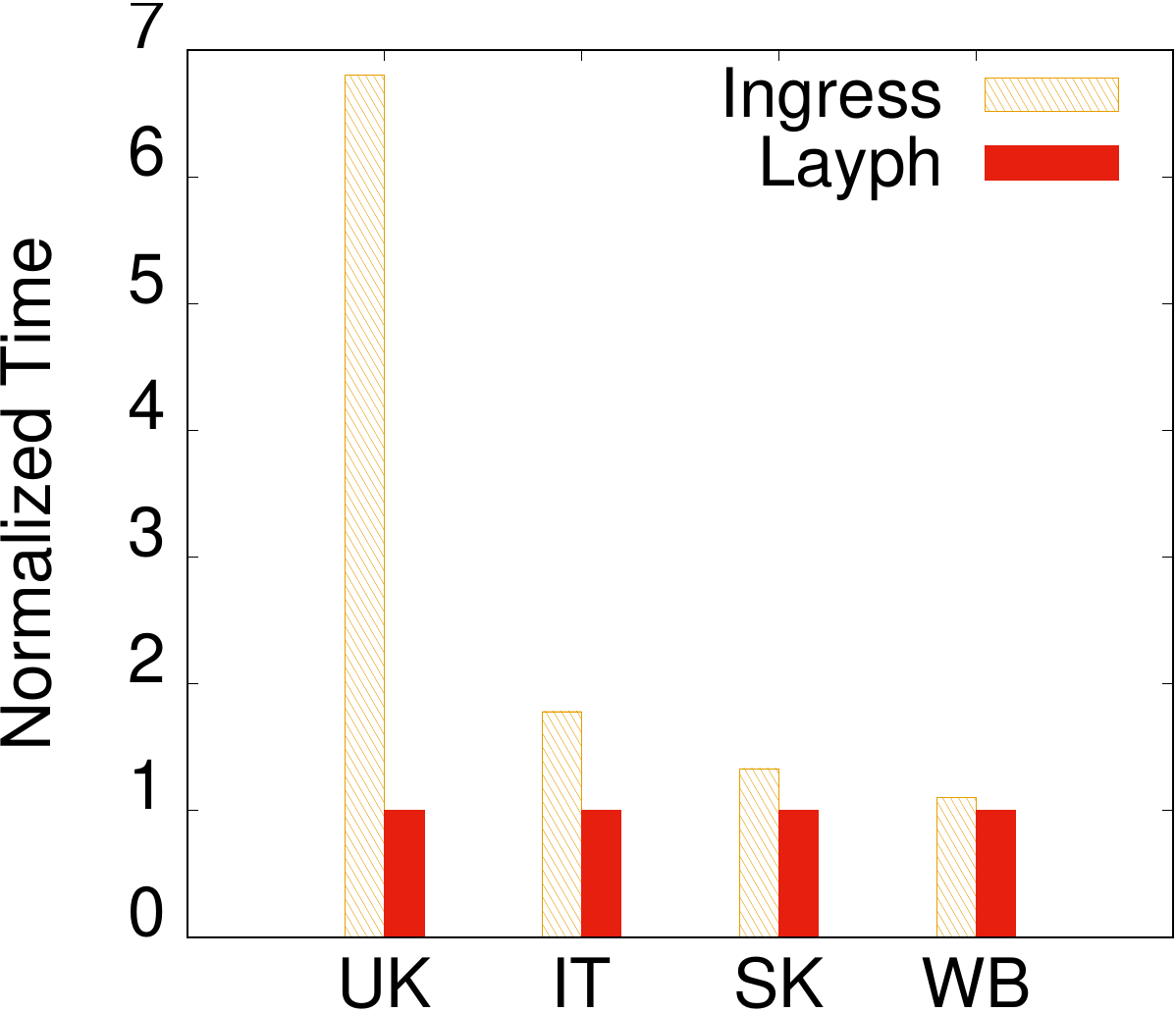}\label{fig:runtime_pr_vertex}} 
  %\end{minipage}
  \vspace{-0.08in}
    \caption{%Convergence
    Response time comparison.}\label{fig:compare_runtime}
\end{figure*}

\begin{figure*}[tbp]
\vspace{-0.3in}
\hspace{-0.2in}
  \centering
 % \begin{minipage}{1.50\textwidth}
  %\flushleft
    \subfloat[SSSP]{\includegraphics[width = \tw]{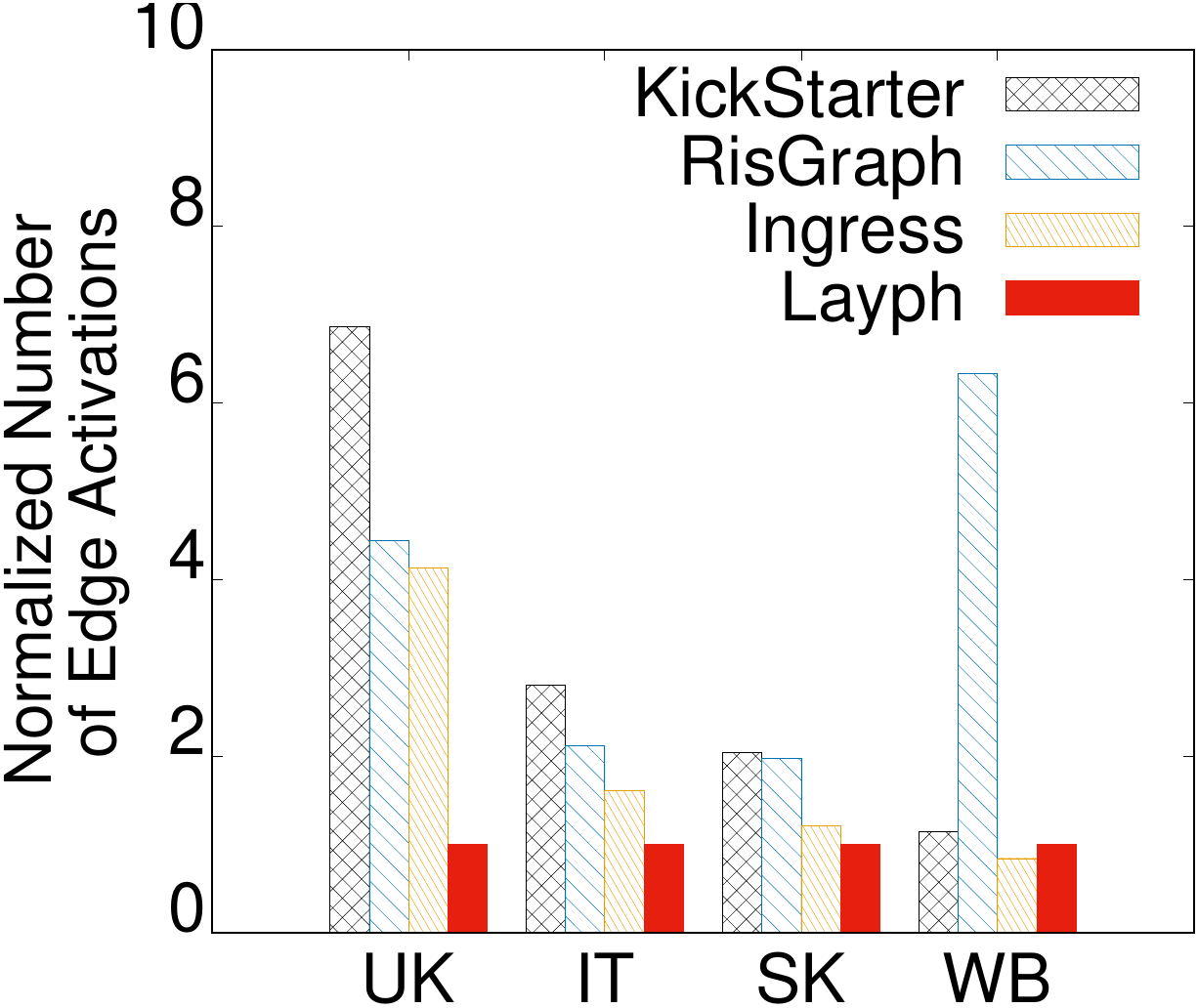}\label{fig:active_sssp}}\hsp
  \subfloat[BFS]{\includegraphics[width = \tw]{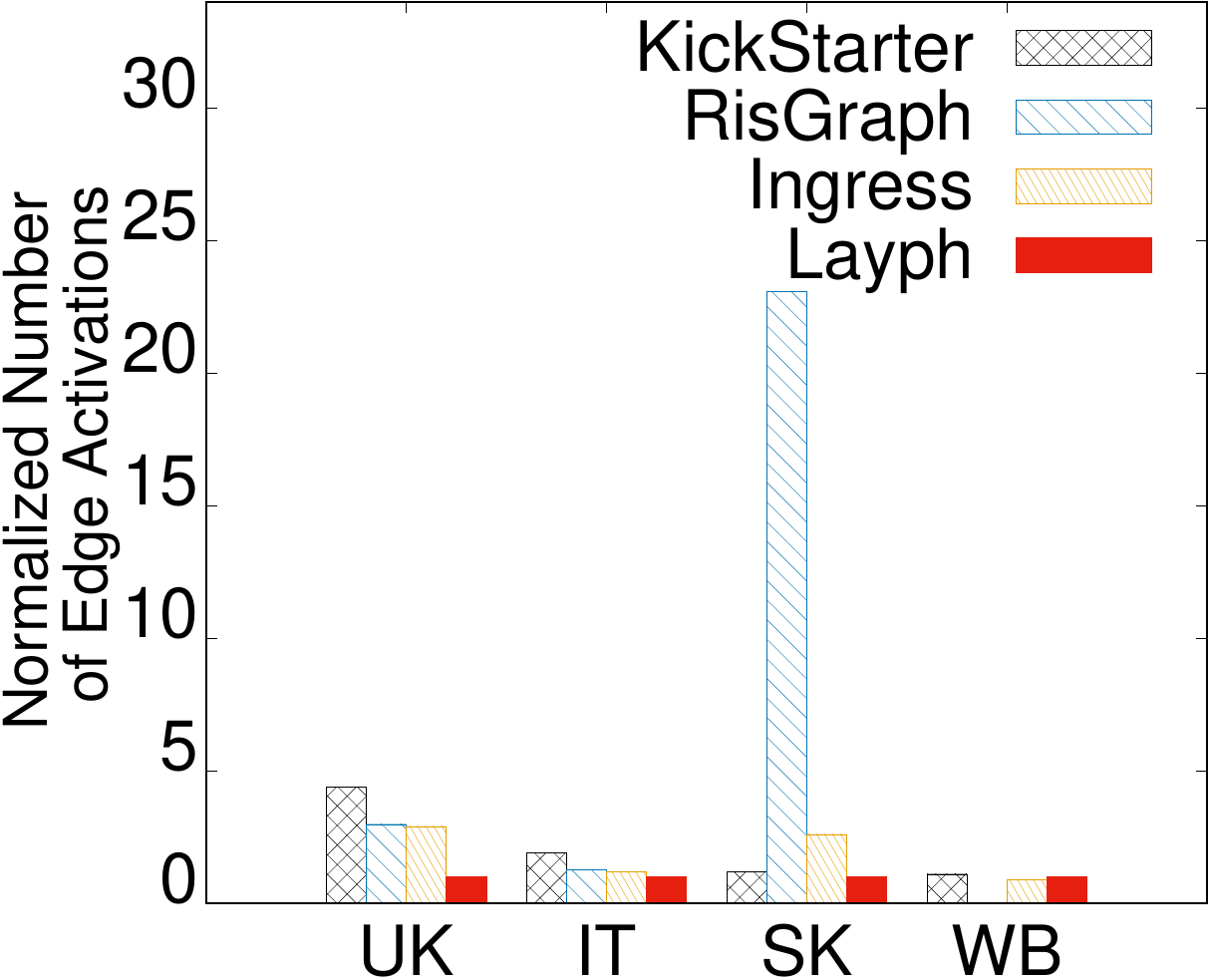}\label{fig:active_bfs}}\hsp
  \subfloat[PageRank]{\includegraphics[width = \tw]{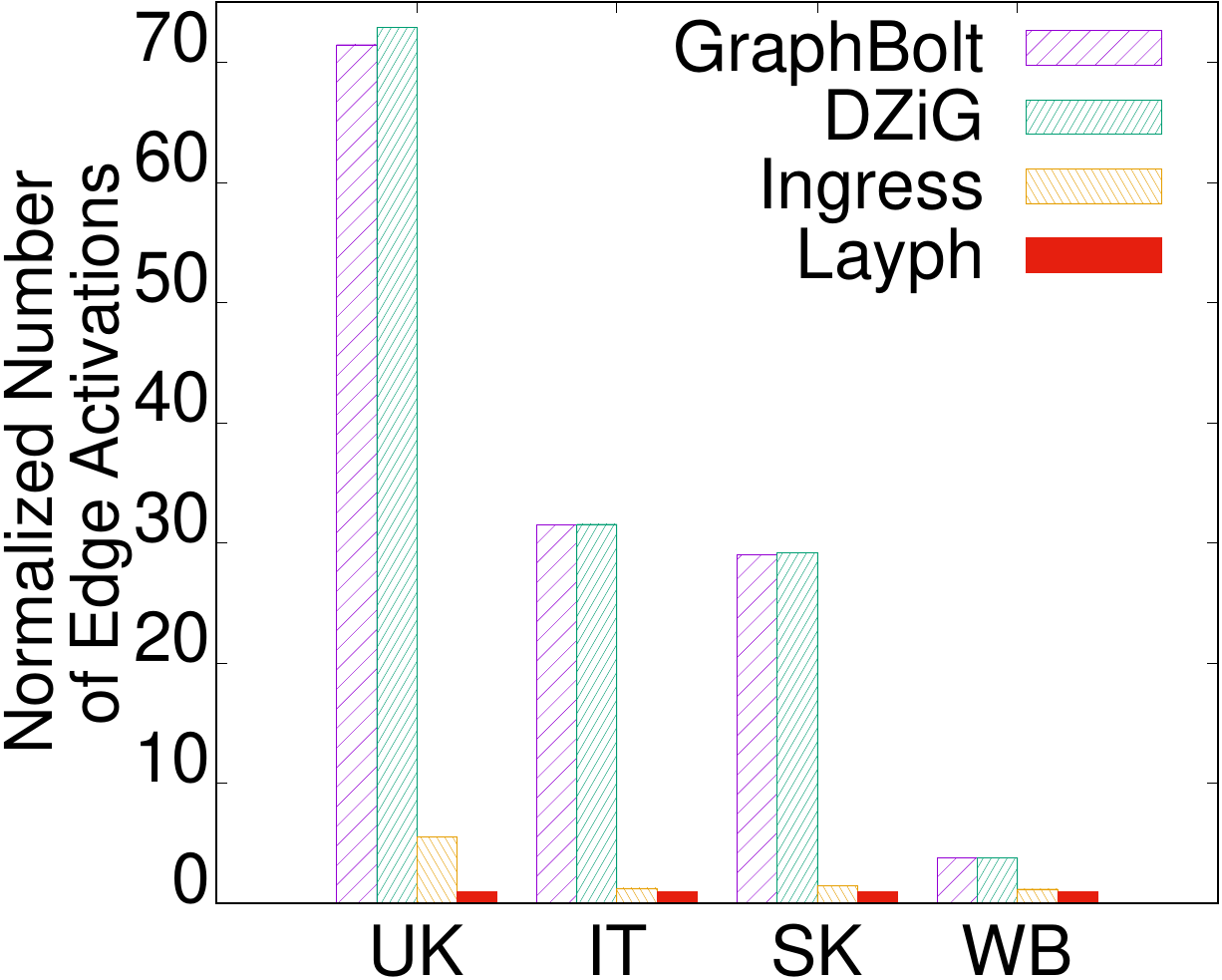}\label{fig:active_pagerank}}\hsp
  \subfloat[PHP]{\includegraphics[width = \tw]{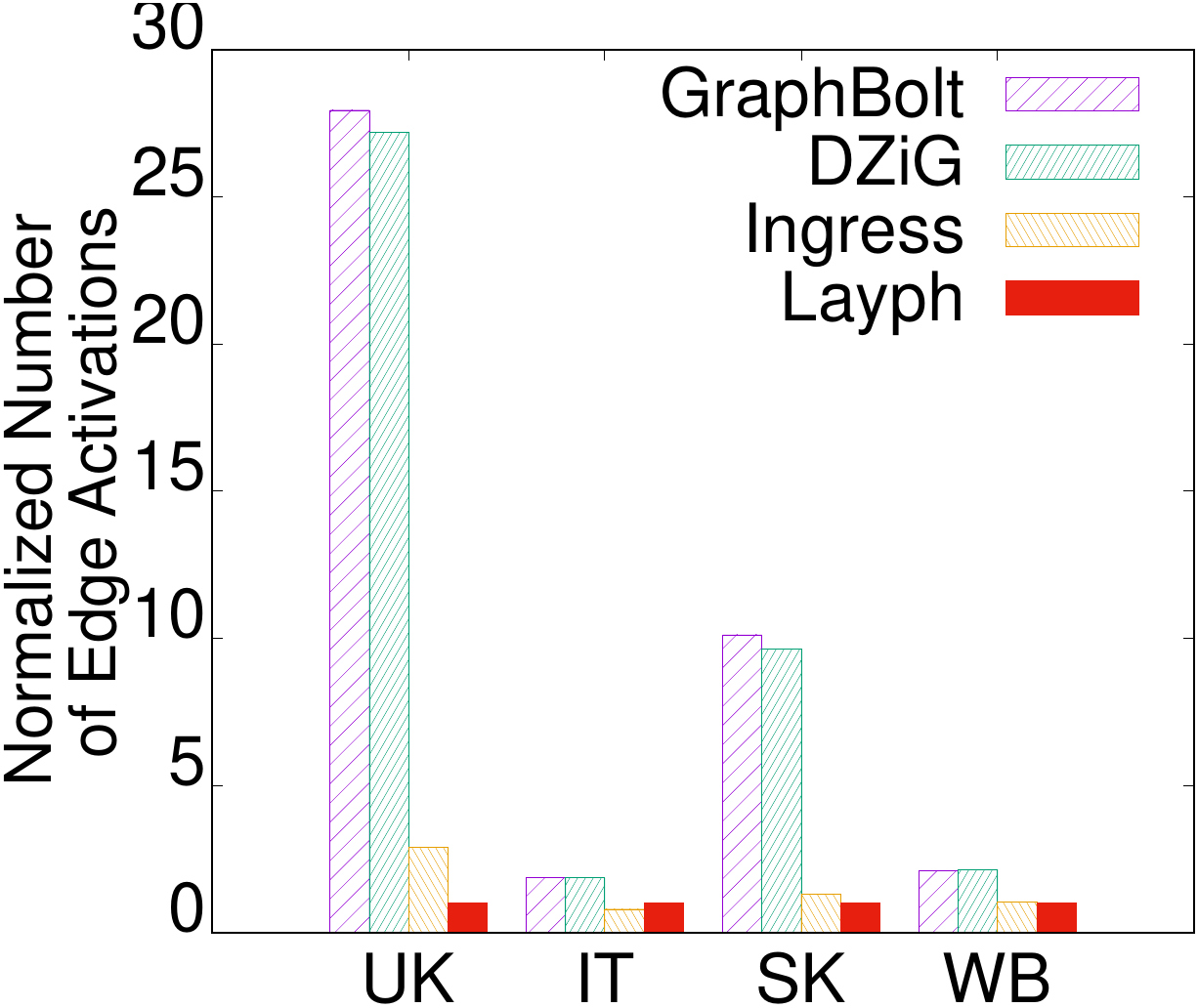}\label{fig:active_php}}\hsp
  \subfloat[\red{PageRank(Vertex update)}]{\includegraphics[width = \tw]{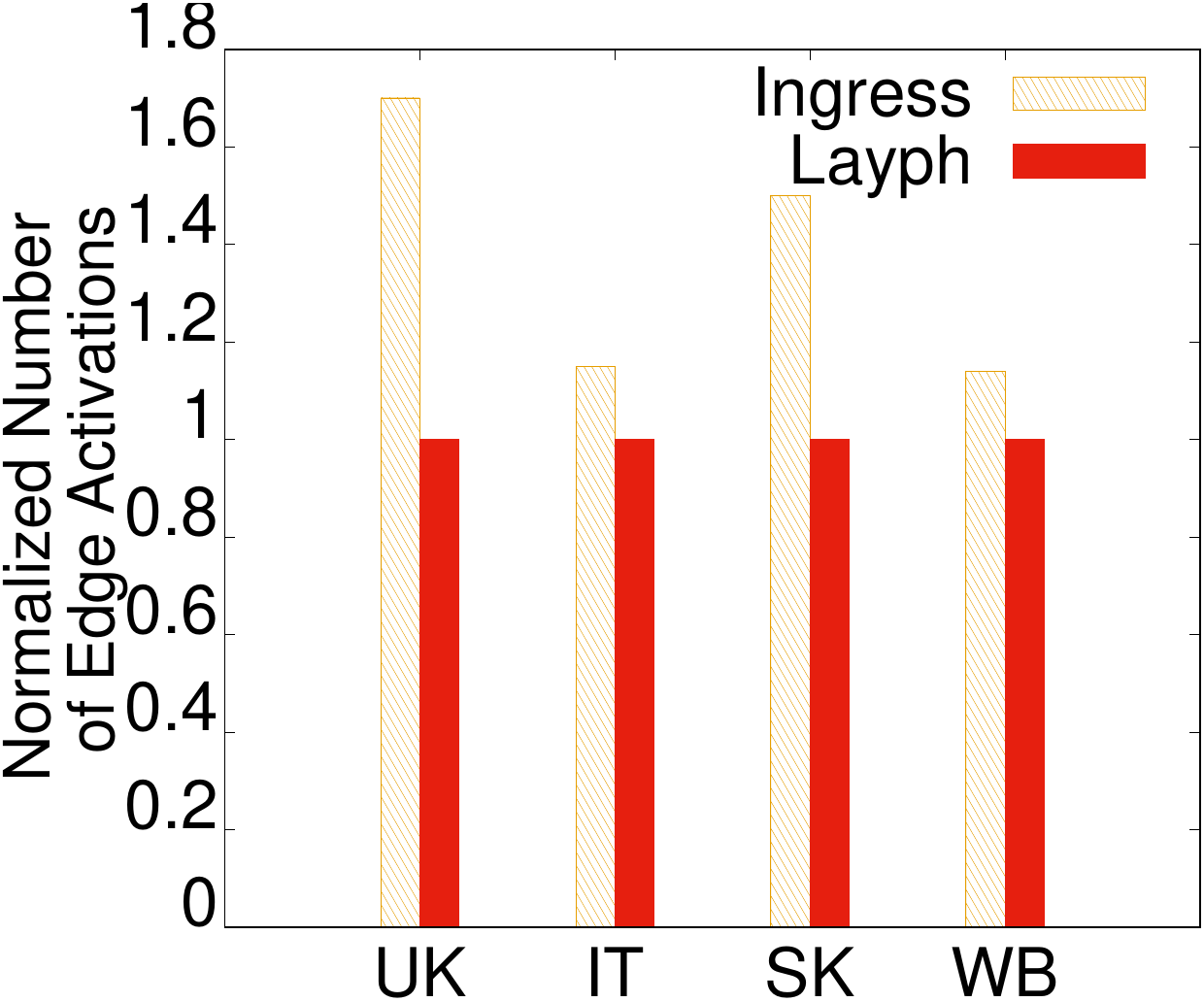}\label{fig:active_pr_vertex}} 
%  \end{minipage}
  \vspace{-0.08in}
    \caption{Number of edge activations comparison.}\label{fig:compare_active}
    \vspace{-0.22in}
\end{figure*}

\begin{figure*}%[tbp]
%\vspace{-0.3in}
% \centering
\hspace{-0.23in}
\begin{minipage}{.3\textwidth}
    % \centering
    % \vspace{-0.1in}
    \begin{center}
    \includegraphics[width=0.7\linewidth]{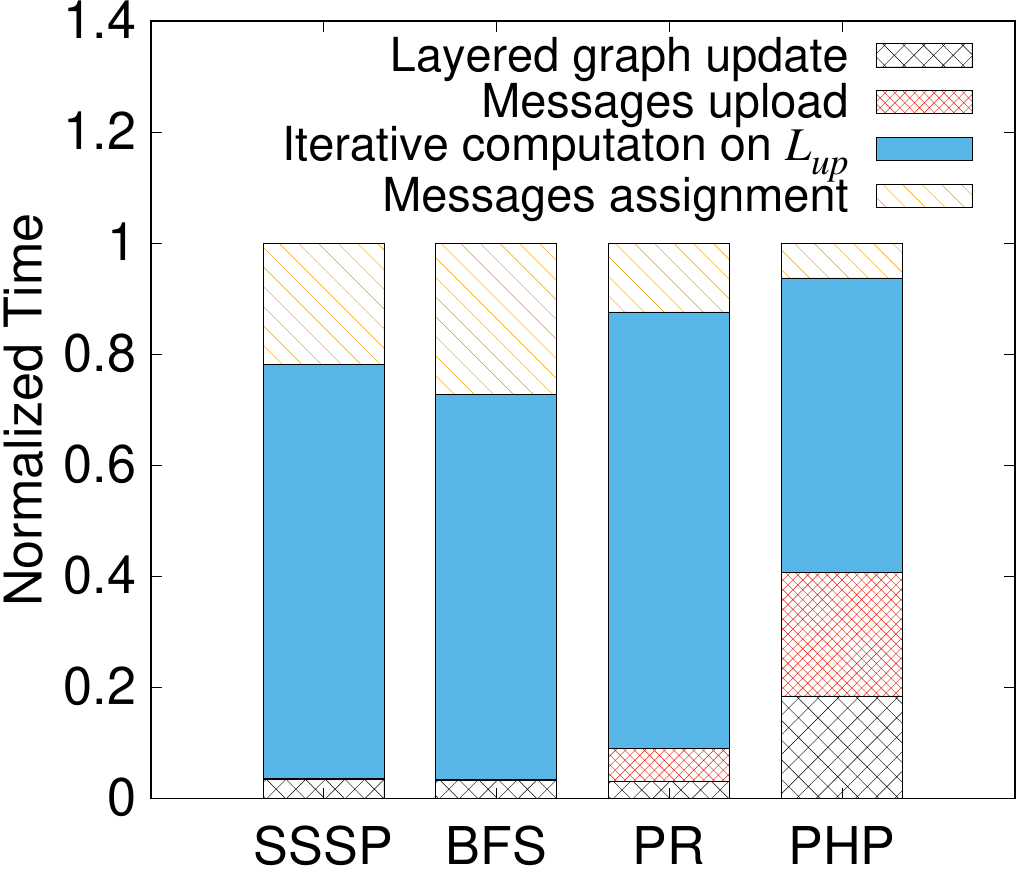}
    % \hspace{-0.18in}
    %\vspace{0.18in}
    \caption{Runtime breakdown. % (UK) 
    % \red{modify phase name}
    } % The red circles are the vertices whose states are updated.
    \label{fig:breakdown}
    \vspace{-0.12in}
    \end{center}
\end{minipage}%
\hspace{-0.23in}
\begin{minipage}{.73\textwidth}
% 		\centering
        % \hspace{-0.28in}
		\subfloat[Graph size]{\label{fig:skeleton_size}\hsp
		\includegraphics[width=0.29\linewidth]{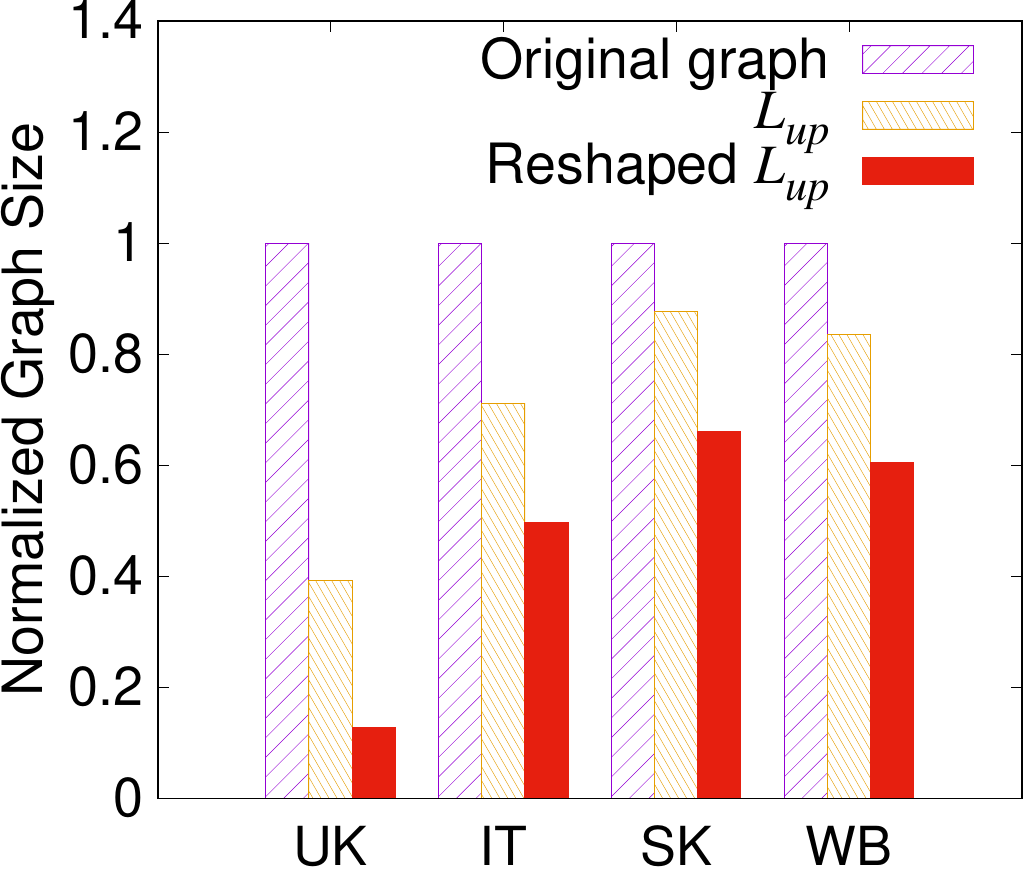}} \ \
		\hspace{0.01in}
		\subfloat[SSSP runtime]{\label{fig:master_mirror_sssp}\hsp
		\includegraphics[width=0.29\linewidth]{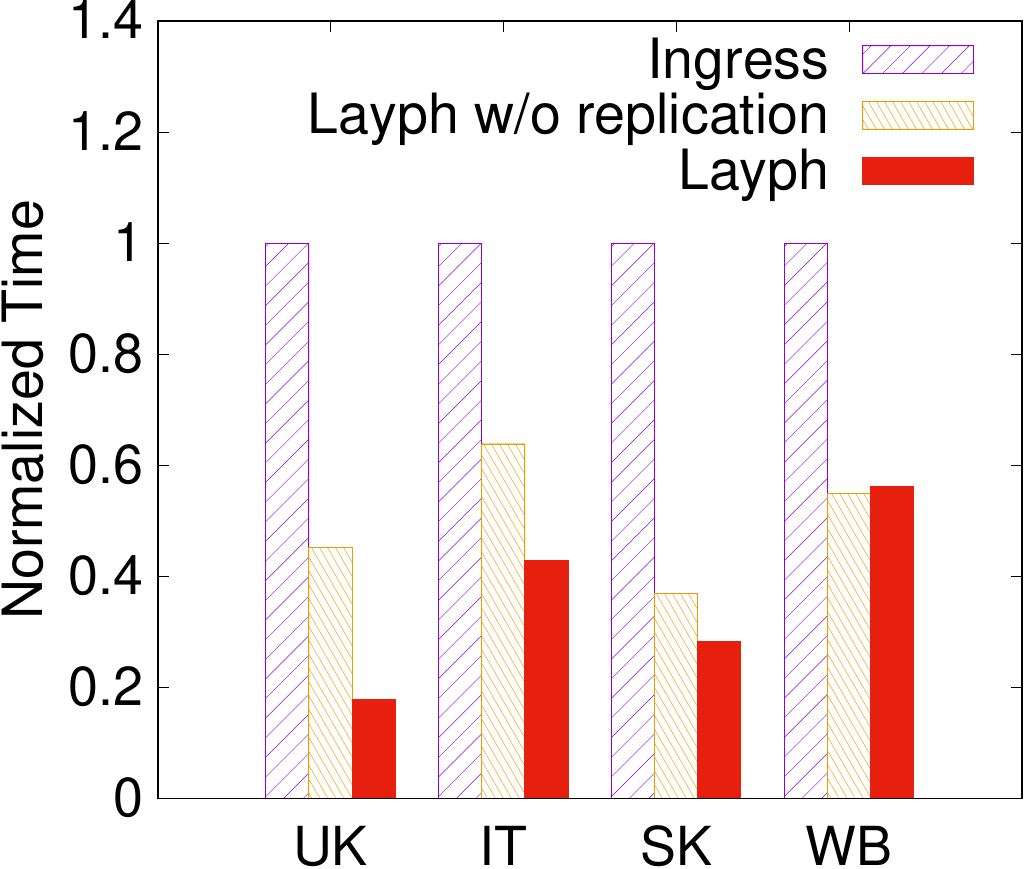}} \ \
         \hspace{0.01in}
		\subfloat[PageRank runtime]{\label{fig:master_mirror_pr}\hsp
		\includegraphics[width=0.29\linewidth]{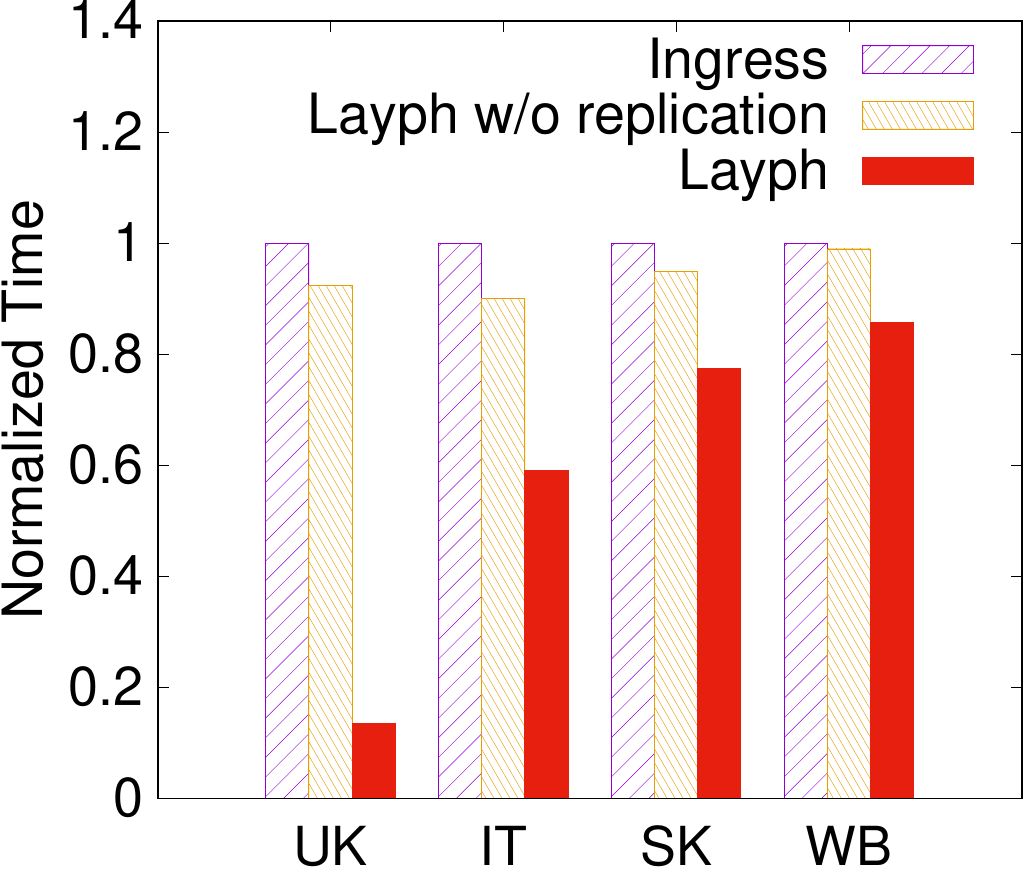}}
		\vspace{-0.08in}
		\label{fig:master-mirror}
		\caption{Effect of vertex replication.
		%\red{(a) use skeleton?}
		}
% 		\vspace{-0.1in}
\end{minipage}
% \vspace{-0.1in}
\end{figure*}
\vspace{-0.15in}

\begin{figure*}%[tbp]
\vspace{-0.2in}
    \centering
    %\hspace{-0.05in}
    \begin{minipage}{.5\textwidth}
    % 	\centering
    % \hspace{-0.05in}
    	\subfloat[SSSP %(UK)
    	]{\label{fig:thread:sssp}\hsp
    	\includegraphics[width=0.42\linewidth]{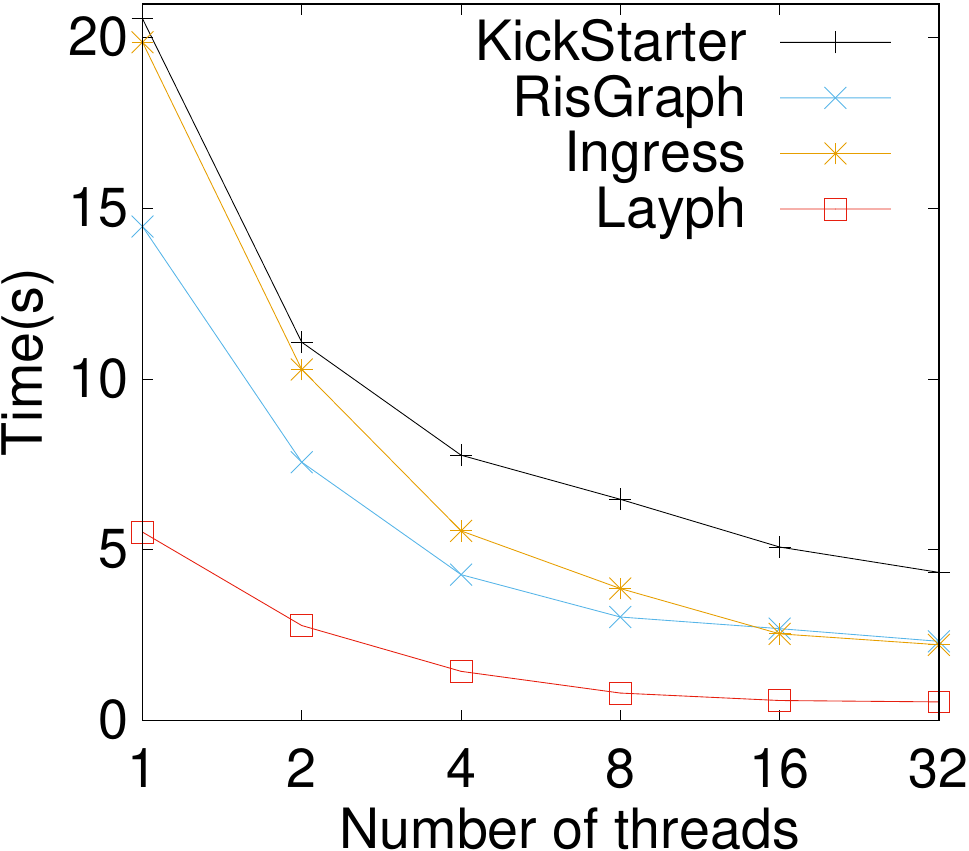}}\ \ \
    % 	\hspace{0.1in}
    	\subfloat[PageRank %(UK)
    	]{\label{fig:thread:pagerank}\hsp
    	\includegraphics[width=0.42\linewidth]{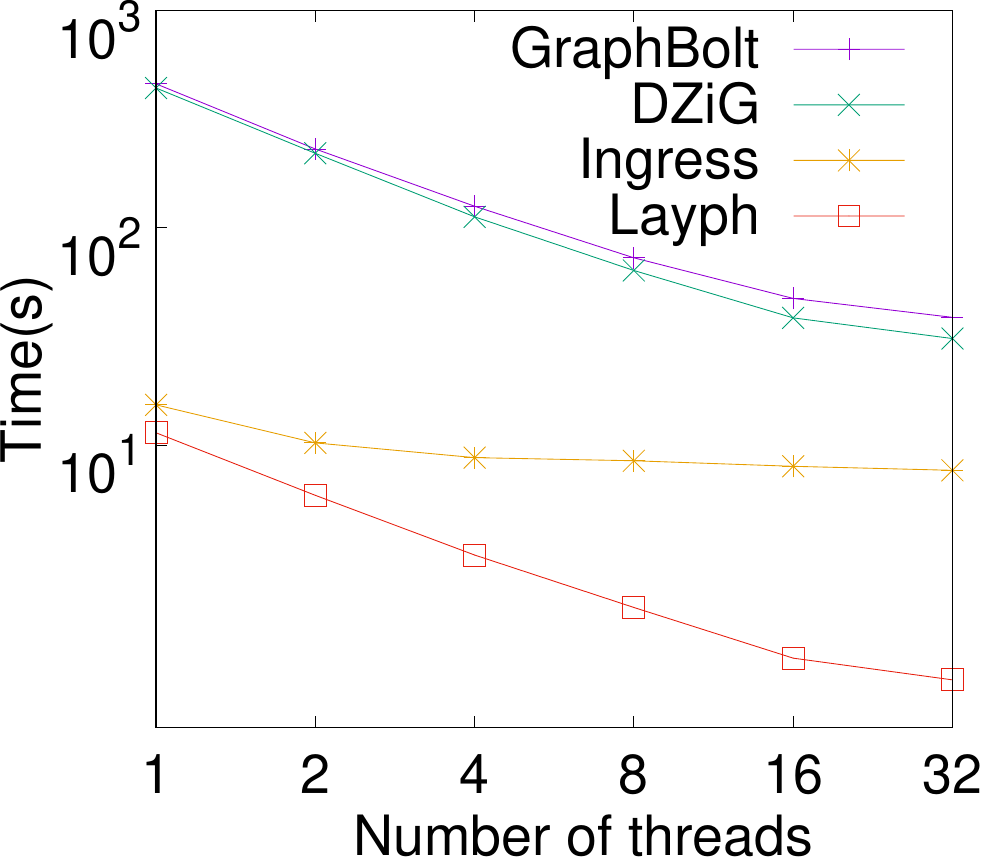}}
    	\vspace{-0.05in}
    	\caption{Scaling number of threads from 1 to 32.}
    	\label{fig:thread}
    % 	\vspace{-0.1in}
    \end{minipage}%
    \hspace{-0.1in}
    \begin{minipage}{.50\textwidth}
    	\centering
    	\subfloat[SSSP %(UK)
    	]{\label{fig:vary_update_sssp}\hsp
    	\includegraphics[width=0.42\linewidth]{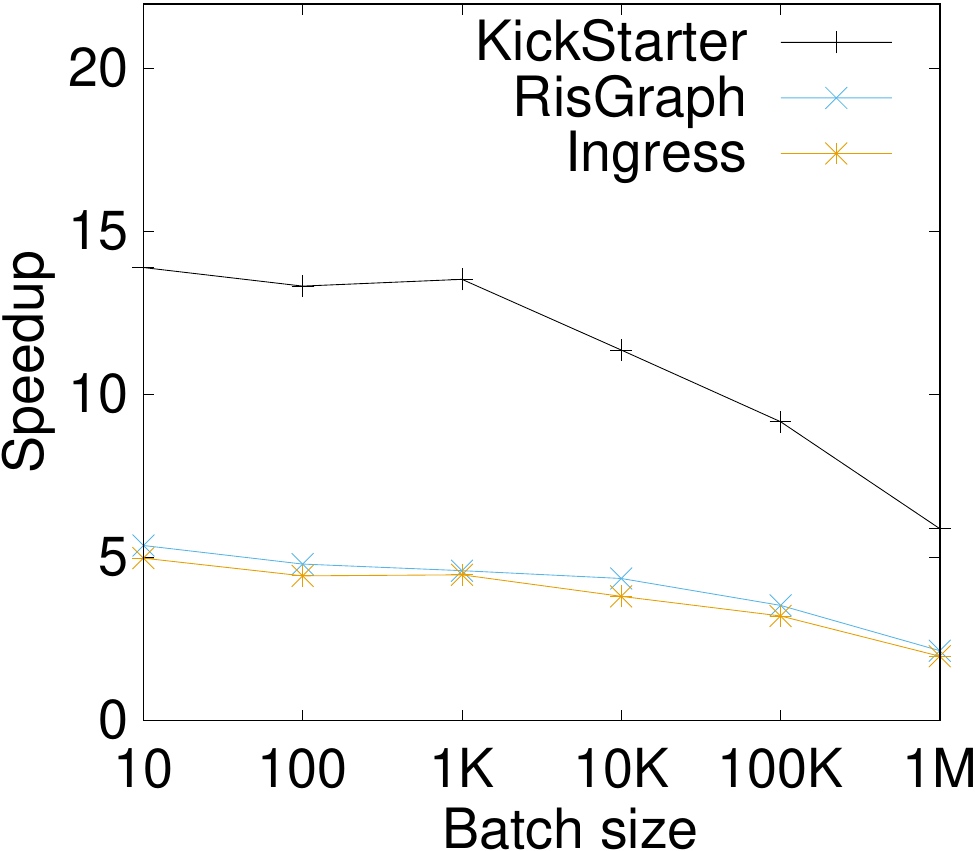}} \ \ \
    	\subfloat[PageRank %(UK)
    	]{\label{fig:vary_update_pr}\hsp
    	\includegraphics[width=0.42\linewidth]{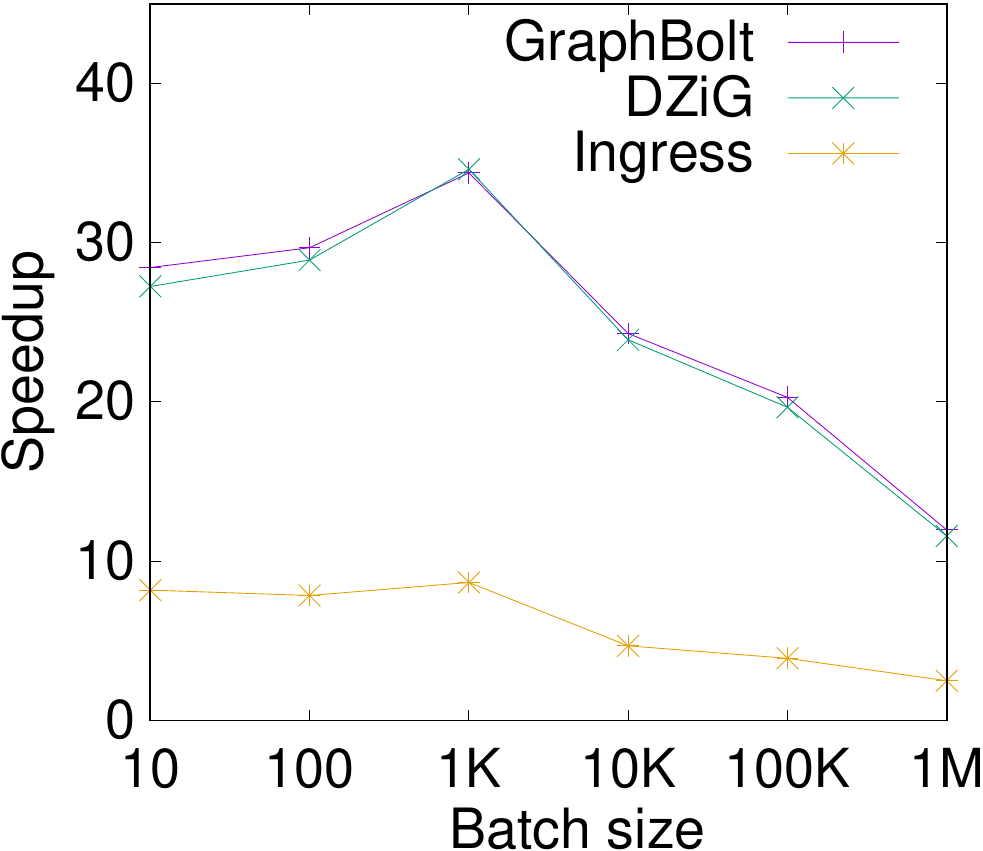}}
    	\vspace{-0.05in}
    	\caption{Speedup over competitors when varying batch size.
        }
    	\label{fig:vary_update}
    % 	\vspace{-0.1in}
    \end{minipage}
\vspace{-0.2in}
\end{figure*}

\vspace{0.05in}
\subsection{Overall Performance}

% time和activation times一起介绍，我们首先比较了各个系统的response time针对图变化后。
%\etitle{Response time}. 
We first compare {\oursys} with the competitors in response time of each workload executed on different datasets. The {\em Normalized} results are reported in Figure \ref{fig:compare_runtime}, 
where the response time of \oursys is treated as the baseline, {\ie} {\oursys} finishes in unit time 1. 
In particular, Figure 5e reports the response time for processing vertex updates, while the rest is used for edge updates. We can see that the improvement in handling vertex changes in \oursys is consistent with the improvement in handling edge changes. When updating vertices, the other systems meet runtime errors, thus we only compare Ingress with \oursys.
It is shown that \oursys consistently outperforms others in most cases. Specifically, {\oursys} achieves 3.13-15.82$\times$ (8.49$\times$ on average) speedup over KickStarter, 2.54-8.49$\times$ (4.49$\times$ on average) speedup over RisGraph, 2.99-36.66$\times$ (18.99$\times$ on average) speedup over GraphBolt, 2.92-32.93$\times$ (17.53$\times$ on average) speedup over DZiG, and 1.06-7.22$\times$ (%2.49
2.54$\times$ on average) speedup over Ingress. To explain the reason for the above results, we also report the total number of edge activations in Figure \ref{fig:compare_active}. An edge activation represents an $\GE$ operation. In most graph workloads, the cost of $\GE$ is much greater than that of $\AGG$ operation, because the number of $\GE$ and the unit cost of $\GE$ are often both larger than that of $\AGG$. From Figure \ref{fig:compare_runtime} and Figure \ref{fig:compare_active}, we can see that the normalized number of edge activations is a similar trend to the normalized response time of each system. %, except the \SSSP and \BFS in WB.
%为了解释我们效果的原因，我们也report了增量执行过程中激活边的总次数。一次激活就代表一次f操作，在图计算中，f操作的开销远大于G操作的开销，因为f操作次数和单位开销往往比g大。从图中可以看出，激活边次数与系统的response时间大致相同，除了WBSSSP和WBBFS，我们的系统主要通过减少F的次数来减少responsetime。

Regarding \SSSP and \BFS, RisGraph is faster than KickStarter since it allows more parallelism during incremental updates and allows for localized data access. Ingress and RisGraph are comparable because the memoization-path engine in Ingress follows a similar idea. % of allowing parallelization. 
\oursys outperforms the other competitors by leveraging the layered graph. Note that, when performing BFS on WB, RigGraph is slower than \oursys but with fewer edge activations. This is because that RisGraph can identify the safe and unsafe updates to reduce edge activations. It just so happens that most of the updates on WB are safe for \BFS. However, the additional cost of identifying the safe or unsafe is relatively expensive since WB is very small. While in SSSP, compared with Ingress, \oursys also requires less response time but with more edge activations. This is because there are some large dense subgraphs in WB, 
% which require more shortcut updates. More shortcut updates increase the number of edge activations. 
requiring more shortcut updates, which increase the number of edge activations. 
Since \oursys is parallel-friendly for shortcut updates, it will only have a small effect on the response time.    %WB中的子图都比较大，需要更多的shortcut更新，额外增加了edgeactivation的数量，但是由于shortcut更新具有良好的并行性在oursys中，因此我们的

Regarding \PR and \PHP, DZiG is faster than GraphBolt since DZiG has a sparsity detection mechanism, based on which it can adjust the incremental computation scheme. %to adapt to different execution stages. 
Besides, Ingress is faster than DZiG and GraphBolt. This can be attributed to its memoization-free engine %with selective scheduling optimization,
which is more efficient than others.  
% \oursys is built on top of Ingress and can further limit the change propagation scope by graph sketching as shown in Figure \ref{fig:update-prop}.\todo 
\oursys is built on top of Ingress, and can further limit the iterative computation scope with the layered graph, which reduces the number of activation edges, as shown in Figure \ref{fig:compare_active}. 
% \eat{The improvement over Ingress is more evident in the UK graph when running \PR. This is because there are more high degree vertices in the UK dataset, which results in more atomic conflicts on the original graph. The \oursys can alleviate this problem %by making the graph smaller.
% by having the upper layer smaller and the lower layer with good parallelism. }
We find that \oursys exhibits less improvement on WB. %dataset when running \PR,
% 应该删了或者加PHP
The reason is that the subgraphs in WB are much larger than that in other graphs, which increases costs and weakens gains.
%. This leads to significant update costs for shortcuts, which may overweight the benefits of the layered graph. 
The reason will be further explained in Section \ref{sec:expr:reshape}.

\subsection{Runtime Breakdown}
\label{expr:breakdown}
% batch processing: runtime of different phases
% incremental processing: runtime

During incremental computation, %our approach
\oursys consists of %contains
four phases: the layered graph update, %the initiation of vertex states on skeleton,
revision messages upload, 
iterative computation on the upper layer, and messages assignment. To study the time spent in each phase, we run four algorithms on UK and record the runtime of each phase. The proportion of runtime for different phases is shown in Figure \ref{fig:breakdown}. We can see that the iterative computation takes up most of the runtime. The messages assignment is the second most expensive phase. The layered graph update and revision messages upload are both very fast except in \PHP. This is because the iterative computation of \PHP is very fast, say 418 ms, which makes those two phases relatively longer. The results indicate that the additional cost in our system, \ie the maintenance of the layered graph, is lightweight. Based on the above experimental analysis, it is worth adopting the layered graph in incremental graph processing.

\subsection{Varying Number of Threads}

% 我们接下来测试了{\oursys}在多线程下的可拓展性。我们评估了KickStarter, RisGraph, Ingress和{\oursys}在UK下运行SSSP的性能，我们通过将线程数从1增加到了32。实验结果如图\ref{fig:Scalability}所示，展示的四个系统均随着线程数的增加,运行时间都在逐步下降. 其中在1-8个线程时,每个系统运行时间下降相对较快,而之后增将线程则都是平滑下降,这是由于以上系统中均采用了原子操作来保证并行条件下结果的正确性,随着线程数的增加,并行冲突加剧,导致运行时间不能完全与线程数成比例下降. 此外,可以看出我们的系统在不同线程数下均为最优的性能. 我们统计了KickStarter, RisGraph, Ingress和{\oursys}在一个线程和32线程时的时间比值, 分别为4.7, 6.2, 9.0, 10.1, 可以看出我们的系统在多线程下拓展下具有更大优势.

We vary the number of execution threads from 1 to 32 to see the runtime reduction. We run \SSSP on UK and compare \oursys with KickStarter, RisGraph, and Ingress. The results are shown in Figure \ref{fig:thread:sssp}. We can see that as the threads increase, the runtime decreases steadily in all systems as expected. The reduction is smoother when the number of threads is larger than 8. This is because all these systems use atomic operations to guarantee correctness, hence threads will lead to more write-write conflicts which will hurt parallelism. Compared with the runtime with 1 thread, \oursys with 32 threads can achieve 10.1$\times$ speedup, which is higher than KickStarter (4.7$\times$ speedup), RisGraph (6.2$\times$ speedup), and Ingress (9.0$\times$ speedup). We also run \PR on UK and compare \oursys with GraphBolt, DZiG, and Ingress. The results are reported in Figure \ref{fig:thread:pagerank} where a base-10 log scale is used for the Y axis. We can observe that GraphBolt, DZiG, and {\oursys} show better scaling performance than Ingress. The reason is that the problem of the write-write conflict in \PR is more serious than that in \SSSP. In GraphBolt and DZiG, vertex states need to be recorded during each iteration, which can alleviate the conflict problem with massive space cost in sacrifice. In \oursys, both the shortcut update process and the local assignment process contain many independent local computations, making \oursys more parallel-friendly. Therefore, \oursys can benefit more from parallelism.

%In the figure, it can be seen that except for a data fluctuation in KickStarter, the running time of all systems can be smoothly reduced with the number of threads in other cases, and {\oursys} in different threads in the case of the number, it is better than other systems.

% 
%\vspace{-0.03in}
\subsection{Varying Amount of Updates}
%\vspace{-0.03in}

To study the performance with different amounts of updates, we vary the size of the updates set (\aka batch size) from 10 to 10 million on UK and compare \oursys with the competitors when running \SSSP and \PR. Figure \ref{fig:vary_update} shows the speedup results of \oursys over the competitors. The speedup is more significant with a smaller batch size because \oursys utilizes the layered graph to effectively reduce the scope of global iterations. %while having less overhead of updating the layered graph. 
In \PR, if the batch size is too small, \eg 10, the effects of these updates might only be applied within subgraphs, thus the iterative computations are constrained in affected subgraphs. %tiny enough to disappear after a small number of iterations, so it will not hurt the performance of the competitor systems too much. 
However, the speedup is less significant when the batch size gets larger. This is because more updates are likely to affect more subgraphs in our system, which increases the shortcut update cost and undermines the benefits of the layered graph. However, large batches of updates will prolong the response time and lose the real-time property, so smaller batches of updates are preferable for delay-sensitive applications or online applications.

\subsection{Effect of %Skeleton Reshaping
Vertex Replication}
\label{sec:expr:reshape}
% non-mirror vs. mirror

% \begin{figure}[t]

% \begin{figure}[t]
% \vspace{-0.15in}
%     \centering
%     \includegraphics[width=1.7in]{fig/breakdown_uk.pdf}
%     \vspace{-0.05in}
%     \caption{Runtime breakdown (UK).} % The red circles are the vertices whose states are updated.
%     \label{fig:breakdown}
%     \vspace{-0.1in}
% \end{figure}

% \begin{figure}
% \vspace{-0.2in}
% 		\centering
% 		\subfloat[Skeleton size]{\label{fig:skeleton_size}\hsp
% 		\includegraphics[width=1.1in]{fig/master_mirror_rate.pdf}}
% % 		\hspace{0.1in}
% 		\subfloat[SSSP runtime]{\label{fig:master_mirror_sssp}\hsp
% 		\includegraphics[width=1.1in]{fig/master_mirror_sssp.pdf}}
% % 		\hspace{0.1in}
% 		\subfloat[PageRank runtime]{\label{fig:master_mirror_pr}\hsp
% 		\includegraphics[width=1.1in]{fig/master_mirror_pr.pdf}}
% 		\vspace{-0.05in}
% 		\caption{Effect of skeleton reshaping.}
% 		\label{fig:master-mirror}
% 		\vspace{-0.1in}
% \end{figure}

% 正如 \ref{sec:system:reshape}中所介绍的{\oursys}采用了vertex Replication 优化来进一步提高系统的压缩率。为了验证我们优化方法的有效性，我们在table\ref{table}中所展示的五个数据集上分别测试了{\oursys}带节点复制优化和不带复制优化下的压缩率，结果如图\ref{fig:master-mirror}所示，我们优化方案在其中四个数集上均提高了压缩率，例如在UK上压缩率从60%提高到了87%，提高了27%。同时我们注意到在ER数据集上，我们的优化方案没有导致压缩率提高，原因正如\ref{sec:system:reshape}中所分析的，ER中没有大量共享同一源/目标节点，所以无法通过复制节点来进一步获得收益。有趣的是即使路网图的压缩率没有被进一步优化，但是其本身的压缩率很高，如ER的压缩率已经达到了58%。

%As described in Section \ref{sec:layer:build:key-vertex}, {\oursys} can reduce the upper layer ($L_{up}$) size by %reshapes skeleton by 
%replicating vertices. % to reduce skeleton size. 
To verify the effectiveness of vertex replication proposed in Section \ref{sec:layer:build:key-vertex}, we measure the sizes of the original graphs, the original upper layers, and the reshaped upper layers as shown in Figure \ref{fig:skeleton_size}. We can see that the sizes of the original graphs are greatly reduced (by 12\%-60\%) by using the layered graph, and the sizes of the original upper layers are further reduced (by 34\%-87\%) through vertex replication. %We noticed that on the road network dataset ER, our vertex replication 
% method does not lead to a reduced upper layer size. This is because the vertices in road networks are usually not with high degrees and there is a limited number of shared source/destination vertices in the graph. But we can see that vertex replication effectively reduces the size of other graphs with the power-law property.
%为了进一步评估我们优化方案对性能带来的影响，我们在上述的五个数据集中测试了SSSP算法。实验结果如图\ref{fig:master_mirror_sssp}所示，可以发现在除了ER数据集以外的四个数据集上，{\oursys}均获得了性能的提升。其中在UK数据集上更是获得了2.5倍的加速比，在五个数据集上平均加速比为1.57。在ER上性能反而有略微的降低，这是由于我们并没有提升压缩率，但是加入的优化模块会带来一定的开销，通过图上数据也可以看出，即使没有提升压缩率，该优化模块带来的额外开销很低, 不超过3%。
% 
We also run \SSSP and \PR on the original graph with Ingress, the original upper layer with \oursys (without vertex replication), and the reshaped upper layer with \oursys. The runtime results are reported in Figure \ref{fig:master_mirror_sssp} and Figure \ref{fig:master_mirror_pr}, respectively. We can see that most of the runtime results are proportional to their graph sizes or the upper layer sizes. It is noticeable that the runtime of \SSSP on WB by {\oursys} is longer
%than that by Ingress. 
with vertex replication than without vertex replication. 
By digging into the graph property, we find that the sizes of subgraphs in WB are very large. With vertex replication, an edge update could incur multiple local recomputations on multiple subgraphs that are correlated to this updated edge. 
Therefore, if many large subgraphs are affected, the layered graph update cost for shortcut calculations is evident, which may overweigh the benefits. 
On the contrary, if the size of the affected subgraph is small, this will not impact performance as the shortcut calculation will be very fast.

\vspace{-0.05in}
\subsection{Analysis of Additional Overhead}
\label{sec:expr:offline_time}

% \eat{
\begin{figure}[h]
%\vspace{-0.3in}
		\centering
		\subfloat[Space cost]{\label{fig:space_cost}
		\includegraphics[width=0.45\linewidth]{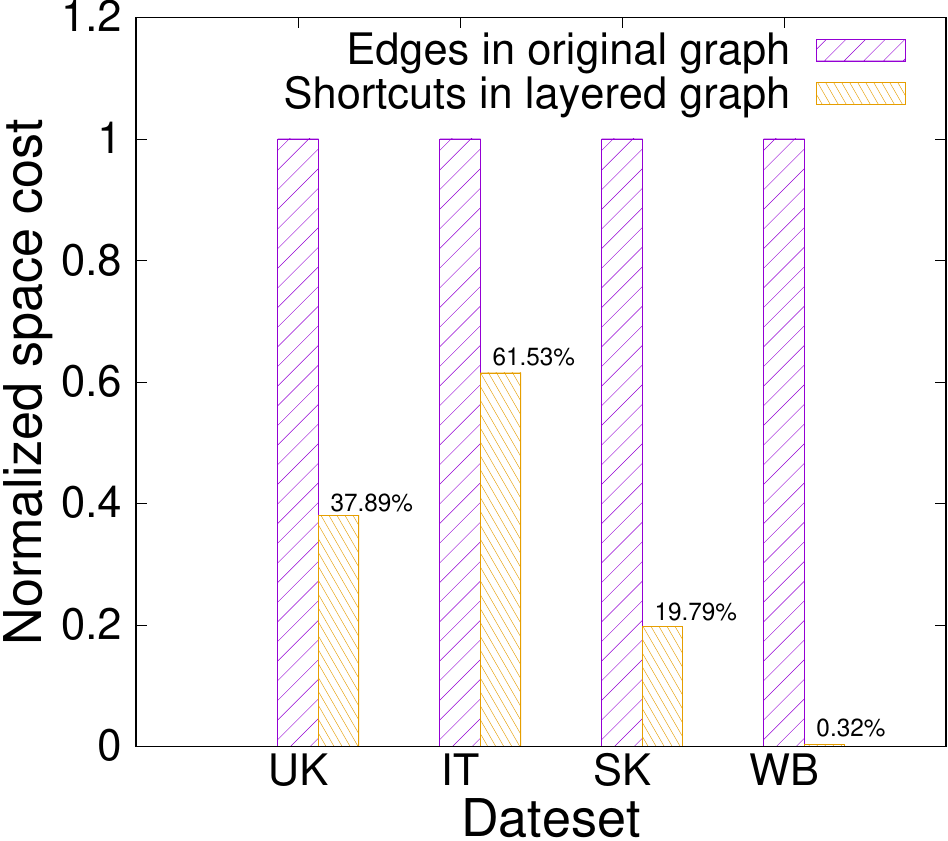}}
		\subfloat[Offline preprocessing time]{\label{fig:time_cost}

        \hspace{0.05in}
		\vspace{-0.2in}\includegraphics[width=0.45\linewidth]{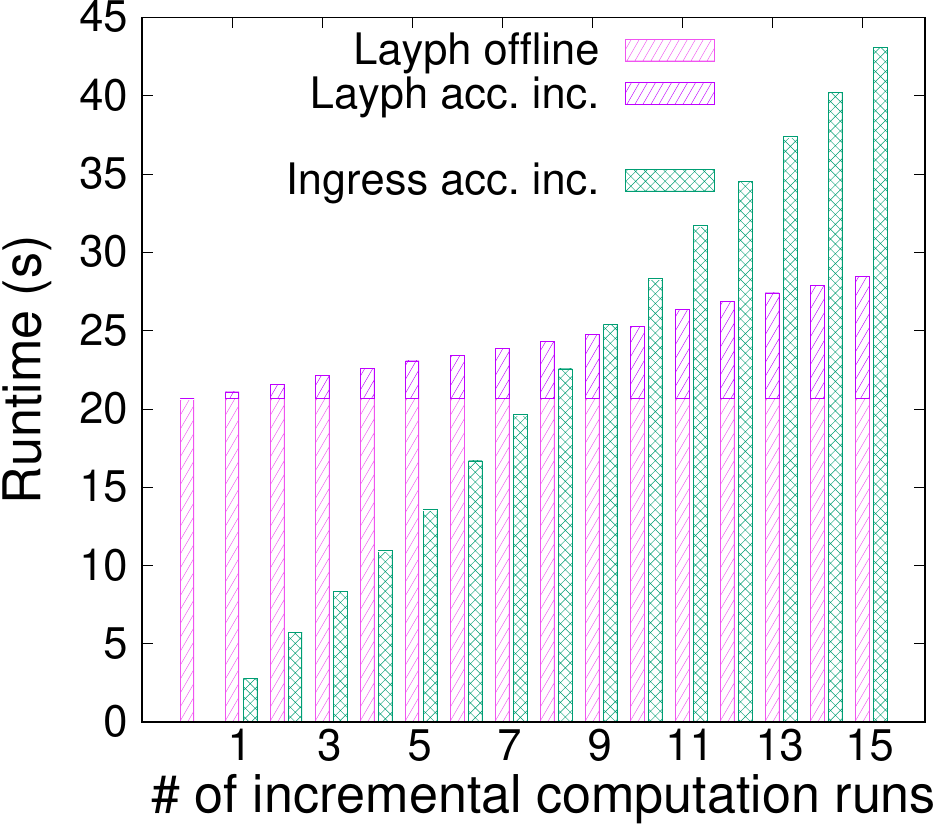}}
		\vspace{-0.05in}
        \caption{Additional space cost and offline preprocessing time.}
        \label{fig:space_time_cost}
		\vspace{-0.12in}
\end{figure}

To evaluate the effect of additional space and offline operations on \lframe, %on incremental computing systems, 
we first report the additional space cost of \lframe in Figure \ref{fig:space_cost}. %We can see that the additional space overhead is always smaller than the original graph size, which is acceptable. 
We can see that the additional space cost brought by the layered graph is 37.89\%, 61.53\%, 19.79\%, and 0.32\% of the original graph, %they always smaller than the original graphs, 
which is acceptable. 
We then report the offline preprocessing time (\oursys offline), the accumulative incremental computation time of \oursys (\oursys acc. inc.), and that of Ingress (Ingress acc. inc.) in Figure \ref{fig:time_cost} when performing SSSP on UK. It is shown that after %9 times graph updates, 
9 runs of incremental computation, the runtime of \oursys, including the offline time and the accumulative incremental computation time, becomes less than Ingress. This is because the offline operation is performed only once but can bring a significant performance gain on each incremental computation.
%As shown in Figure \ref{fig:overview}, like all incremental graph processing systems, \oursys needs to start an offline static graph analysis before incremental graph analysis. Figure \ref{fig:space_time_cost} shows the additional space and time overhead of this offline stage when \oursys runs the SSSP algorithm under different datasets. 
%The extra space of \oursys is caused by shortcuts, as shown in Figure \ref{fig:space_cost}, we can see that the number of shortcuts between the two layers and on $L_{up}$ are very small compared to the size of the original graph. Moreover, it can also be seen from the figure that the shortcuts (on $L_{up}$) used for global iterative calculation is much less than the number of shortcuts between the two layers. This also meets the expectation that $L_{up}$ is smaller, since expensive iterative computations only need to be performed on $L_{up}$.
%In addition, we also analyzed the time to build the layered graph offline in Figure \ref{fig:time_cost}, that is, the dense subgraph discovery and shortcut establishment. Although there are additional space and time consumptions offline, they are only executed once during the startup phase of the system, so the overhead is thoroughly worthwhile for continuous and efficient incremental analysis.

%% file: 7related.tex
\section{Related Work}

%DZiG, feng2020risgraph, jiang2021tripoline， JetStream, iTurboGraph

% DZiG: DZiG是一个基于稀疏感知的增量图处理系统，相对于至前的系统，它能够更捕捉计算过程中稀疏计算阶段，能够自动切换增量策略，提高性能，此外，它能够处理更高吞吐量的更新流。

% RisGraph: RisGraph是一个针对单调图计算的增量图计算系统，通过全局维护一个依赖数，在更新流来时，更具依赖树将更新流识别为安全更新和非安全更新，并行的调整依赖树得到最终结果，RisGraph具有非常高的吞吐量。

% tripoline: tripoline关注到目前针对单调图算法都是基于依赖实现的增量，例如KickStarter和RisGraph等，它们有效的条件取决于用户每次查询的源点是不变。tripline针对切换源点的情况设计了一个增量图系统，它要求算法满足三角不等式性质，然后利用基于三角形不等式的性质用旧源点的查询结果推导出一个新查询源点的初始状态，相对于默认初始化获得了一定收益。

% JetStream: JetStream是一个硬件流图加速器，它通过一个事件驱动来处理累积图算法和单调图算法。分别针对两类图算法设计了增量方案，以支持流图计算。

% iTurboGraph: 该工作提出了一个特定于邻居为中心图法分析问题的编成语言LNGA, 并设计了iTuboGraph系统，支持用于的直观编程、自动查询增量化和查询优化。

\etitle{Incremental Graph Processing Systems}. Incremental processing for evolving graphs has attracted great attention in recent years \cite{gong2021ingress,shi2016tornado,sengupta2016graphin,vora2017kickstarter,mariappan2019graphbolt,DZiG,feng2021risgraph,iTurboGraph,jiang2021tripoline,ZakianCH19,Naiad,mcsherry2013differential,VaziriV21minimal,chen2022graphfly,zhang20152,zhang2013i2mapreduce,graphinc,Wickramaarachchi15}. 
Tornado~\cite{shi2016tornado} provides loop-based incrementalization support for the fix-point graph computations.
KickStarter~\cite{vora2017kickstarter} maintains a dependency tree to memorize the critical paths for converged states and performs necessary adjustments to accommodate changes.
RisGraph~\cite{feng2021risgraph} deduces safe approximation results
upon graph updates and fixes %the approximation errors 
these results via iterative computation.
%Both KickStarter and RisGraph only maintain dependency tree from a single source (or root). If the query changes, they cannot work.
% Tripoline \cite{jiang2021tripoline} addresses this problem by supporting the switch of the query source node.
% Tripoline \cite{jiang2021tripoline} can support queries that switch the source vertices. 
GraphBolt~\cite{mariappan2019graphbolt}
keeps track of the dependencies via the memorized intermediate states among iterations and adjusts the dependencies iteration-by-iteration to achieve incremental computation.
i$^2$MapReduce \cite{zhang20152,zhang2013i2mapreduce} extends Hadoop MapReduce to support incremental iterative graph computations by memorizing the intermediate map/reduce output. 
Similarly, many other works, e.g., DZiG~\cite{DZiG} %, GraphInc~\cite{graphinc}, 
and HBSP model~\cite{Wickramaarachchi15}, also memorize
and reuse the previous computations to minimize useless re-execution. 
% JetStream \cite{rahman2021jetstream} is a hardware-accelerator for graph streams that relies on an event model. 
% Apart from the commonly used memorization method, \oursys goes further by constraining change propagation through graph sketching.
Ingress \cite{gong2021ingress} can automatically select the best memoization scheme according to algorithm property. 
% 上述系统在增量处理过程中，总是在原图上进行传播图更新的影响，这会导致大量的点和边被激活，并最终导致大量的计算。
The above systems propagate the effects of graph updates over the whole graph, which causes a large number of vertices and edges to be activated, and ultimately leads to a large number of computations.

% 硬件加速器
% JeStream[2021],  GraSU[2021], TDGraph[2022]
% Improving Streaming Graph Processing Performance using Input Knowledge [2021]
% 最近提出了许多基于新硬件的支持流图处理的解决方案。
% GraSU是一个基于FPGA的动态图系统，其利用空间相似性进行快速图更新，且其能够从任何现成的静态图形加速器轻松构建基于 FPGA 的动态图形加速器。
% JetStream是一个异步图处理加速器，它利用事件驱动的执行模型来处理累积图算法和单调图算法在流图上。
% [ABR2021]根据输入批次的度分布自适应地对输入的更新边进行排序。为了提高图计算效率，它提出了输入感知工作聚合，它根据批次间的局部性特征自适应地调节计算粒度。
% TDGraph发现受图更新影响的顶点的新状态会沿着图的拓扑结构不规则地传播，从而导致较大的冗余计算开销和不规则内存访问。其在加速器设计中提出了一种有效的拓扑驱动的增量执行方法，以实现更规则的状态传播和更好的数据局部性。

\etitle{Hardware Accelerators for Incremental Graph Processing}. A number of solutions based on new hardware to accelerate dynamic graph processing have been proposed recently \cite{wang2021grasu,rahman2021jetstream,ABR2021,TDGraph,chen2022graphfly}. GraSU \cite{wang2021grasu} provides the first FPGA-based high-throughput graph update library for dynamic graphs. It accelerates graph updates by exploiting spatial similarity. %GraSU mainly focuses on graph updates instead of incremental graph workloads, \eg SSSP and \PR.
% GraSU \cite{GraSU} is an FPGA-based dynamic graph system that exploits spatial similarity for fast graph updates, and it can easily build FPGA-based dynamic graph accelerators from any off-the-shelf static graph accelerators. 
JetStream \cite{rahman2021jetstream} extends the event-based accelerator \cite{rahman2020graphpulse} for graph workloads to support streaming updates. It works well on both accumulative and monotonic graph algorithms. \cite{ABR2021} proposes input-aware software and hardware solutions to improve the performance of incremental graph updates and processing. 
%To update the graph efficiently, \cite{ABR2021} proposes input-aware batch reordering to adaptively reorder input batches based on their degree distributions. %To improve incremental graph computation efficiency, it proposes input-aware work aggregation which adaptively modulates the computation granularity based on inter-batch locality characteristics. %proposes input-aware software and hardware solutions to improve the performance of incremental graph workloads. It optimizes the performance of streaming graph processing by knowledge-driven software and hardware co-design. And it can adaptively order the update edges according to the degree distribution of the input batch. 
TDGraph \cite{TDGraph} %finds that new states of vertices affected by graph updates propagate irregularly along the graph topology, resulting in large redundant computation overhead and irregular memory accesses. 
% It 
proposes efficient topology-driven incremental execution methods in accelerator design for more regular state propagation and better data locality. 

\etitle{Incremental Graph Algorithms}. 
There are also a number of incremental methods proposed for specific algorithms, \eg regular path queries~\cite{FanHT17},
strongly connected components~\cite{HolmLT01},
subgraph isomorphism~\cite{KimSHLHCSJ18},
k-cores~\cite{LiYM14}, graph
partitioning~\cite{FanLTXZ20,fan2019dynamic} and triangle counting~\cite{McGregorVV16}.
In contrast to these algorithm-specific methods, our \oursys framework extends Ingress \cite{gong2021ingress}, which can automatically deduce incremental
algorithms from the batch counterparts by a generic approach. It supports a series of incremental graph algorithms with different computation patterns, i.e., traversal-based (e.g., SSSP and BFS) and iteration-based (e.g., PageRank and PHP).

\etitle{Partition-based Methods}. 
Some partition-based methods have been proposed to improve graph processing, %in which they employ a block-centric framework to process graphs,  %and reduce the information flow cost, %are mainly applied in static graph processing systems, 
such as Blogel \cite{blogel}, Giraph++ \cite{Giraph++}, Grace \cite{grace}, GRAPE \cite{fan2017grape}. They employ a block-centric (or subgraph-centric) framework to process graphs and try to reduce the communication overhead between threads or processors (reducing the information flow between subgraphs). %Furthermore, some graph-level optimization can be applied on these systems. 
However, these systems are designed for static graph processing. 
Different from these existing approaches, the novelty of \lframe lies in that we propose a layered graph structure to improve the incremental graph processing for dynamic graphs, %The update messages in the incremental computation are often propagated from the vertices/edges where the graph changes, while some batch workloads may require graph processing systems to perform iterative computation on the entire graph data, such as PageRank. Therefore, 
which aims to reduce the computation caused by massive message propagation. %We avoid massive message propagation and iterative computation within dense subgraphs by building shortcuts in dense subgraphs.

\section{Conclusions}

We have proposed \oursys, a framework to accelerate incremental graph processing by layering graph. It relies on limiting global iterative computations on the original graph to a few independent small-scale local iterative computations on the lower layer, which is used to update shortcuts and upload messages, and a global computation on the upper layer graph skeleton. %where local \sr{iterative} computation is used to update shortcuts and upload messages.
%\eat{It relies on limiting expensive global iterative computations to small-scale graph skeleton (\ie the upper layer), and doing some cheap small-scale local computations to update shortcuts, upload and assign revision messages.} 
This greatly fits incremental computation for evolving graphs %that are consistently updated with a small number of local updates, 
since the number of vertices and edges involved in iterative computations is effectively limited by our layered graph. Specifically, only the dense subgraphs affected by $\Delta G$ on the lower layer and the graph skeleton on the upper layer perform iterative computations. \oursys is implemented on top of our previous work Ingress to support message-driven incremental computation. 
% It is also implemented with a set of necessary optimizations for supporting graph sketching approach. 
Our experimental study verifies that \oursys can greatly improve incremental processing performance for dynamic graphs.

%We have introduced an incrementalization framework. It incorporates four types of memoization policies that can be adopted to incrementalize various vertex-centric algorithms. We have shown that there exist sufficient conditions for deciding whether a given batch algorithm can be correctly incrementalized with a policy. We have also shown that \autoinc can largely automate the incrementalization based on this framework.

% \looseness=-1
\balance
\section*{Acknowledgment}
% \vspace{-0.05in}
  % We thank the anonymous reviewers for their insightful comments. 
  The work is supported by the National Natural Science Foundation of China (62072082, U2241212, U1811261, 62202088, 62202301), the National Social Science Foundation of China (21\&ZD124), the Fundamental Research Funds for the Central Universities (N2216012, N2216015), the Key R\&D Program of Liaoning Province (2020JH2/10100037), and a research grant from Alibaba Innovative Research (AIR) Program.